\documentclass[preprints,review,accept,moreauthors,pdftex]{mdpi}

\newcommand{\av}[1]{\left\langle #1 \right\rangle}
\newcommand{\comm}[2]{\left[ #1,#2 \right]}
\newcommand{\ket}[1]{\left| #1 \right\rangle}
\newcommand{\bra}[1]{\left\langle #1\right|}

\firstpage{1} 
\pubvolume{xx}
\issuenum{1}
\articlenumber{5}
\pubyear{2019}
\copyrightyear{2019}
\history{Received: date; Accepted: date; Published: date}

\pdfoutput=1

\Title{Phase-Coherent Dynamics of Quantum Devices With Local Interactions}

\Author{Michele Filippone $^{1}$\orcidA{}, Arthur Marguerite $^2$\orcidB{}, Karyn Le Hur $^{3}$, Gwendal F\`eve $^{4}$\orcidD{} and Christophe Mora $^{5}$\orcidC{}}

\AuthorNames{Michele Filippone, Arthur Marguerite, Karyn Le Hur, Gwendal F\`eve and Christophe Mora}

\address{%
$^{1}$ \quad Department of Quantum Matter Physics, University of Geneva\\
$^{~}$ \quad 24 Quai Ernest-Ansermet, CH-1211 Geneva, Switzerland\\
$^{2}$ \quad Department of Condensed Matter Physics, Weizmann Institute of Science, Rehovot 7610001, Israel\\
$^{3}$ \quad CPHT, CNRS, Institut Polytechnique de Paris, Route de Saclay, 91128 Palaiseau, France \\
$^{4}$ \quad 
Laboratoire de Physique de l’Ecole Normale Sup\'erieure, ENS, Universit\'e PSL, CNRS, Sorbonne Universit\'e, Universit\'e de Paris, F-75005 Paris, France  \\
$^{5}$ \quad Université de Paris, Laboratoire Matériaux et Phénomènes Quantiques, CNRS, F-75013, Paris, France }

\corres{Correspondence: michele.filippone@unige.ch}

\abstract{This review illustrates how Local Fermi Liquid (LFL) theories describe the strongly correlated and coherent low-energy dynamics of quantum dot devices. This approach consists in an effective elastic scattering theory, accounting exactly for strong correlations. 
Here, we focus on the mesoscopic capacitor and recent experiments achieving Coulomb-induced quantum state transfer. 
Extending to out-of-equilibrium regimes, aiming at triggered single electron emission, we illustrate how inelastic effects become crucial, requiring approaches beyond LFLs, shedding new light on past experimental data, by showing clear interaction effects in the dynamics of mesoscopic capacitors.  }


\begin{document}

\tableofcontents


\section{Introduction}
The manipulation of local electrostatic potentials and electron Coulomb interactions has been pivotal to control  quantized charges in solid state devices. Coulomb blockade~\cite{averin_coulomb_1986,grabert_single_1993,aleiner_quantum_2002} has revealed a formidable tool to trap and manipulate single electrons in localized regions behaving as highly tunable artificial impurities, so called quantum dots. Beyond a clear practical interest, which makes quantum dots among the promising candidates to become the building block of a quantum processor \cite{loss_quantum_1998,vandersypen_interfacing_2017,vinet_towards_2018}, hybrid~\cite{xiang_hybrid_2013} quantum dot systems became also a formidable platform to address the dynamics of many-body systems in a controlled fashion, and a comprehensive theory, which could establish the role of Coulomb interactions when these systems are strongly driven out of equilibrium, is still under construction.

\begin{figure}[t]
\centering
\includegraphics[width=\linewidth]{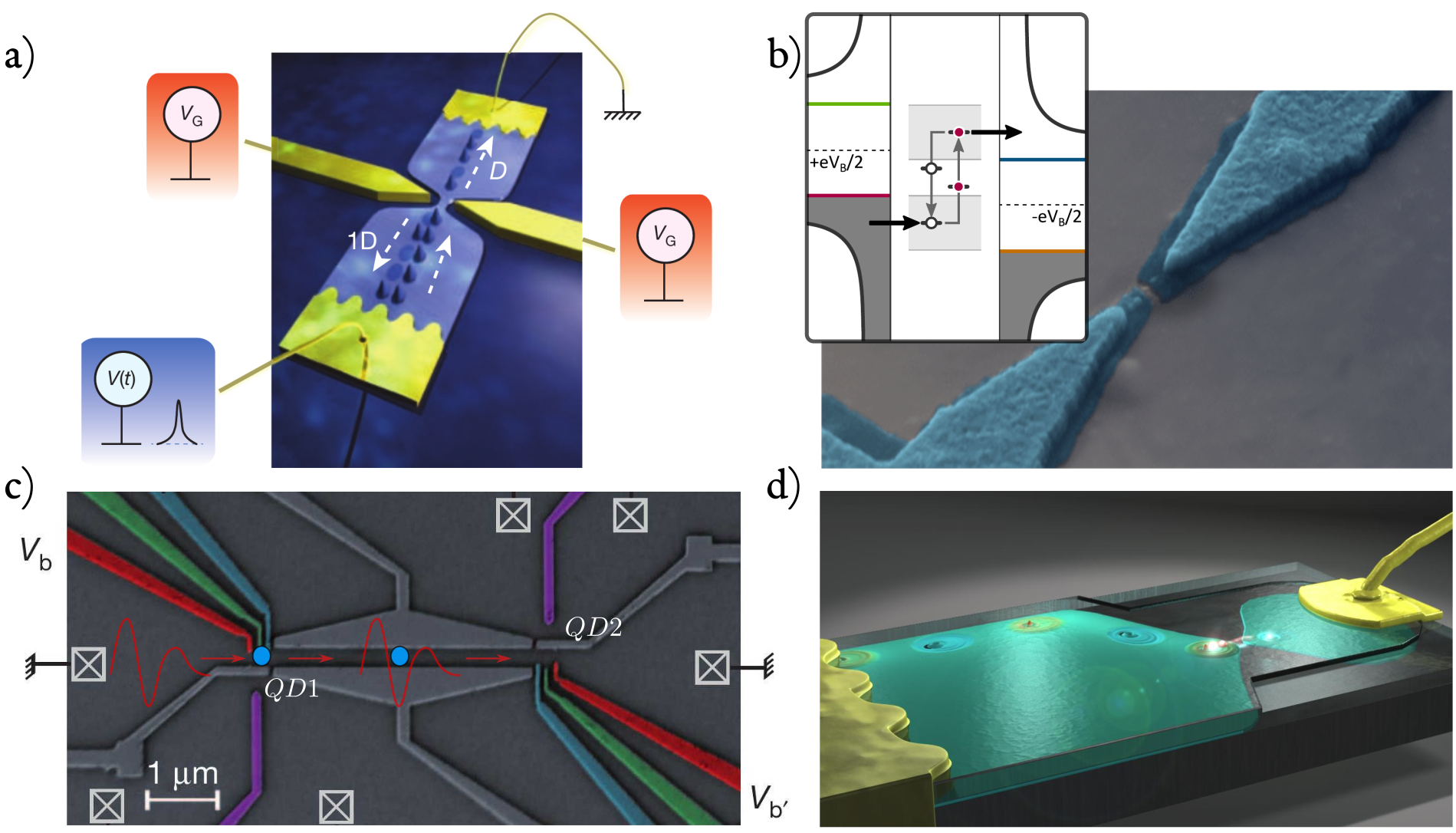}
\caption{Some recent experiments achieving real-time control of single electrons.  a) Leviton generation by a Lorentzian voltage pulse in metallic  contacts, generating a noiseless wave-packet carrying the electron charge $e$~\cite{levitov_electron_1996,ivanov_coherent_1997,keeling_minimal_2006,dubois_minimal-excitation_2013,jullien_quantum_2014,rech_minimal_2017}. This wave-packet is partitioned on a quantum point contact (QPC), whose transmission $D$ is controlled by the split-gate voltage $V_{\rm G}$. b) Single quantum level electron turnstile~\cite{van_zanten_single_2016,basko_landau-zener-stueckelberg_2017}. Two superconductors, biased by a voltage $V_{\rm B}$, are connected by a single-level quantum dot. Inset -- Working principle of the device: a gate voltage controls the orbital energy of the quantum  dot, which is filled by the left superconductor and emptied in the right one. c) Long-range single-electron transfer via a radio-frequency pulse between two distant quantum dots QD1 and QD2~\cite{hermelin_electrons_2011,bertrand_fast_2016,bertrand_injection_2016,takada_sound-driven_2019}. The electron ``surfs'' along the moving potential generated by the radio-frequency source and is transferred along a one-dimensional channel from QD1 to QD2. d) The mesoscopic capacitor~\cite{gabelli_violation_2006,gabelli_coherentrccircuit_2012,feve_-demand_2007,mahe_current_2010,parmentier_current_2012,freulon_hong-ou-mandel_2015,marguerite_decoherence_2016}, in which a gate-driven quantum dot emits single electrons through a QPC in a two-dimensional electron gas. This platform will be extensively discussed in this review.  }\label{fig:exps}
\end{figure}

Beyond theoretical interest, this question is important for ongoing experiments with mesoscopic devices aiming at the full control of single electrons out of equilibrium. Figure~\ref{fig:exps} reports  some of these experiments~\cite{levitov_electron_1996,ivanov_coherent_1997,keeling_minimal_2006,dubois_minimal-excitation_2013,jullien_quantum_2014,rech_minimal_2017,hermelin_electrons_2011,bertrand_fast_2016,bertrand_injection_2016,takada_sound-driven_2019,van_zanten_single_2016,basko_landau-zener-stueckelberg_2017}, in addition to the mesoscopic capacitor~\cite{gabelli_violation_2006,gabelli_coherentrccircuit_2012,feve_-demand_2007,mahe_current_2010,parmentier_current_2012,freulon_hong-ou-mandel_2015,marguerite_decoherence_2016}, which will be extensively discussed in this review. These experiments and significant others~\cite{leicht_generation_2011,battista_spectral_2012,fletcher_clock-controlled_2013,waldie_measurement_2015,kataoka_time--flight_2016,johnson_ultrafast_2017,roussely_unveiling_2018} have a common working principle: a fast~\cite{schoelkopf_radio-frequency_1998} time-dependent voltage drive $V(t)$, applied either on metallic or gating contacts, triggers emission of well defined electronic excitations. Remarkably,  these  experiments achieved to generate, manipulate and detect single electrons on top of a complex many-body state such as the Fermi sea.  A comprehensive review of these experiments can be found in Ref.~\cite{bauerle_coherent_2018}.

In this context, interactions are usually considered detrimental, as they are responsible for inelastic effects leading to diffusion and dephasing~\cite{akkermans_mesoscopic_2007}. Interaction screening or, alternatively, the disappearance of such inelastic effects at low driving energies or temperatures \cite{altshuler_effects_1982,pierre_dephasing_2003,huard_effect_2005,mallet_scaling_2006,saminadayar_electron_2007,niimi_quantum_2010} is thus crucial to identify single-electron long-lived excitations (quasi-particles) close to the Fermi surface. The possibility to identify such excitations, even in the presence of strong Coulomb interactions, is the core of the Fermi liquid theory of electron gases in solids~\cite{pines_theory_nodate,coleman_introduction_2015}, usually identified with the $\propto T^2$ suppression of resistivities in bulk metals. It is the validity of this theory for conventional metals which actually underpins the success of  Landauer-B\"uttiker elastic scattering  theory~\cite{landauer_electrical_1970,landauer_spatial_1988,buttiker_generalized_1985} to describe coherent transport in mesoscopic devices.

The aim of this review is to show how a similar approach can be also devised to describe transport in mesoscopic conductors involving interacting artificial quantum impurities. In these systems, electron-electron interactions are only significant in the confined and local quantum dot regions, and not in the leads for instance, therefore we use the terminology of a {\it Local Fermi Liquid} theory (LFL) in contrast to the conventional Fermi liquid approach for bulk interactions. Originally, the first LFL approach~\cite{nozieres_fermi-liquid_1974} was introduced to derive the low energy thermodynamic and transport properties of Kondo local scatterers in materials doped with magnetic impurities~\cite{hewson_kondo_1993}. In this review, we will show how LFLs provide the unifying framework to describe both elastic scattering  and strong correlation phenomena in the out-of-equilibrium dynamics of mesoscopic devices. This approach makes also clear how inelastic effects, induced by Coulomb interactions, become visible and unavoidable as soon as such systems are strongly driven out of equilibrium. We will discuss how extensions of LFLs and related approaches describe such regimes as well.  

As a paradigmatic example, we will focus on recent experiments showing electron transfer with Coulomb interactions~\cite{duprez_transmitting_2019}, see Fig.~\ref{fig:pierre}, and, in more detail, on the mesoscopic capacitor~\cite{gabelli_violation_2006,gabelli_coherentrccircuit_2012,feve_-demand_2007,mahe_current_2010,parmentier_current_2012,freulon_hong-ou-mandel_2015,marguerite_decoherence_2016}, see Fig.~\ref{fig:emission}. The mesoscopic capacitor does not support  DC transport, and it makes possible the direct investigation and control of the coherent dynamics of charge carriers. The LFL description of such devices entails the seminal results relying on  self-consistent elastic scattering approaches by B\"uttiker and collaborators~\cite{buttiker_dynamic_1993,buttiker_mesoscopic_1993,pretre_dynamic_1996,nigg_mesoscopic_2006,buttiker_mesoscopic_2006,buttiker_role_2009},  but it  also allows to describe effects induced by strong Coulomb correlations, which remain nevertheless elastic and coherent. The intuition provided by the LFL approach is a powerful lens through which explore various out-of-equilibrium phenomena, which are coherent in nature but are governed by Coulomb interactions. As an example, we will show how a bold treatment of Coulomb interaction unveils originally overlooked strong dynamical effects, triggered by interactions, in  past experimental measurements  showing fractionalization effects in out-of-equilibrium charge emission from a driven mesoscopic capacitor~\cite{freulon_hong-ou-mandel_2015}.

This review is structured as follows. In Section~\ref{sec:coherence}, we give a simple example showing how  Coulomb interactions trigger phase-coherent electron state transfer in experiments as those reported in Ref.~\cite{duprez_transmitting_2019}, Fig.~\ref{fig:pierre}. Section~\ref{part:overview} discusses how the effective LFL approach~\cite{mora_theory_2009,mora_fermi-liquid_2009,mora_fermi-liquid_2015,oguri_higher-order_2018,oguri_higher-order_2018-1,oguri_higher-order_2018-2,filippone_at_2018,teratani_fermi_2020} provides the unified framework describing such coherent phenomena. In Section~\ref{sec:meso}, we consider the study of the low-energy dynamics of the mesoscopic capacitor, in which the LFL approach has been fruitfully applied~\cite{mora_universal_2010,filippone_fermi_2012,filippone_giant_2011,filippone_admittance_2013,dutt_strongly_2013}, showing novel quantum coherent effects. Section~\ref{sec:ooe} extends the LFL approach out of equilibrium and describes signatures of interactions in measurements of strongly driven mesoscopic capacitors~\cite{freulon_hong-ou-mandel_2015}.


\section{Phase-coherence in quantum devices with local interactions}\label{sec:coherence}

\begin{figure}[t]
\begin{center}
\includegraphics[width=\linewidth]{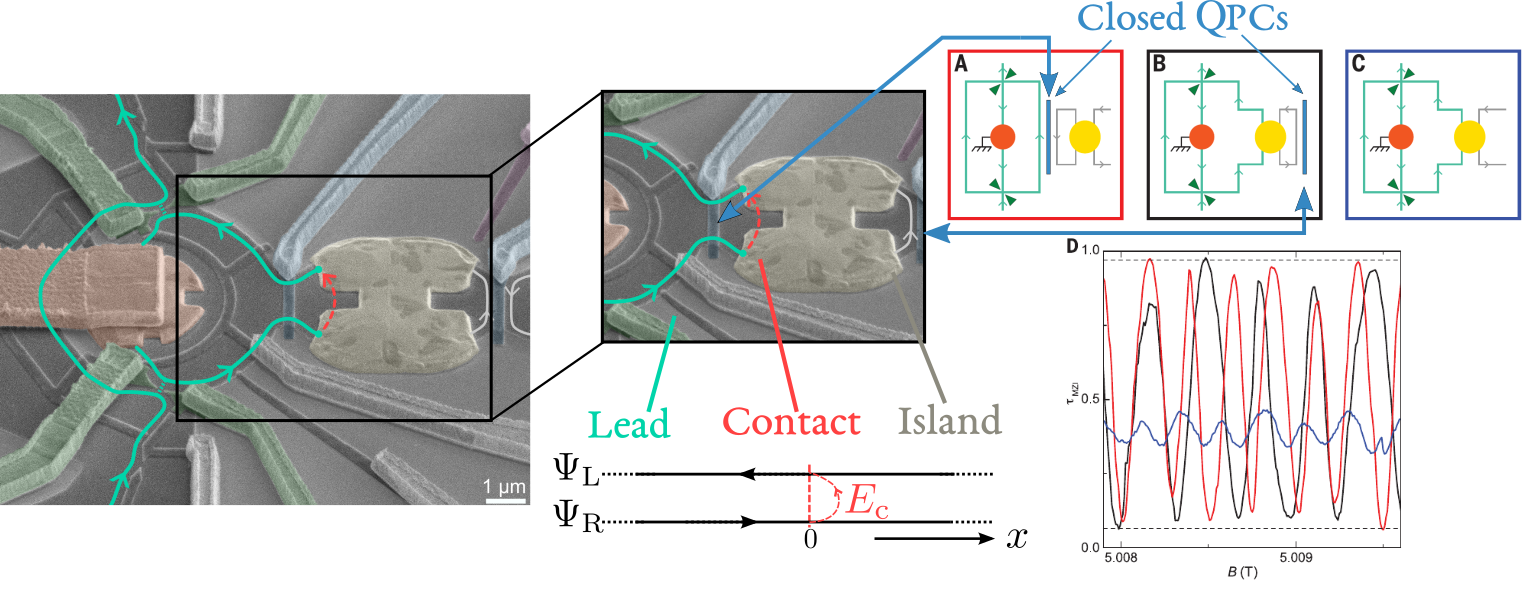}
\caption{{Left -- }Mach-Zehnder interferometer with a floating metallic island (colored in yellow)~\cite{duprez_transmitting_2019}. The green lines denote chiral quantum Hall edge states, which can enter the floating island  passing through a gate-tunable QPC (in blue). An additional QPC separates the floating island from an additional reservoir on its right. Center -- The floating island is described by two infinite counter-propagating edges, exchanging electrons coherently thanks to the charging energy $E_{\rm c}$ of the island (red arrow). Right -- Mach-Zehnder visibility of the device as a function of magnetic field $B$. Oscillation of this quantity as function of $B$ signal quantum coherent interference between two paths encircling an Aharonov-Bohm flux. In the situation sketched in box A, the first QPC is closed and the interferometer is disconnected from the island and visibility oscillations are observed, as expected (red line).  Remarkably, the oscillations persist (black line) in the situation sketched in box B, where the leftmost QPC is open and one edge channel enters the floating island. The visibility oscillations  are only suppressed in the situation sketched in box C, where the rightmost QPC is also open and the island is connected to a further reservoir (blue line). }\label{fig:pierre}
\end{center}
\end{figure}

To illustrate the restoration of phase coherence at low temperatures in the presence of interactions, we consider two counter-propagating edge states entering a metallic quantum dot, or cavity. Such system was recently realized  as  a constitutive element of the Mach-Zender interferometer of Ref.~\cite{duprez_transmitting_2019}, reported in Fig.~\ref{fig:pierre}. In that experiment, the observation of fully preserved Mach-Zehnder oscillations, in a system in which a quantum Hall edge state penetrates a metallic floating island,  demonstrates such, interaction-induced, restored  phase coherence~\cite{clerk_interaction-induced_2001,idrisov_dephasing_2018}.

The dominant electron-electron interactions in the cavity have the form of a charging energy \cite{averin_coulomb_1986,grabert_single_1993,aleiner_quantum_2002}
\begin{equation}\label{eq:charging}
 \mathcal  H_{\rm c} =E_{\rm c} [N-{\cal N}_{\rm g}(t)]^2\,,
\end{equation}
in which $N$ is the number of electrons in the island, $C_{\rm g}$ the geometric capacitance, and ${\cal N}_{\rm g} = C_{\rm g} V_{\rm g}(t)/e$  the dimensionless gate voltage, which corresponds to the number of charges that would set in the cavity if $N$ was a classical, non-quantized, quantity. We also define the  charging energy $E_{\rm c}=e^2/2C_{\rm g}$: the energy cost required to add one electron in the isolated cavity. For the present discussion, we neglect the time-dependence of the gate-potential $V_{\rm g}$, which will be reintroduced to describe driven settings.  In the linear-dispersion approximation, the right/left-moving fermions $\Psi_{\rm R,L}$ in Fig.~\ref{fig:pierre}, moving with Fermi velocity $v_F$, are described by the Hamiltonian
\begin{equation}\label{eq:chiral}
\mathcal H_{\rm kin}=v_F\hbar\sum_{\alpha=\rm R/L}\int_{-\infty}^\infty dx\, \Psi^\dagger_\alpha(x)(-i\alpha\partial_x)\Psi_\alpha(x)\,,
\end{equation}
with the sign $\alpha=+/-$ multiplying the $\partial_x$ operator for right- and left-movers respectively. The floating island occupies the semi-infinite one-dimensional space located at $x>0$ with the corresponding charge $N=\sum_\alpha\int_0^\infty dx \Psi^\dagger_\alpha(x)\Psi_\alpha(x)$. It is important to stress that  the model~(\ref{eq:charging}-\ref{eq:chiral}) is general and effective in describing different quantum dot devices. It was originally suggested by Matveev to describe  quantum dots connected to leads through a single conduction channel~\cite{matveev_coulomb_1995} and it equally describes the mesoscopic capacitor, see Sections~\ref{sec:meso} and~\ref{sec:ooe}.

The model~(\ref{eq:charging}-\ref{eq:chiral}) characterizes an open-dot limit in the sense that it does not contain an explicit backscattering term coupling the $L$ and $R$ channels. It can be solved exactly relying on the bosonization formalism \cite{haldane_luttinger_1981,haldane_effective_1981, delft_bosonization_1998,giamarchi_quantum_2004}, which, in this specific case,  maps interacting fermions onto non-interacting bosons \cite{matveev_coulomb_1995,aleiner_mesoscopic_1998,brouwer_nonequilibrium_2005}. Using this mapping, one can show that the charging energy $E_{\rm c}$ perfectly converts, far from the contact, right-movers into left-movers. This fact is made apparent by the  ``reflection'' Green function~$G_{LR}$~\cite{aleiner_mesoscopic_1998}~\footnote{$\mathcal T_\tau$ is the usual time-ordering operator defined as $\mathcal T_\tau A(\tau)B(\tau')=\theta(\tau-\tau')A(\tau)B(\tau')\pm\theta(\tau'-\tau)B(\tau')A(\tau)$, in which the sign $+/-$ is chosen depending on the bosonic/fermionic statistics of the operators $A$ and $B$~\cite{altland_condensed_2006} and  $\theta(\tau)$ is the Heaviside step function.}:
\begin{align}\label{eq:GLR}
G_{\rm LR}=\left\langle \mathcal T_\tau\Psi^\dagger_{\rm L}(x,\tau)\Psi_{\rm R}(x',0)\right\rangle \simeq e^{-i2 \pi\mathcal N_{\rm g}}\frac{T/2v_F}{\sin\left[\frac {\pi T}\hbar\left(\tau+i(x+x')-i \frac{\pi\hbar}{E_{\rm c}e^{\mathcal C}}\right)\right]}\,,
\end{align}
which we consider at finite temperature $T$.  As first noted by Aleiner and Glazman~\cite{aleiner_mesoscopic_1998}, the form of $G_{LR}$ at large (imaginary) time $\tau$ corresponds to the elastic reflection of electrons incident on the dot, with a well-defined scattering phase $\pi \mathcal N_g$. The correlation function~\eqref{eq:GLR} would be identical if the interacting dot was replaced with
a non-interacting wire of  length $v_F \pi\hbar/E_{\rm c}\gamma$ (with $\ln\gamma=\mathcal C\simeq0.5772$  is Euler's  constant), imprinting a phase $\pi \mathcal N_g$ when electrons are back-reflected at the end of the wire~\cite{clerk_aspects_2011}. 

The physical picture behind Eq.~\eqref{eq:GLR} is that an electron entering and thereby charging the island violates energy conservation at low temperature and must escape on a time scale $\hbar/E_c$ fixed by the uncertainty principle. The release of this incoming electron can happen either elastically, in which case the electron keeps its energy, or inelastically via the excitation of electron-hole pairs. As we discuss in Sec.~\ref{sec:fermi},  inelastic processes are suppressed by the phase space factor $(\varepsilon/E_c)^2$, $\varepsilon$ being the energy of the incoming electron, and they die out at low energy or large distance (time), reestablishing purely elastic scattering despite a nominally strong interaction.

\begin{figure}[t]
\begin{center}
\includegraphics[width=\linewidth]{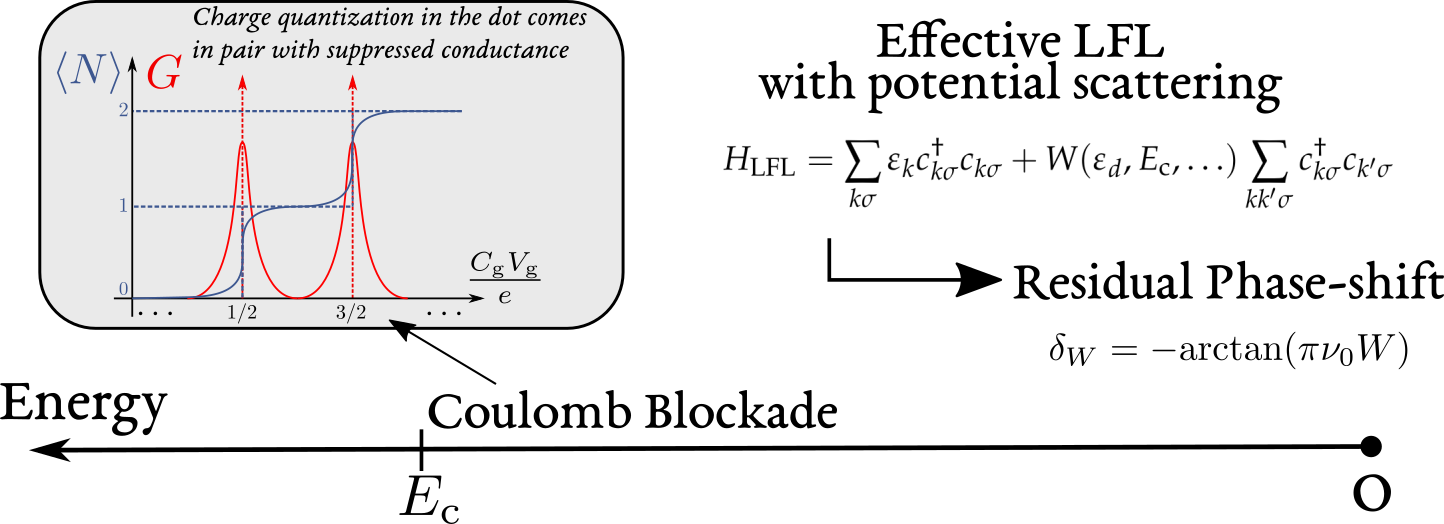}
\caption{Coulomb blockade and emergent LFL behavior. When the typical energy of the system (temperature, bias-voltage, \ldots) is smaller than the charging energy $E_{\rm c}$, charge quantization $Q=e\av N$ in the dot suppresses the conductance $G$ of the system. Degeneracy between different charge occupations lead to conductance peaks, which become larger the stronger the tunnel exchange of electrons with the leads. Conductance peaks and charge quantization disappear in the open-dot limit. For any tunneling strength, the dot behaves as an elastic scatterer described by the LFL theory~\eqref{eq:potscatt}, with potential scattering of strength $W$, inducing a phase-shift $\delta_W$ on lead electrons set by the dot occupation $\langle N\rangle$.   }\label{fig:CBM}
\end{center}
\end{figure}

Equation~\eqref{eq:GLR} is thus a remarkable example of how interactions trigger coherent effects in mesoscopic devices. It has been derived here for an open dot, a specific limit in which the charge quantization of the island is fully suppressed. However, the restoration of phase coherence at low energy is more general and applies for an arbitrary lead-island transmission, in particular in the tunneling limit where the charge states of the quantum dot are well quantized~\cite{averin_coulomb_1986,grabert_single_1993,aleiner_quantum_2002}. This quantization is known to induce Coulomb blockade in the conductance of the device, see Fig.~\ref{fig:CBM}. Nevertheless, a Coulomb blockaded dot acts at low energy as an elastic scatterer imprinting a phase $\delta$~\cite{schuster_phase_1997,edlbauer_non-universal_2017} related to its average occupation $\av N$ via the Friedel sum rule, see Sec.~\ref{sec:friedel}. For weak transmissions, $\av N$ strongly deviates from the classical value  $\mathcal N_{\rm g}$. These features constitute the main characteristics of the local Fermi liquid picture detailed in the forthcoming sections.


\section{What are local Fermi liquids and why are they important to understand quantum-dot devices?}\label{part:overview}

In this Section, we introduce the local Fermi liquid theory and discuss its application to quantum transport devices. The general system considered in this paper is a central interacting region, such as a quantum dot, connected to leads described as non-interacting electronic reservoirs. The Hamiltonian takes the general form     
\begin{equation}\label{eq:gen}
\mathcal   H = \mathcal H_{\rm res}+ \mathcal H_{ \rm res-dot}+\mathcal H_{\rm dot }+\mathcal H_{\rm c}\,.
\end{equation}
The first term describes the lead reservoir, which could be either a normal metal \cite{hermelin_electrons_2011}, a chiral edge state in the quantum Hall regime \cite{feve_-demand_2007,fletcher_clock-controlled_2013}, a superconductor \cite{van_zanten_single_2016}. In the case of a normal metal, it is given by
\begin{equation}\label{eq:hres}
\mathcal H_{\rm res}=\sum_k\varepsilon_kc^\dagger_kc_k\,,
\end{equation}
in which $c_k$ annihilates a fermion in the eigenstate state $k$ of energy $\varepsilon_k$ in the reservoir. For instance, in  Fig.~\ref{fig:pierre}, the reservoir modes correspond to the $x<0$ components of the operators $\Psi_{\rm R/L}$. The field $\Psi_{\rm res}(x)=\theta(-x)\Psi_{\rm R}(x)+\theta(x) \Psi_{\rm L}(-x)$, with $\theta(x)$ the Heaviside step function, {\it unfolds} the chiral field onto the interval $x\in[-\infty,\infty]$  and its Fourier transform $c_k=\int_{-\infty}^\infty dxe^{-ikx} \Psi_{\rm res}(x)$ recovers Eq.~\eqref{eq:hres} from Eq.~\eqref{eq:chiral}, with $\varepsilon_k =\hbar v_Fk$. 

The single particle physics of the quantum dot is described instead by 
\begin{equation}\label{eq:cbm}
\mathcal H_{\rm dot}=\sum_l(\varepsilon_d+\varepsilon_l)n_l
\end{equation}
in which $n_l=d^\dagger_ld_l$ counts the occupation of the orbital level  $l$ and $d_l$ annihilates fermions in that state. The spectrum can be either discrete for a finite size quantum dot or dense for a metallic dot as in the case of Fig.~\ref{fig:pierre}.  We also introduced the orbital energy $\varepsilon_d$ as a reference. $\mathcal H_{\rm res-dot}$ describes the exchange of electrons between dot and reservoir. It has generally the form of a tunneling Hamiltonian
\begin{equation}\label{eq:hamtunnel}
\mathcal H_{\rm res-dot}=t\sum_{k,l}\Big[c^\dagger_k d_l+d^\dagger_lc_k\Big]\,,
\end{equation}
in which we neglect, for simplicity, any $k$ dependence of the tunneling amplitude $t$. The charging energy $\mathcal H_{\rm c}$ is given in Eq.~\eqref{eq:charging} with  the dot occupation operator $N=\sum_ln_l$. 

\begin{figure}[t]
\begin{center}
\includegraphics[width=\linewidth]{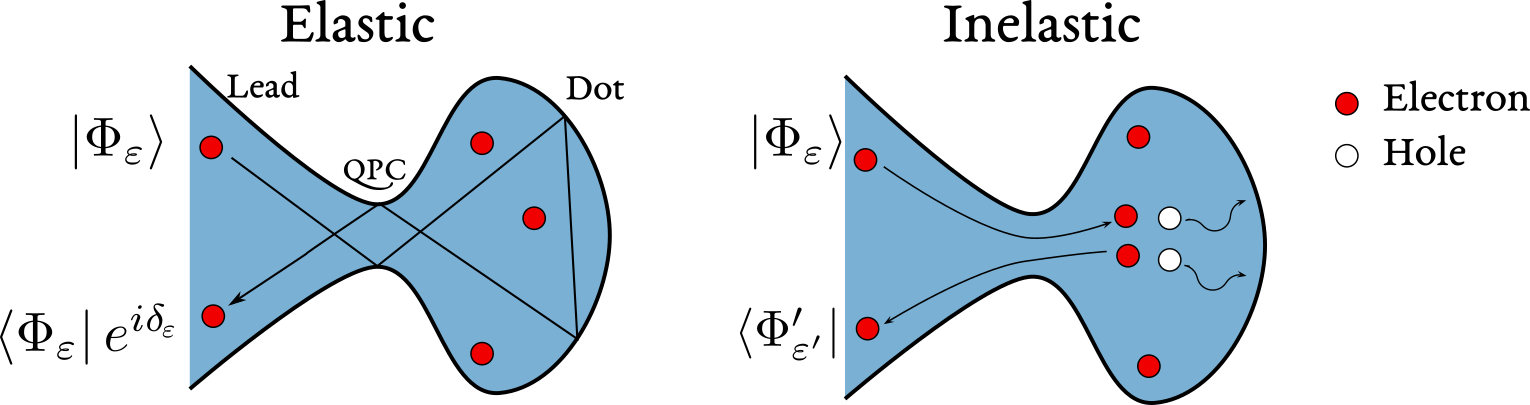}
\caption{Difference between elastic (left) and inelastic (right) events for electrons scattering on a quantum dot. In the elastic case, electrons do not change energy $\varepsilon$. The wave function is preserved and the only residual effect of scattering is a phase-shift $\delta_\varepsilon$. In the inelastic case, many-body interactions trigger the creation of particle-hole pairs. Outgoing electrons are then emitted in a  state $\left|\Phi'_{\varepsilon'}\right \rangle$ of energy $\varepsilon'$ different from the initial $\varepsilon$ and phase coherence is gradually lost.  }\label{fig:elastic}
\end{center}
\end{figure}

Without any approximation, deriving the out-of-equilibrium dynamics of interacting models such as Eq.~\eqref{eq:gen} is a formidable task. The presence of  local interactions leads to inelastic scattering events, creating particle-hole pairs when electron scatter on the dot, see Fig.~\ref{fig:elastic}. From a technical point of view, such processes are difficult to handle and, even if these difficulties are overcome, one has to identify the dominant physical mechanisms governing the charge dynamics.  In our discussion, interactions are usually controlled by the charging energy $E_{\rm c}$, which cannot be treated perturbatively in Coulomb blockade regimes. The possibility to rely on Wick's theorem~\cite{wick_evaluation_1950}, when performing perturbative calculations in the exchange term $\mathcal H_{\rm res-dot}$, is also denied. Thus, one has to look for a more efficient theoretical approach.

\subsection{The local Fermi liquid }\label{sec:fermi}		

The local Fermi liquid approach is justified by the physical picture already presented in Sec.~\ref{sec:coherence}, namely that an incoming reservoir electron  with an energy much smaller than the charging energy of the quantum dot is effectively scattered in a purely elastic way~\cite{aleiner_mesoscopic_1998}. At temperatures well below the charging energy $E_{\rm c}$, energy conservation prevents any permanent change in the charge of the quantum dot and each electron entering the dot must be compensated by an electron leaving it within the (short) time  $\hbar/{E_{\rm c}}$ fixed by the uncertainty principle. The electron escape can occur via elastic or inelastic processes, sketched in Fig. \ref{fig:elastic}, depending on whether the electron energy is preserved or not. Inelastic processes cause decoherence and call for a many-body approach to be properly evaluated.

At low energy $\varepsilon$ of the incoming electron, the inelastic processes are typically suppressed by the ratio $(\varepsilon/E_{\rm FL})^2$~~\cite{pines_theory_nodate,coleman_introduction_2015}. $E_{\rm FL}$ is a  Fermi liquid energy scale, typically of the order of the charging energy $E_{\rm c}$. Nevertheless, in the presence of spin-fluctuations, the emergence of strong Kondo correlations, to be discussed in  Sec.~\ref{sec:aim}, can sensibly reduce $E_{\rm FL}$  down to the Kondo energy scale $T_{\rm K}$, Eq.~\eqref{eq:tkoverview}. Therefore, in the  $\varepsilon\rightarrow 0$ limit, inelastic processes are ignored and the scattering is purely elastic. It is described within a single-particle formalism where the scattering by the quantum dot imprints a phase shift $\delta_W$ to the outgoing electronic wavefunctions. This phase shift alone incorporates all interaction and correlation effects.

The simplest model entailing these features is a free Fermi gas in which a delta barrier located at $x=0$ (the entrance of the dot) scatters elastically quasi-particles. In the language of second quantization, the delta barrier is described by  the electron operator $\Psi^\dagger_{\rm res}(x=0)\Psi_{\rm res}(x=0)$, and its strength $W$ has to depend on the parameters of the parent model (such as the orbital energy $\varepsilon_d$, the charging energy $E_{\rm c}$, etc). Switching to momentum space, such model is a free Fermi gas with a potential scattering term
\begin{equation}\label{eq:potscatt}
\mathcal H_{\rm LFL}=\sum_{k\sigma}\varepsilon_k c^\dagger_{k\sigma}c_{k\sigma}+W(\varepsilon_d,E_{\rm c},\ldots)\sum_{kk'\sigma}c^\dagger_{k\sigma}c_{k'\sigma}+\mathcal O\left(\frac\varepsilon{E_{\rm FL}}\right)^2\,.
\end{equation}
where the scattering potential leads, as shown in Appendix~\ref{app:scattering}, to the quasi-particle phase shift
\begin{equation}\label{eq:deltapotscatt}
\delta_W=-\arctan\left(\pi\nu_0W\right)\,,
\end{equation}
in which $\nu_0$ is the density of states of the lead electrons  at the Fermi energy, see also Eq.~\eqref{eq:gamma} in Appendix~\ref{app:scattering} for its rigorous definition. In this {\it Local Fermi Liquid} Hamiltonian, $\sigma$ labels either a spin polarization or a channel. The number of channels in the lead can be controlled by the opening of a quantum point contact~\cite{buttiker_quantized_1990}.
The potential strength $W$  can be cumbersome to compute, but it is nevertheless related to the occupancy of the quantum dot via the Friedel sum rule, as explained in the next Sec.~\ref{sec:friedel}.

The simplicity of the local Fermi liquid Hamiltonian~\eqref{eq:potscatt} makes it powerful to evaluate low energy properties. Being non-interacting, it also includes the restoration of phase coherence in the scattering of electrons seen in Sec.~\ref{sec:coherence}. An important assumption that we made is that the system exhibits a Fermi liquid ground state, or Fermi liquid fixed point in the language of renormalization group. Non-Fermi liquid fixed points exist and cannot be described by such Hamiltonian~\cite{mora_probing_2013}, but they are generally fine-tuned and unstable with respect to perturbations. Also, Eq.~\eqref{eq:potscatt} is not applicable to genuine out-of-equilibrium regimes, when the perturbations are too strong or vary too fast with respect to the Fermi liquid energy scale $E_{\rm FL}$ (typically of the order of the charging energy $E_{\rm c}$).

		 
\subsection{The role of the Friedel sum rule in Local Fermi Liquid theories}\label{sec:friedel}

In the process of electron backscattering by the quantum dot, or by the interacting central region, the phase shift relates the incoming and outgoing electronic wavefunctions $\Psi_{\rm L}(0^-)=e^{2i\delta_\varepsilon}\Psi_{\rm R}(0^-)$, see Appendix~\ref{app:scattering} for an explicit illustration on the resonant level non-interacting model. The Friedel sum rule~\cite{friedel_electrical_1956} establishes the relation between the average charge occupation of the dot $\av {N}$ and this phase-shift $\delta$. Its form in the case of $M$ conducting channels reads
\begin{equation}\label{eq:friedel}
\av{N}=\frac 1\pi\sum_{\sigma=1}^M \delta_\sigma \,.
\end{equation}
The Friedel sum rule has been proven rigorously for interacting models~\cite{langreth_friedel_1966,rontani_friedel_2006}. It is valid as long the ground state has a Fermi liquid character. Physically, it can be understood in the following way: the derivative of the phase shift $\delta_\varepsilon$ with respect to energy defines (up to $\hbar$) the Wigner-Smith scattering time~\cite{ringel_delayed_2008}, see Eq.~\eqref{eq:wigner}, that is the time delay experienced by a scattered electron. In the presence of a continuous flow of electrons, a time delay implies that some fraction of the electronic charge has been (pumped) deposited in the quantum dot~\cite{buttiker_dynamic_1993,brouwer1998}. Therefore the phase shift amounts to a left-over charge and it does not matter that electrons are interacting on the quantum dot as long as they are not in the leads - which is the essence of the local Fermi liquid approach.

The Friedel sum rule~\eqref{eq:friedel} combined with Eq.~\eqref{eq:deltapotscatt} relates the dot occupancy to the potential scattering strength. For the single-channel case ($M=1$), one finds 
\begin{equation}\label{eq:npotscatt}
\av {N}=-\frac1\pi\arctan\left(\pi\nu_0W\right)\,.
\end{equation}
This is an important result because the dot occupation $\av N$ is a thermodynamic quantity, which can be also accessed in interacting models,  allowing us to address quantitatively the close-to-equilibrium dynamics of driven settings, as we will discuss in Section~\ref{sec:quasistatic}. 

We emphasize that the local Fermi liquid approach of Eq.~\eqref{eq:potscatt} can be extended to perturbatively include inelastic scattering and higher-order energy corrections, and relate these terms to thermodynamic observables. This program has been realized in detail for the Anderson and Kondo models~\cite{mora_theory_2009,mora_fermi-liquid_2009,mora_fermi-liquid_2015,oguri_higher-order_2018,oguri_higher-order_2018-1,oguri_higher-order_2018-2,filippone_at_2018,teratani_fermi_2020,karki_two-color_2018}.


\subsection{Derivation of the LFL theory in the Coulomb blockade and Anderson model}\label{sec:specific}
We show now how the effective theory~\eqref{eq:potscatt} can be explicitly derived from realistic models describing Coulomb blockaded quantum dot devices~\cite{filippone_fermi_2012}.  
We focus on the Coulomb blockade model (CBM)~\cite{grabert_single_1993,aleiner_quantum_2002}
\begin{equation}\label{eq:cbm}
\mathcal H_{\rm CBM}=\sum_{k}\varepsilon_{k}c^\dagger_kc_k+t\sum_{k,l}\Big[c^\dagger_k d_l+d^\dagger_l c_k\Big]+\sum_l(\varepsilon_d+\varepsilon_l) d^\dagger_ld_l+E_{\rm c}\Big(N-\mathcal N_{\rm g}\Big)^2\,.
\end{equation}
and the Anderson impurity model (AIM), which, in its standard form\footnote{Adding $-eV_{\rm g}N+E_{\rm c}\mathcal N_{\rm g}^2$ to the AIM  and for $U=E_{\rm c}$, the charging energy~\eqref{eq:charging} becomes apparent in Eq.~\eqref{eq:aim}, as in  Eqs.~(\ref{eq:gen}-\ref{eq:cbm}).}, reads~\cite{anderson_localized_1961,pustilnik_kondo_2004}
\begin{equation}\label{eq:aim}
\mathcal H_{\rm AIM}=\sum_{k,\sigma}\varepsilon_{k,\sigma}c^\dagger_{k,\sigma}c_{k,\sigma}+t\sum_{k,\sigma}\Big[c^\dagger_{k,\sigma} d_\sigma+d^\dagger_\sigma c_{k,\sigma}\Big]+\varepsilon_d\sum_\sigma d^\dagger_\sigma d_\sigma+Un_\uparrow n_\downarrow\,,
\end{equation}
The CBM coincides with the Hamiltonian~\eqref{eq:gen} and describes the mesoscopic capacitor in the Quantum Hall regime: a reservoir of spinless fermions $c_k$ of momentum $k$ is tunnel coupled to an island  with discrete spectrum $\varepsilon_l$. The AIM includes the spin degree of freedom and considers a single interacting level in the quantum dot. This model  encompasses Kondo correlated regimes~\cite{schrieffer_relation_1966,hewson_kondo_1993} and  well describes experiments~\cite{goldhaber-gordon_kondo_1998-1,bruhat_scaling_2018}.

To derive the LFL Hamiltonian~\eqref{eq:potscatt}, we rely on the Schrieffer-Wolff (SW) transformation \cite{bruus_many-body_2004,coleman_introduction_2015}, first devised to map the AIM~\cite{schrieffer_relation_1966} onto the Coqblin-Schrieffer model \cite{coqblin_exchange_1969}, and that we extend here to the CBM.   Far from the charge degeneracy points, in the $t\ll E_{\rm c}$ limit, the ground-state charge configuration $n=\av N$ is fixed by the gate potential $V_{\rm g}$ and fluctuations to $n\pm 1$  require  energies of order  $E_{\rm c}$. For temperatures much lower than $E_{\rm c}$, the charge degree of freedom of the quantum dot is frozen, acting but virtually on the low energy behavior of the system. The SW transformation is a controlled procedure to diagonalize perturbatively in $t$ the Hamiltonian. The Hamiltonian is separated in two parts $\mathcal H=\mathcal H_0+\mathcal H_{\rm res-dot}$, in which $\mathcal H_0$ is diagonal in the charge sectors labeled by the eigenvalues $n$ of the dot occupation $N$, which are mixed by the tunneling Hamiltonian $\mathcal H_{\rm res-dot}$, involving the tunneling amplitude $t$. The perturbative diagonalization consists in finding the  Hermitian operator  $S$ (of order $t$) generating the unitary  $U=e^{iS}$ rotating   the Hamiltonian in the diagonal form $\mathcal H'=U^\dagger\mathcal HU$. 
To leading order in $S$ we find 
\begin{equation}\label{eq:partialsw}
\mathcal H'=\mathcal H_0+\mathcal H_{\rm res-dot}+i\comm{S}{\mathcal H_0}+O\left(t^2\right).
\end{equation}
This Hamiltonian is  block diagonal if the condition  
\begin{equation}\label{eq:condsw}
i\mathcal H_{\rm res-dot}=\comm{S}{\mathcal H_0}.
\end{equation}
is fulfilled and Eq.~\eqref{eq:partialsw} becomes
\begin{equation}\label{eq:swt2}
\mathcal H'=\mathcal H_0+\frac i2\comm{S}{\mathcal H_{\rm res-dot}},
\end{equation}
which is then projected on separated charge sectors.


\subsubsection{Coulomb blockade model}

To derive the effective low-energy form of the CBM model, it is useful, following Grabert~\cite{grabert_charge_1994,grabert_rounding_1994}, to decouple the charge occupancy of the dot from the fermionic degree of freedom of the electrons. This is achieved by adding the operator $\hat n=\sum_n |n\rangle\langle n|$, measuring to the dot occupation number. The fermionic operators $d_l$ in Eq.~\eqref{eq:cbm} are replaced by new operators describing a non-interacting electron gas in the dot. The Hamiltonian~\eqref{eq:cbm} acquires then the form
\begin{equation}
\mathcal H_{\rm CBM}=\sum_{k}\varepsilon_{k}c^\dagger_kc_k+t\sum_{n,k,l}\left[d^\dagger_lc_k\ket{n+1}\bra n+\mbox{h.c.}\right]+\sum_l\varepsilon_l d^\dagger_ld_l+\varepsilon_d\hat n+E_{\rm c}\Big(\hat n-\mathcal N_{\rm g}\Big)^2\,. 
\end{equation}
The operator $S=s+s^\dagger$ fulfilling the condition~\eqref{eq:condsw} reads 
\begin{align}
  s=&it\sum_{k,l,n} s_{kln} c^\dagger_k d_l \ket{n-1} \bra n, &
  s_{kln}=&\frac1{\varepsilon_l-\varepsilon_k+E_c(2n-1)+\varepsilon_d}.
\end{align}
This operator, when inserted into Eq.~\eqref{eq:swt2}, also generates higher order  couplings between sectors of charge  $n$ and $n\pm2$, which we neglect in the present discussion. The Hamiltonian becomes then block diagonal in the sectors given by different values of $n$. For $(\mathcal N_{\rm g}-\varepsilon_dC_{\rm g}/e)\in [-1/2,1/2]$, the lowest energy sector corresponds to $n=0$ and the effective Hamiltonian reads $\mathcal H'_{\rm{CBM}}=\mathcal H_0+\mathcal H_{\rm B}$, with
\begin{equation}\label{eq:swgrabert}
\mathcal H_{\rm B} = \frac{t^2}2\sum_{kk'll'}
  \left(s_{kl0}d^\dagger_{l'}c_{k'}c^\dagger_kd_l-s_{kl1}c^\dagger_kd_ld^\dagger_{l'}c_{k'} 
+ {\rm h.c.}\right).
\end{equation}
This interaction can be simplified by a mean-field treatment 
\begin{equation}\label{eq:meanfield}
d^\dagger_l c_k c^\dagger_{k'} d_{l'}=\av{d^\dagger_l d_{l'}}c_k c^\dagger_{k'}+\av{c_k c^\dagger_{k'}}d^\dagger _l d_{l'}=\delta_{ll'}\theta(-\varepsilon_l)c_kc^\dagger_{k'}+\delta_{kk'}\theta(\varepsilon_k)d^\dagger_l d_{l'}\, ,
\end{equation}
allowing to  carry out part of the sums in Eq.~\eqref{eq:swgrabert}. Notice that the orbital energy $\varepsilon_d$ does not appear in Eq.~\eqref{eq:meanfield} as it is now only associated to the charge degree of freedom $n$, while the Fermi gases corresponding to  $c_k$ and $d_l$ have the same Fermi energy $E_F=0$. One thus finds the effective low energy model, which,  to leading order, reads~\cite{filippone_fermi_2012}
\begin{equation}\label{eq:effintmat}
\mathcal H_{\rm CBM}' = \mathcal H_0+  \frac{g}{\nu_0} \ln\left(\frac{E_c-\varepsilon_d}{E_c+\varepsilon_d}\right)\left[\sum_{ll'}d^\dagger_l d_{l'}-\sum_{kk'}c^\dagger_kc_{k'}\right]\,.
\end{equation}
in which we have introduced  the dimensionless conductance $g=(\nu_0 t)^2$, corresponding to the conductance of the QPC connecting dot and lead in units of $e^2/h$. This Hamiltonian describes two decoupled Fermi gases, but affected by potential scattering with opposite amplitudes. Equation~\eqref{eq:effintmat} coincides with the LFL Hamiltonian~\eqref{eq:potscatt} for the lead electrons. The phase-shift $\delta_W$~\eqref{eq:deltapotscatt} allows to calculate the charge occupation of the dot to leading order by applying the Friedel sum rule Eq.~\eqref{eq:friedel}
\begin{equation}\label{eq:swmat}
 \av {N}=\frac{\delta_W}\pi= g \, \ln
\left(\frac{E_c-\varepsilon_d}{E_c+\varepsilon_d}\right).
\end{equation}
This result  reproduces the direct calculation of the dot occupation~\cite{matveev_quantum_1990,grabert_charge_1994,grabert_rounding_1994}, showing the validity  of the LFL model~\eqref{eq:potscatt}, with Friedel sum rule for the CBM. The extension to $M$ channels is obtained by replacing $g\rightarrow M(\nu_0t)^2$ in Eq.~\eqref{eq:swmat}. The extended proof to next-to-leading order in $g$ is given in Ref.~\cite{filippone_fermi_2012}.

\subsubsection{Anderson impurity model}\label{sec:aim}

\begin{figure}[t]
\begin{center}
\includegraphics[width=\textwidth]{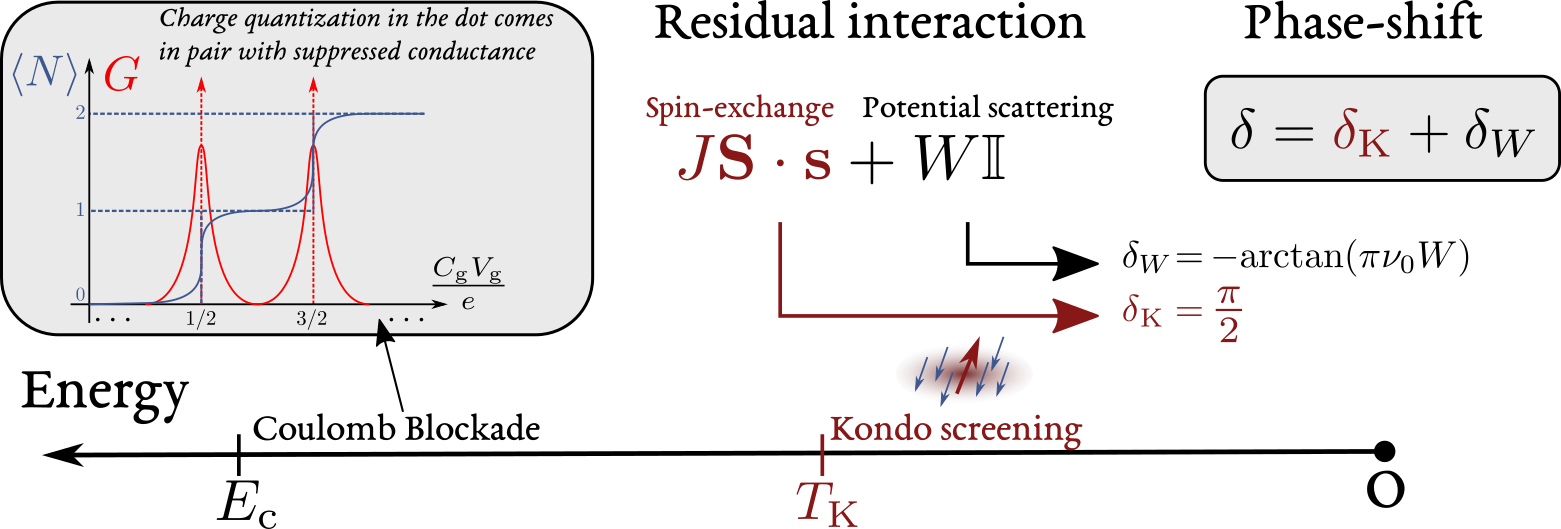}
\caption{Modification of the physical scenario of  Fig.~\ref{fig:CBM} in the presence of spin-exchange interactions between the dot and lead electrons in the AIM. Spin-exchange interactions trigger the formation of the Kondo singlet below the Kondo temperature $T_{\rm K}$, which is responsible for an additional elastic $\delta_{\rm K}=\pi/2$ phase-shift of lead electrons in the effective LFL theory. }\label{fig:renormscheme}
\end{center}
\end{figure}

Considering the internal spin degree of freedom in the AIM~\eqref{eq:aim} does not fundamentally affect the effective LFL  behavior. Nevertheless,  in the case where a single electron is trapped in the quantum dot ($-U\ll\varepsilon_d\ll0$), the derivation of Eq.~\eqref{eq:potscatt} is more involved, and  sketched in Fig.~\ref{fig:renormscheme}.  The SW transformation maps the AIM  onto a Kondo Hamiltonian including a potential scattering term~\cite{schrieffer_relation_1966}
\begin{equation}\label{eq:kondo}
 \mathcal H_{\rm AM}' =\mathcal H_0
+J \mathbf S\cdot\mathbf s+W\sum_{kk'\sigma}c^\dagger_{k\sigma}c_{k'\sigma}\,.
\end{equation}		        
The spin of the electron in the quantum dot $\mathbf S$ is coupled anti-ferromagnetically to the local spin of the lead electrons $\mathbf s=\sum_{kk'\tau\tau'}c^\dagger_{k\tau}\frac{\sigma_{\tau\tau'}}2c_{k'\tau'}$, with $\sigma_{\tau\tau'}$ the vector composed of the Pauli matrices, and
\begin{align}\label{eq:sww}
J&= \frac {2\Gamma}{\pi\nu_0}\left(\frac1{\varepsilon_d+U}-\frac1{\varepsilon_d}\right)\,, & W&=-\frac{\Gamma}{2\pi\nu_0}\left(\frac1{\varepsilon_d+U}+\frac1{\varepsilon_d}\right)\,,
\end{align}
in which we introduced the hybridization energy $\Gamma=\pi\nu_0t^2$,  corresponding to the width acquired by the orbital level when coupled to the lead and which depends on  the density of states of the lead electrons  at the Fermi energy  $\nu_0$, see also Eq.~\eqref{eq:gamma} in Appendix~\ref{app:scattering} for its rigorous definition. Neglecting for the moment the Kondo anti-ferromagnetic coupling controlled by $J$, the LFL Hamiltonian~\eqref{eq:potscatt} is directly recovered.  Nevertheless, 
 the potential scattering term is absent ($W=0$)  at the particle-hole symmetric point $\varepsilon_d=-U/2$. At this point,  the charge on the dot is fixed to one by symmetry, and the absence of potential scattering allows to derive various rigorous results, for instance concerning the ground state properties relying on Bethe ansatz~\cite{tsvelick_exact_1983,wiegmann_exact_1983}. It is a well established fact that the system described by Eq.~\eqref{eq:kondo} behaves as a LFL at low energies~\cite{nozieres_fermi-liquid_1974,krishna-murthy_renormalization-group_1980,krishna-murthy_renormalization-group_1980-1} and that the Friedel sum rule applies~\cite{langreth_friedel_1966}. As a consequence, the Kondo coupling is responsible for the phase-shift of the low energy quasi-particles.  Particle-hole symmetry, spin degeneracy and Friedel sum rule fix the Kondo phase-shift to $\delta_K=\pi/2$. The Friedel sum rule states that 
\begin{equation}\label{eq:pi2}
\av{N}=2\frac {\delta_{\rm K}}\pi\,,
\end{equation}
$\av{N}=1$ because of particle-hole symmetry and the factor 2 signals spin degeneracy, fixing  $\delta_{\rm K}=\pi/2$. The detailed description of the Kondo effect is far beyond the scope of this review and we direct the interested reader to Ref.~\cite{hewson_kondo_1993} for a comprehensive review and to Refs.~\cite{affleck_conformal_1995,cardy_boundary_1989,affleck_critical_1991,affleck_exact_1993,lesage_perturbation_1999,lesage_strong-coupling_1999} for the description of the low energy fixed point relying on boundary conformal field theory.  For the  scopes of this review it is enough to mention that below the Kondo temperature
\cite{tsvelick_exact_1983,wiegmann_exact_1983,haldane_theory_1978,haldane_scaling_1978}
\begin{equation}\label{eq:tkoverview}
T_{\rm K}=\frac{e^{\frac14}\gamma}{2\pi}\sqrt{\frac{2U\Gamma}{\pi }}e^{\frac{\pi\epsilon_d(\epsilon_d+U)}{2U\Gamma}}\,,
\end{equation}
the spin-exchange coupling $J$ in Eq.~\eqref{eq:kondo} flows to infinity in the renormalization group sense. The relevance of this interaction brings the itinerant electron to  screen  the local spin-degree of freedom of the quantum dot and phase-shifts the resulting quasi-particles by $\delta_{\rm K}$, see Fig.~\ref{fig:renormscheme}.  The phase-shift $\pi/2$  acquires thus a simple interpretation in one dimension~\cite{affleck_conformal_1995,coleman_introduction_2015}: writing $\Psi_{\rm R} = e^{2 i \delta} \Psi_{\rm L}$ for a given spin channel at the impurity site, $\delta=\delta_{\rm K} =\pi/2$ leads to $\Psi_{\rm R} + \Psi_{\rm L} =0$. The fact that the wave-function is zero at the impurity site, corresponds to the situation in which an electron screens the impurity spin, leading to Pauli blockade (due to Pauli principle), thus preventing other electrons with the same spin to access the impurity site. This dynamical screening of the impurity spin forms the so called ``Kondo cloud''~\cite{sorensen_scaling_1996,barzykin_kondo_1996,affleck_detecting_2001,affleck_friedel_2008}, see Fig.~\ref{fig:renormscheme}. It is responsible for increasing the local density of states and leads to the Abrikosov-Suhl resonance~\cite{suhl_dispersion_1965}, which causes the increase, below $T_{\rm K}$, of the dot conductance in Coulomb blockaded regimes~\cite{glazman_resonant_1988,pustilnik_kondo_2004,goldhaber-gordon_kondo_1998}. Remarkably, the Kondo phase-shift $\delta_{\rm K}=\pi/2$ and the the Kondo screening cloud have been also directly observed in two recent distinct experiments~\cite{takada_transmission_2014,v_borzenets_observation_2020}. In Section~\ref{sec:rckondo}, we illustrate  how such phenomena  also affect the dynamical properties of the mesoscopic capacitor in a non-trivial way.

It remains to establish the combined role of spin-exchange and potential scattering on the low-energy quasi-particles. Remarkably, the phase-shift $\delta_W$, caused by potential scattering,  is \emph{additive} to $\delta_{\rm K}$~\cite{cragg_potential_1978,lloyd_nozieres-wilson_1979,cragg_universality_1979}  
\begin{equation}\label{eq:additivity}
\delta=\delta_{\rm K}+\delta_W,
\end{equation}
and can thus be calculated independently. The validity of the above expression is demonstrated by comparison with the exact Bethe ansatz solution of the AIM~\cite{wiegmann_exact_1983,tsvelick_exact_1983}. Inserting the expression~\eqref{eq:deltapotscatt} for  the phase-shift caused by the potential scattering in the Friedel sum rule one finds
\begin{equation}\label{eq:nsw}
\av {N}=\frac2\pi\left[\delta_K-\arctan(\pi\nu_0 W)\right]=1+\frac\Gamma\pi\left(\frac1{\varepsilon_d+U}+\frac1{\varepsilon_d}\right)\,.
\end{equation}
This expression is consistent with the condition $\av N=1$, imposed by particle-hole symmetry, but also with the static charge susceptibility $\chi_{\rm c}$, which was derived with the Bethe ansatz~\cite{horvatie_equivalence_1985}
\begin{equation}\label{eq:chicba}
\chi_{\rm c}=\left.-\frac{\partial \av {N}}{\partial \varepsilon_d}\right|_{\varepsilon_d=-\frac U2}=\frac{8\Gamma}{\pi U^2}\left(1+\frac {12\Gamma}{\pi U}\,+\ldots\right).
\end{equation}
The extension of this proof to next-to-leading order in $t$, is given in Refs.~\cite{filippone_fermi_2012,filippone_admittance_2013} and it shows how LFL approaches are effective in providing analytic predictions also out of particle-hole symmetry, extending Bethe ansatz results.

This discussion concludes our demonstration of the persistence of elastic and coherent effects triggered by interactions at equilibrium. Local Fermi liquids provide a general framework to describe interacting and non-interacting systems at low energy, within an effective elastic scattering theory.  Nevertheless, it is important to stress that LFL theories can fail in specific cases, such as overscreened Kondo impurities~\cite{nozieres_kondo_1980,cox_exotic_1998}, and that their validity is limited to close-to-equilibrium/low-energy limits. It is thus expected that interactions become crucial as soon as such systems are driven out of equilibrium. We will illustrate now how the LFL theory allows to describe exotic, but still coherent in nature, dynamical effects in a paradigmatic setup  such as the mesoscopic capacitor.


\begin{figure}[t]
\centering
\includegraphics[width=.5\textwidth]{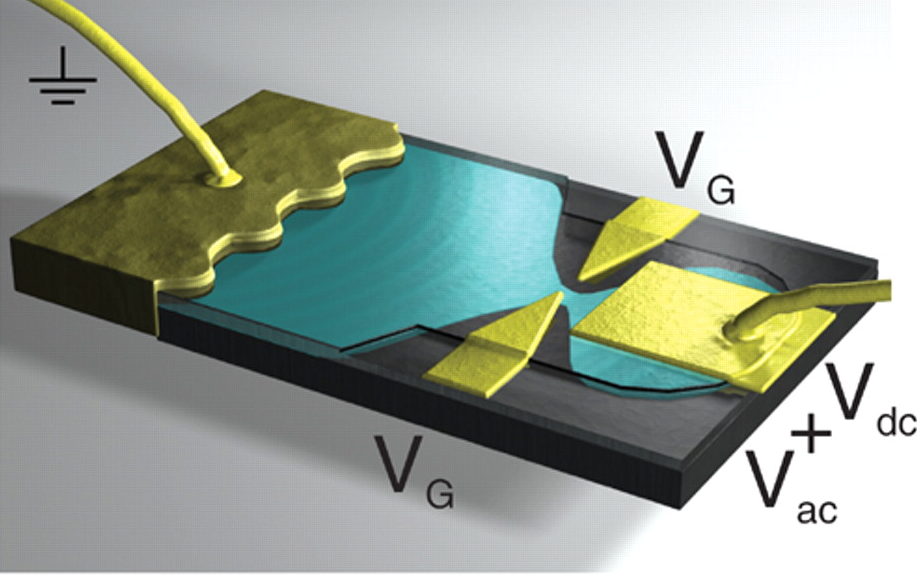}\vspace{2mm}
\includegraphics[width=.8\textwidth]{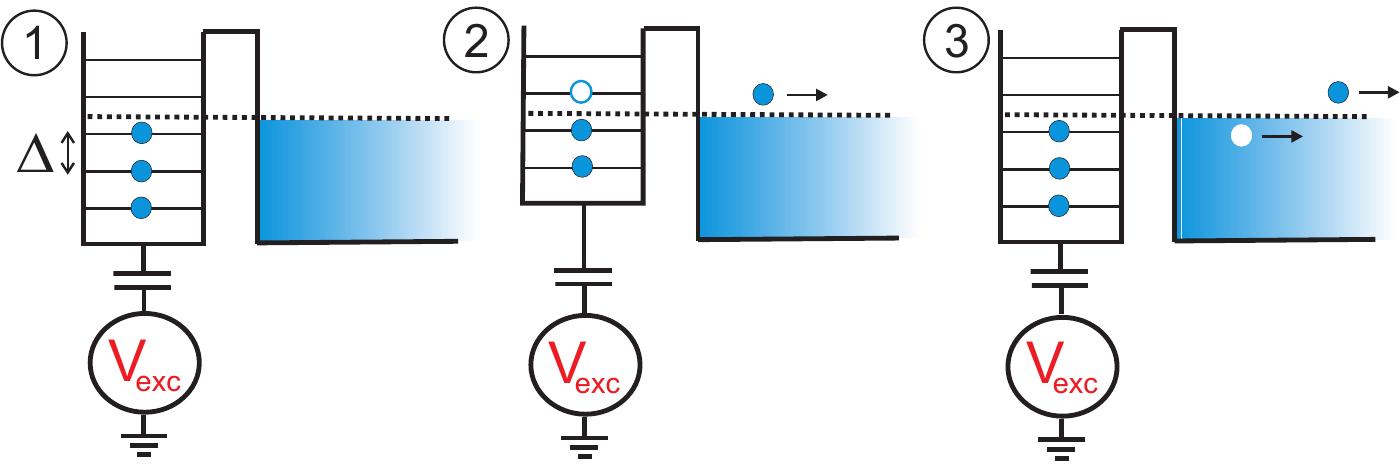}
\caption{ Top -- First realization of the mesoscopic capacitor~\cite{gabelli_violation_2006}: a 2DEG in the Quantum Hall regime is coupled to a quantum cavity via a gate-controlled QPC.  Bottom -- Working principle of single electron emission~\cite{feve_-demand_2007,mahe_current_2010,parmentier_current_2012,freulon_hong-ou-mandel_2015,marguerite_decoherence_2016}. A gate potential moves the quantized levels of the cavity above and below the Fermi surface of the coupled reservoir. Electron/hole emission in steps 2 and 3 follows from  moving occupied/empty orbitals above/below the Fermi surface. }\label{fig:emission}
\end{figure}

\section{The mesoscopic capacitor}\label{sec:meso}

The mesoscopic capacitor in Fig.~\ref{fig:emission} plays a central role in the quest to achieve full control of scalable coherent quantum systems \cite{loss_quantum_1998,petta_coherent_2005,koppens_driven_2006}. A mesoscopic capacitor is an electron cavity  coupled to a lead via a QPC and capacitively coupled to a metallic gate \cite{buttiker_dynamic_1993,buttiker_mesoscopic_1993,pretre_dynamic_1996}.  The
interest in this device stems from the absence of DC transport, making possible the investigation and
control of the coherent dynamics of single electrons. The first experimental realization of this system was a two-dimensional cavity in the quantum Hall regime \cite{gabelli_violation_2006,gabelli_coherentrccircuit_2012}, exchanging electrons with the edge of a bulk two-dimensional electron gas (2DEG). Operated out of equilibrium and in the weak tunneling limit, this system allows the triggered emission of single electrons \cite{feve_-demand_2007,mahe_current_2010,parmentier_current_2012}, and paved the way to the realization of single-electron quantum optics experiments~\cite{bocquillon_electron_2012,bocquillon_coherence_2013,bocquillon_electron_2014,grenier_single-electron_2011}, as well as probing electron fractionalization~\cite{bocquillon_separation_2013,freulon_hong-ou-mandel_2015},  accounted by  the  scattering of charge density waves (plasmons) in the conductor~\cite{safi_transport_1995,safi_transport_1996,safi_dynamic_1999,oreg_interedge_1995,blanter_interaction_1998,fazio_anomalous_1998,sukhorukov_resonant_2007,grenier_electron_2011},  and their relaxation~\cite{marguerite_decoherence_2016}. On-demand single-electron sources were also recently realized with real-time switching of tunnel-barriers \cite{leicht_generation_2011,battista_spectral_2012,fletcher_clock-controlled_2013,waldie_measurement_2015,kataoka_time--flight_2016,johnson_ultrafast_2017}, electron sound-wave surfing \cite{hermelin_electrons_2011,bertrand_long-range_2015,bertrand_injection_2016},  generation of levitons \cite{levitov_electron_1996,ivanov_coherent_1997,keeling_minimal_2006,dubois_minimal-excitation_2013,rech_minimal_2017} and superconducting turnstiles \cite{van_zanten_single_2016,basko_landau-zener-stueckelberg_2017}. We direct again the interested reader to Ref.~\cite{bauerle_coherent_2018} for a comprehensive review of these experiments.

The key question concerning the dynamics of a mesoscopic capacitor is which electronic state, carrying a current $\mathcal I$, is emitted from the cavity  following a change in the gate voltage $V_{\rm g}$. The linear response is characterized by the admittance ${\cal A}(\omega)$,
\begin{equation}  \label{eq:admittance}
  \mathcal I(\omega) = {\cal A}(\omega) V_{\rm g}(\omega) + {\cal O}(V_{\rm g}^2)\,.
\end{equation}
In their seminal work, B\"uttiker and coworkers showed that the low-frequency admittance of a mesoscopic capacitor reproduces the one of a classical {\it RC} circuit \cite{buttiker_dynamic_1993,buttiker_mesoscopic_1993,pretre_dynamic_1996},
\begin{equation}
  {\cal A}(\omega)= -i\omega C (1+ i\omega R_{\rm q} C) + \mathcal{O}(\omega^3)\,,
\label{eqn:rccircuit2}
\end{equation}
in which both the capacitance $C$ and the \textit{charge relaxation resistance} $R_{\rm q}$ probe novel coherent  dynamical quantum effects. The capacitance $C$ was originally interpreted as an \textit{electro-chemical capacitance} $1/C=1/C_{\rm g}+1/C_{\rm q}$, series  of a \textit{geometric} ($C_{\rm g}$) and a \textit{quantum} ($C_{\rm q}$) contribution~\cite{buttiker_dynamic_1993,buttiker_mesoscopic_1993,pretre_dynamic_1996,gabelli_coherentrccircuit_2012}. The geometric contribution is classical and  depends on the shape of the capacitive contact between gate and quantum dot. The quantum contribution is a manifestation of Pauli exclusion principle and was found proportional to the local density of states in the cavity, see Fig.~\ref{fig:c0}. Remarkably, the charge relaxation resistance $R_{\rm q} = h/2 e^2$ was predicted to be universally equal to half of the resistance quantum in the case of one conducting channel~\cite{gabelli_violation_2006}, independently of the transparency of the QPC connecting  cavity and lead.  This result is in striking contrast with the resistance measured in DC experiments and was originally labeled as a \textit{Violation of Kirchhoff's Laws for a Coherent {\it RC} Circuit}~\cite{gabelli_violation_2006}. Reference~\cite{gabelli_coherentrccircuit_2012} extensively reviews the original theoretical predictions  and their experimental confirmation, in a non-interacting and self-consistent setting, which we also review and put in relation with their Hamiltonian formulation in Appendix~\ref{app:nonint}.

Below, we discuss how the LFL approach challenges and extends the above studies. In particular:

\begin{enumerate}
\item The total capacitance $C$ is given by  the static charge susceptibility $\chi_{\rm c}=-e^2\partial \av N/\partial V_{\rm g}$ of the cavity and does not generally correspond to a series of a geometric and quantum contribution, proportional to the density of states in the cavity. For instance,  in Kondo regimes, the charge susceptibility of the cavity remains small, because of frozen charge fluctuations, while the density of states increases below the Kondo temperature~\cite{hewson_kondo_1993}. This effect was directly probed in a recent experiment with a quantum dot device embedded in a circuit-QED architecture~\cite{desjardins_observation_2017}. 
\item A LFL low energy behavior implies universality of the charge relaxation resistance in the single channel case. In particular, the universality of $R_{\rm q}$ stems from a Korringa-Shiba (KS) relation~\cite{shiba_korringa_1975}
\begin{equation}\label{eq:KS}
\mbox{Im}\left[\chi_{\rm c}(\omega)]\right|_{\omega\rightarrow0}=\omega\hbar \pi \chi_{\rm c}^2(\omega=0)\,,
\end{equation}
in which $\chi_{\rm c}(\omega)$ is the Fourier transform of the dynamical charge susceptibility~\eqref{eq:dyncharge} . 
\item The LFL approach shows various non-trivial dissipative effects triggered by strong correlations. In particular, it predicts a mesoscopic crossover between two universal regimes in which $R_{\rm q}=h/2e^2\rightarrow h/e^2$~\cite{mora_universal_2010} by increasing the dot size, also at charge degeneracy, in which the CBM maps on the Kondo model~\cite{matveev_quantum_1990}. It also predicts giant dissipative regimes, described by giant universal peaks in $R_{\rm q}$, triggered by the destruction of the Kondo singlet by a magnetic field~\cite{lee_effect_2011,filippone_giant_2011}.  
\item  In proper out-of-equilibrium regimes, interactions and inelastic effects become unavoidable and  circuit analogies, such as Eq.~\eqref{eqn:rccircuit2}, do not capture the dynamic behavior of the mesoscopic capacitor~\cite{litinski_interacting_2017}. We show here how previously published data~\cite{freulon_hong-ou-mandel_2015} also show a previously overlooked signature of non-trivial many-body dynamics induced by interactions. 
\end{enumerate}

\subsection{Hamiltonian description of the quantum {\it RC} circuit: differential capacitance and Korringa-Shiba relation}\label{sec:micro}

Expanding the square in Eq.~\eqref{eq:charging} and neglecting  constant contributions, $\mathcal H_{\rm c}$ renormalizes the orbital energy $\varepsilon_d$ in Eq.~\eqref{eq:cbm} and adds a quartic term in the annihilation/creation operators $d_l$, namely 	
\begin{equation}\label{eq:hexpand}
\mathcal H_{\rm c}=-eV_{\rm g}(t) N+E_c N^2\,.
\end{equation}
The driving gate voltage $V_{\rm g}$  couples to the charge occupation of the quantum dot $ Q=e\av N$. In single-electron emitters, one operates on the time dependent voltage drive $V_{\rm g}(t)$ to bring occupied discrete levels above the Fermi surface and then trigger the emission of charge, see Fig. \ref{fig:emission}.  The current of the  device is a derivative in time of the charge leaving the quantum dot, the admittance reads then, in Fourier frequency representation,
\begin{equation}\label{eq:gcharge}
\mathcal A(\omega)=-i\omega\frac{ Q(\omega)}{V_{\rm g}(\omega)}\,.
\end{equation}
We start by considering small oscillations of amplitude $\varepsilon_\omega$ of the gate voltage:
\begin{equation}\label{eq:vgt}
V_{\rm g}(t)=V_{\rm g}+\varepsilon_\omega\cos(\omega\,t)\,. 
\end{equation}
Close to equilibrium, expression~\eqref{eq:gcharge} is  calculated relying on Kubo's linear response theory~\cite{kubo_statistical_1992}
\begin{equation}\label{eq:gcdyn}
\mathcal A(\omega)=-i\omega e^2\chi_{\rm c}(\omega)\,,
\end{equation}
in which $\chi_{\rm c}(\omega)$ is the Fourier transform of the dynamical charge susceptibility
\begin{equation}\label{eq:dyncharge}
\chi_{\rm c}(t-t')=\frac i\hbar \theta(t-t')\av{\comm{ N(t)}{ N(t')}}_0\,.
\end{equation}
The notation $\langle\cdot\rangle_0$ refers to quantum averages performed at equilibrium, i.e. without the driving term $V_{\rm  g}(t)$ in Eq.~\eqref{eq:hexpand}. The low frequency expansion of $\chi_{\rm c}(\omega)$ reads
\begin{equation}\label{eq:lowfrecgdyn}
\mathcal A(\omega)=-i\omega e^2\left\{\chi_{\rm c}+i\mbox{Im}\left[\chi_{\rm c}(\omega)\right]\right\}+\mathcal O(\omega^2)\,,
\end{equation}
where we relied on the fact that the even/odd part of the response function~\eqref{eq:dyncharge} coincide with its real/imaginary part, see Appendix \ref{app:power}. We also introduce the static charge susceptibility $\chi_{\rm c}=\chi_{\rm c}(\omega=0)$.  The expansion~\eqref{eq:lowfrecgdyn} matches that of  a classical RC circuit~\eqref{eqn:rccircuit2}. Identifying term by term, we find the expression of the charge relaxation resistance and, in particular, that the capacitance $C$ of the mesoscopic capacitor is actually given by a  {\it differential capacitance} $C_0$ 
\begin{align}\label{eq:crchic}
C&=C_0=e^2\chi_{\rm c}=-e^2\frac{\partial \av { N}}{\partial \varepsilon_d}=\frac{\partial Q}{\partial V_{\rm g}}\,,&R_{\rm q}&=\frac1{e^2\chi_{\rm c}^2}\left.\frac{\mbox{Im}\chi_{\rm c}(\omega)}{\omega}\right|_{\omega\rightarrow0}\,.
\end{align}
The differential capacitance is  proportional to the density of states of \textit{charge} excitations on the dot, which, as mentioned above, generally differs from the local density of states in the presence of strong correlations. Equation~\eqref{eq:crchic}  provides also the general condition for the universal quantization of the charge-relaxation resistance $R_{\rm q}=h/2e^2$, namely:
\begin{equation}\label{eq:ks1chintro}
\left.\mbox{Im}\chi_{\rm c}(\omega)\right|_{\omega\rightarrow0}=\hbar\pi \omega\chi_{\rm c}^2\,,
\end{equation}
Such kind of relation is known as a {\it Korringa-Shiba} (KS) relation \cite{shiba_korringa_1975}. The KS relation establishes that the imaginary part of the dynamic charge susceptibility, describing dissipation in the system, is controlled by the static charge fluctuations on the dot, $\chi_{\rm c}$.

Additionally, we mention that the relation~\eqref{eq:ks1chintro} also affects the phase-shift of reflected or transmitted light through a mesoscopic system in the Kondo LFL regime~\cite{schiro_tunable_2014,le_hur_many-body_2016,le_hur_driven_2018}. Such situations have been recently realized with quantum-dot devices embedded in circuit-QED architectures~\cite{delbecq_coupling_2011,delbecq_photon-mediated_2013,liu_photon_2014,bruhat_cavity_2016,mi_circuit_2017,viennot_towards_2016,deng_quantum_2015}, in which the driving input signal can be modeled by an AC potential of the form~\eqref{eq:vgt}.

\subsection{The origin of the differential capacitance as a `quantum' capacitance as far as interactions are neglected}\label{sec:cq}

\begin{figure}[t]
\begin{center}
\includegraphics[width=\textwidth]{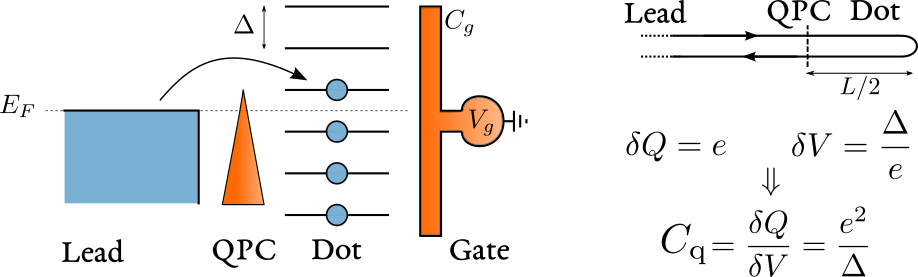}
\caption{Physical origin of the quantum capacitance $C_{\rm q}$. Pauli exclusion forces electrons entering the dot to pay an energy price equal to the local level spacing $\Delta$, resulting in a capacitance $C_{\rm q}=e^2/\Delta$. On the right, a one-dimensional representation of the mesoscopic capacitor, with a dot of size $\ell$.}\label{fig:c0}
\end{center}
\end{figure}

As far as interactions are neglected, the differential  capacitance $C_0$ is  a manifestation of the fermionic statistics of electrons, determined by Pauli exclusion principle. For this reason it has been originally labeled as a `quantum' capacitance $C_{\rm q}$~\cite{buttiker_dynamic_1993,buttiker_mesoscopic_1993,pretre_dynamic_1996}.  When an electron is added to the quantum dot, in which energy levels are spaced by $\Delta$, the Pauli exclusion principle does not allow to fill an occupied energy state, but requires to pay a further energy price $\Delta$, see Fig. \ref{fig:c0}. The capacitance associated to this process is then $C_{\rm q}=\delta Q/\delta V$. For one electron  $\delta Q=e$  and $\delta V=\Delta/e$. Substituting these two expressions, we recover a uniform quantum capacitance 
\begin{equation}\label{eq:c0pauli}
C_{\rm q}=\frac{e^2}\Delta\,.
\end{equation}
This expression establishes that the quantum capacitance is proportional to the density of states in the quantum dot at the Fermi energy $C_{\rm q}=e^2\mathcal N(E_F)$, with $\mathcal N(E_F)=1/\Delta$, to be distinguished from $\nu_0$, the density of states of the lead electrons. 

Additionally, the quantum capacitance is  related to the dwell-time spent by electrons in the cavity. The general relation is derived in Appendix~\ref{app:nonint}, but it also results from simple estimates. Considering the representation of the mesoscopic capacitor of Fig.~\ref{fig:c0}, in the open-dot limit, the time spent by an electron in the cavity coincides with its time of flight $\tau_{\rm f}=\ell/v_F$: the ratio between the size of the cavity $\ell$ and its (Fermi) velocity $v_F$. The level spacing $\Delta$ of an isolated cavity of size $\ell$ is estimated by linearizing the spectrum close to the Fermi level. The distance in momentum between subsequent levels is $h/\ell$, corresponding to $\Delta=hv_F/\ell=h/\tau_{\rm f}$. Substituting in Eq.~\eqref{eq:c0pauli} leads to an equivalent expression for  the quantum capacitance
\begin{equation}\label{eq:cqtf}
C_{\rm q}=\frac{e^2}h\tau_{\rm f}\,.
\end{equation}

On the experimental side, the level spacing of the quantum dot can be actually estimated and, in the experimental conditions of Ref.~\cite{gabelli_violation_2006}, it was established to be of the order of $\Delta\sim15~\mbox{GHz}$, corresponding to a quantum capacitance $C_{\rm q}\sim1 ~\mbox{fF}$. Experimental measurements of $C$, reported in Fig.~\ref{fig:gabellic}, give an estimate also for $C_{\rm g}$, showing that $C_{\rm q}\ll C_{\rm g}$. This implies that the level spacing $\Delta$ was much larger than the charging energy $E_c=e^2/2C_{\rm g}$ , of the order of fractions of the GHz, apparently justifying the mean-field approach to describe experimental results, with the limitations that we are going to discuss in out-of-equilibrium regimes, see Section~\ref{sec:ooe}.

The  argument leading to Eq.~\eqref{eq:c0pauli} implicitly assumes the perfect transparency of the QPC,  namely that the probability amplitude $r$ for a lead electron to be reflected when passing though the QPC to enter the cavity is equal to zero ($r=0$). In this limit, the density of states on the dot is uniform. Finite reflection $r\neq0$ is responsible for resonant tunneling processes, leading to oscillatory behavior of the local density (or the dwell-time) of states as a function of the gate potential $V_{\rm g}$, in agreement with the experimental findings reported in Fig. \ref{fig:gabellic}. In  Appendix~\ref{app:nonint}, we provide a quantitative analysis of this effect by explicitly calculating the differential capacitance $C_0$, Eq.~\eqref{eq:crchic}, by neglecting the term proportional to $E_{\rm c}N^2$ in Eq.~\eqref{eq:hexpand}.

\subsection{The physical origin  of the universal charge relaxation resistance}\label{sec:rqbutt}

\begin{figure}[t]
\begin{center}
\includegraphics[width=.7\textwidth]{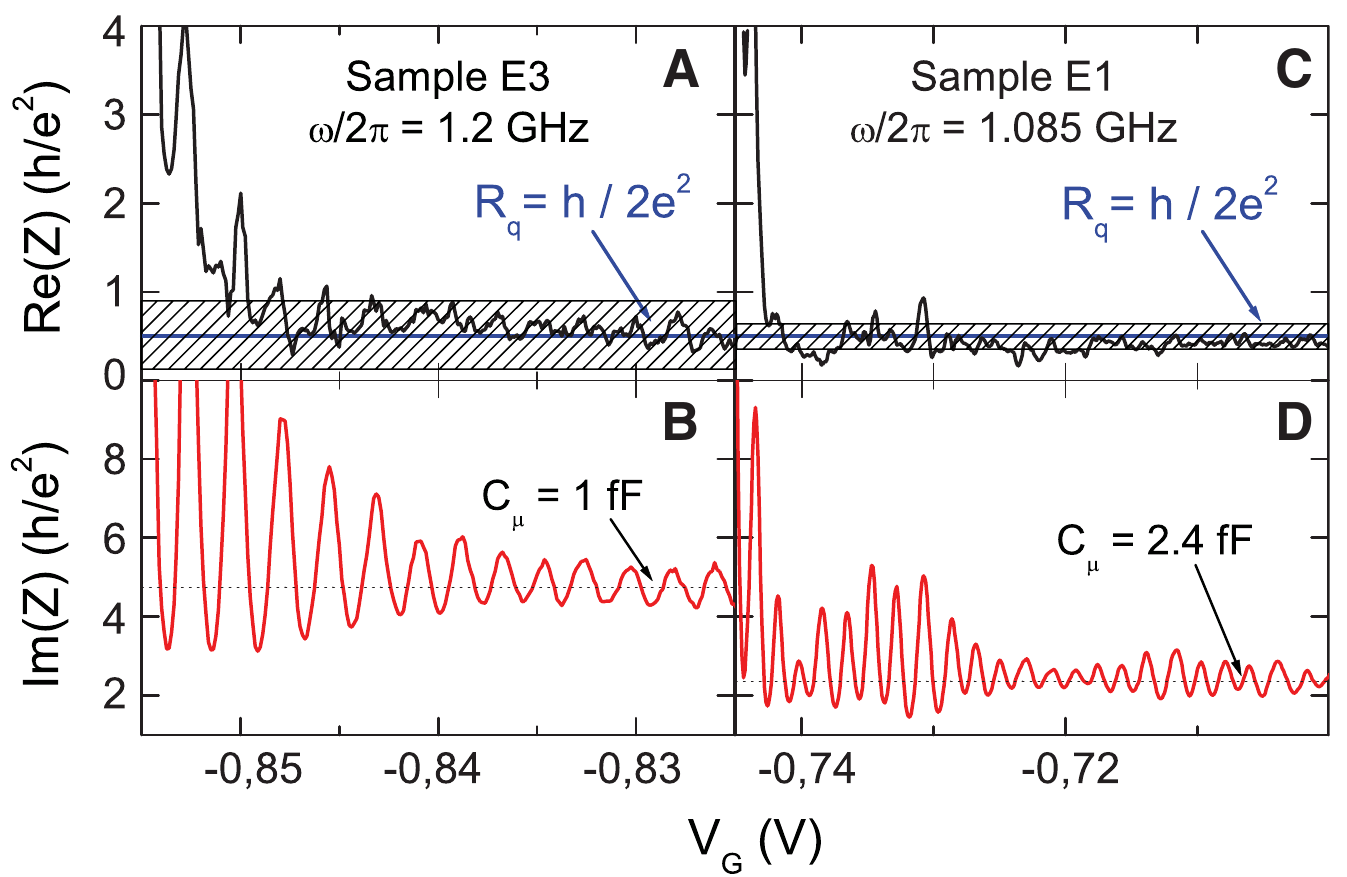}
\caption{Top -- Measurement of the universal charge relaxation resistance from Ref.~\cite{gabelli_violation_2006}. The resistance of two different samples (E3 and E1) is given by the real part of their impedance $Z=1/\mathcal A$ as a function of the QPC potential $V_{\rm G}$, in Fig.~\ref{fig:emission}, which also affected the gate potential $V_{\rm g}$. Measurements were carried out for $T = 30 mK$ and a magnetic field $B=1.3T$ polarizing the electrons, resulting in one conducting channel. Uncertainties are indicated by the hatched areas.  Bottom -- Measurement of the total capacitance $C_{\mu }=C$, through $\mbox{Im(Z)}=-1/\omega C$, for the same samples. The oscillatory behavior is related to resonances in the density of states of the dot.}\label{fig:gabellir}\label{fig:gabellic}
\end{center}
\end{figure}
%
%

The universality of $R_{\rm q}$ was also verified experimentally in Ref.~\cite{gabelli_violation_2006}, see Fig.~\ref{fig:gabellir}. In the quantum coherent regime the charge relaxation resistance is \textit{universal}: it does not depend on the microscopic details of the circuit ($C_{\rm g},\,r,\ldots$), but only on fundamental constants, namely Planck's constant $h$ and the electron charge $e$. This is  surprising. When applying a DC voltage across a  QPC connecting two leads, the   QPC behaves as a resistive element of resistance~\cite{van_wees_quantized_1988,wharam_addition_1988}  
\begin{equation}\label{eq:dcr}
R_{\rm DC}=\frac h{e^2D}\,,
\end{equation}
where $D=1-|r|^2$ is the QPC transparency. This quantity depends on the probability $r$ for electrons to be backscattered when arriving at the QPC, which can be tuned by acting on a gate potential. As we considered spinless electrons, the factor 2 in $R_{\rm q}=h/2e^2$ cannot be related to spin degeneracy and $D=1$ in Eq.~\eqref{eq:dcr}. It is rather related to the fact that the dot is connected to a single reservoir, in contrast to the source-drain reservoirs present in DC transport experiments~\cite{nigg_quantum_2008,buttiker_role_2009}.   In direct transport, each metallic contact is responsible for a quantized contact resistance $R_{\rm c}=h/2e^2$, the Sharvin-Imry resistance \cite{imry_directions_1986,landauer_electrical_1987}. In source-drain experiments, Eq.~\eqref{eq:dcr} could be recast in the form $R_{\rm QPC}+2R_{\rm c}$, with $R_{\rm QPC}=\frac{h}{e^2}\frac{1-D}D$ the resistive contribution proper to the QPC. For the case of a mesoscopic capacitor, there is a single reservoir and one would thus expect for the charge relaxation resistance
\begin{equation}\label{eq:dccontact}
R_{\rm q}^{\rm expected}=R_{\rm QPC}+R_{\rm c}=\frac h{2e^2}+\frac{h}{e^2}\frac{1-D}D\,.
\end{equation}
The fact that  $R_{\rm q}$ does not depend on the transparency $D$, assuming the universal value $h/2e^2$, cannot be attributed to the laws governing DC quantum transport, and this is the reason why one can speak about the \textit{Violation of Kirchhoff's Laws for a Coherent {\it RC} Circuit}~\cite{gabelli_violation_2006}. The universality of $R_{\rm q}$ is rather a consequence of the fact that $R_{\rm q}$, differently from the contact resistance $R_{\rm c}$, is  related to energy (Joule) dissipation. As electrons propagate  coherently within the cavity, they cannot dissipate energy inside it, but only once they reach the lead. We will illustrate in Sec.~\ref{sec:quasistatic} how this phenomenon is a direct consequence of the possibility to excite particle-hole pairs by driven electrons. At low frequency, dissipation is thus only possible  in the presence of a continuum spectrum, accessible in the metallic reservoirs.  The expected resistance~\eqref{eq:dccontact} is recovered only if  electrons loose their phase coherence inside the dot~\cite{nigg_quantum_2008,buttiker_role_2009}, see also Refs.~\cite{rodionov_charge_2009,rodionov_out--equilibrium_2010}, which  take Coulomb blockade effects into account. For instance, in the high temperature limit  $k_BT\gg\Delta$,  Eq.~\eqref{eq:dccontact} is  not recovered. The reason is that, in scattering theory,  temperature is fixed by the reservoirs without affecting the coherent/phase-preserving propagation in the mesoscopic capacitor.

\subsection{The open-dot limit}

We  address now the role of the charging energy $E_{\rm c}$ and come back to our initial example of the open dot limit, considered in Section~\ref{sec:coherence}. The possibility to rely on an exact bosonized solution for  the model~(\ref{eq:charging}-\ref{eq:chiral}), made possible  the derivation of the admittance ${\cal A}(\omega)$ in linear-response theory for a fully transparent point contact ($r=0$) and  a finite-sized cavity \cite{mora_universal_2010}
\begin{equation}\label{eqn:rccircuit1}
  {\cal A}(\omega) =-i\omega C_{\rm g} \left(1- \frac{i\omega\tau_{\rm c}}{1-e^{ i  \omega\tau_{\rm f}}}\right)^{-1}\,.
\end{equation}
This expression is important as it makes possible to study the interplay between two different time-scales, namely the time of flight $\tau_{\rm f}$ of electrons inside the cavity, already present in the previous discussion, and $\tau_{\rm c} =h C_{\rm g}/e^2$ the time scale corresponding to the charging energy $E_{\rm c}$. We mention that, interestingly,  the admittance \eqref{eqn:rccircuit1} was also found to describe the coherent transmission of electrons through interacting Mach-Zehnder interferometers~\cite{ngo_dinh_nonequilibrium_2012,ngo_dinh_analytically_2013}.

What is quite remarkable about the admittance~\eqref{eqn:rccircuit1} is that, to linear order in $\omega$,   the two time scales $\tau_{\rm f}$ and $\tau_{\rm c}$ still combine into the universal charge relaxation resistance $R_{\rm q} = h/2e^2$ and a series of a geometrical and quantum capacitance $C_{\rm q}=e^2\tau_{\rm f}/h$~\cite{buttiker_dynamic_1993,buttiker_mesoscopic_1993,pretre_dynamic_1996} (see also Eq.~\eqref{eq:cqtf})
\begin{equation}
  \frac 1{C_0} = \left[ \frac{1}{C_{\rm g}} + \frac{h}{e^2 \tau_{\rm f}} \right]\,.
  \label{eq:C}
\end{equation}
The low-frequency behavior of Eq.~\eqref{eqn:rccircuit1} illustrates how interacting systems behave as if interactions were absent at low energies. What is then also implicit in Eq.~\eqref{eqn:rccircuit1} is that, to observe separate effects on the charge dynamics, induced by free  propagation ($\tau_{\rm f}$) or interactions ($\tau_{\rm c}$), one has to consider proper out-of-equilibrium/high-frequency regimes. These regimes  will be addressed in Section~\ref{sec:ooe}. 

Nevertheless, interactions still matter even in low-frequency regimes. Consider the infinite-size (metallic) limit for the cavity, $\tau_{\rm f}\rightarrow\infty$. In this limit, also describing the experiment in Fig.~\ref{fig:pierre}, one implicitly assumes that the driving frequency $\omega$  is larger than the internal level spacing of the dot $\Delta$,
\begin{equation}\label{eq:condition}
\hbar\omega\gg \Delta\,.
\end{equation}
The discrete spectrum of the dot can thus be treated as a continuum, which allows for energy dissipation also inside the cavity, see Fig.~\ref{fig:series}. In particular, averaging the admittance~\eqref{eqn:rccircuit1} over a finite bandwidth $\delta\omega$, such that $\omega\gg\delta\omega\gg\Delta$, one exactly recovers the admittance of a classical RC circuit of capacitance $C_{\rm g}$ and charge-relaxation resistance $R_q=h/e^2$~\cite{mora_universal_2010}
\begin{equation}
\mathcal A(\omega)=\frac{-i\omega C_g}{1-i\omega C_g \frac{h}{e^2}}\,.
\end{equation}
The mesoscopic crossover $R_{\rm q}=h/2e^2\rightarrow h/e^2$ is an exquisite coherent effect triggered by interactions. This phenomenon   has  fundamentally the same origin of the elastic electron transfer exemplified by the correlation function~\eqref{eq:GLR}, considered at the very beginning of this review. 

Remarkably, the universality of the charge-relaxation resistance holds in the presence of backscattering  at the dot entrance, without  affecting the mesoscopic crossover $R_{\rm q}=h/2e^2\rightarrow h/e^2$~\cite{mora_universal_2010}. Nevertheless, the possibility to interpret the differential capacitance as a series of two separate geometric and quantum term as in Eq.~\eqref{eq:C}, is lost. If we locate the entrance of the dot at $x=0$, backscattering corrections to the model~(\ref{eq:charging}-\ref{eq:chiral}) read
\begin{equation}
\mathcal H_r=-\hbar rv_F\Big[\Psi_ {\rm R}(0)^\dagger\Psi_{\rm L}(0)+\Psi_{\rm L}(0)^\dagger\Psi_{\rm R}(0)\Big]\,,
\end{equation}
and  compromise a non-interacting formulation of the problem, even in its bosonized form~\cite{matveev_coulomb_1995,mora_universal_2010,litinski_interacting_2017}.

It becomes then important to understand why and to which extent quantities such as the charge-relaxation resistance show universal coherent behavior even  in the presence of interactions. The extension of the LFL theory in the quasistatic approximation provides the unified framework to understand the generality of such phenomena.


\subsection{The tunneling limit and the quasi-static approximation}\label{sec:quasistatic}

In Section~\ref{part:overview}, we showed that a large class of models of the form~\eqref{eq:gen}, are effectively described, in the  low-energy limit, by a LFL theory~\eqref{eq:potscatt}, in which the potential scattering coupling constant $W$ depends on the orbital energy of the dot $\varepsilon_d$. The expansion of the charging energy Hamiltonian~\eqref{eq:hexpand} made apparent that this energy is renormalized by the gate potential $\varepsilon_d\rightarrow\varepsilon_d-eV_g(t)$. For an AC bias voltage, we consider then a periodic function of time oscillating at the frequency $\omega$
\begin{equation}\label{eq:edt}
\varepsilon_d(t)=\varepsilon_d^0+\varepsilon_\omega\cos\big( \omega t\big)\,.
\end{equation}
The \emph{quasi-static approximation} consists in substituting Eq. (\ref{eq:edt}) directly in Eq. (\ref{eq:potscatt}). This condition assumes that the low energy Hamiltonian~\eqref{eq:potscatt}, derived for the equilibrium problem, follows, without any delay, the orbital oscillations expected from the parent, high-energy, model. The quasi-static approximation is then a statement about a behavior close to adiabaticity.

We consider then the linear response regime and  expand the coupling $W(\varepsilon_d)$ in $\varepsilon_\omega$. Focusing on the single channel case, the extension to multiple channels being straightforward, Eq.~\eqref{eq:potscatt} becomes
\begin{equation}\label{eq:potscattpert}
\mathcal H=\sum_{k}\varepsilon_kc^\dagger_k c_k+\left[W(\varepsilon_d^0)+W'(\varepsilon_d^0)\varepsilon_\omega\cos\big( \omega t\big)\right] \sum_{kk'}c^\dagger_kc_k'\,.
\end{equation}
We diagonalize the time independent part of this Hamiltonian~\cite{krishna-murthy_renormalization-group_1980}
\begin{equation}\label{eq:potscattdiag}
\mathcal H=\sum_{kk'}\varepsilon_k a^\dagger_k a_k +\frac{W'(\varepsilon_d^0)}{1+\left[\pi\nu_0W(\varepsilon_d^0)\right]^2}\varepsilon_\omega\cos\big(\omega t\big)\sum_{kk'} a^\dagger_k a_{k'}\,, 
\end{equation}
where the operators $a$ and $a^\dagger$ describe the new quasi-particles diagonalizing the time independent part of the Hamiltonian (\ref{eq:potscattpert}).  The Friedel sum rule~\eqref{eq:npotscatt} establishes that
\begin{equation}
\chi_{\rm c}=\frac{\nu_0W'(\varepsilon_d^0)}{1+\left[\pi\nu_0W(\varepsilon_d^0)\right]^2}\,, 
\end{equation} 
and the Hamiltonian~\eqref{eq:potscattdiag} can be cast in the more compact and transparent form 
\begin{equation}\label{eq:lowmodel}
\mathcal H=\sum_{kk'}\varepsilon_k a^\dagger_k a_k+\frac{\chi_{\rm c}}{\nu_0}\varepsilon_\omega\cos\big( \omega t \big)\sum_{kk'}a^\dagger_k a_{k'}\,.
\end{equation}
This Hamiltonian shows the mechanism responsible for energy dissipation at low energy for the rich variety of strongly  interacting systems satisfying the Friedel sum rule and LFL behavior at low energy. The time dependent term  pumps energy in the system, which is then dissipated by  the creation of particle-hole pairs. Crucially, this term is controlled by the static charge susceptibility $\chi_{\rm c}$ of the quantum dot. The non-interacting Hamiltonian~\eqref{eq:lowmodel}  explains why non-interacting results hold for the universal charge relaxation resistance also in the presence of interactions on the quantum dot.

We now illustrate how the Hamiltonian~\eqref{eq:lowmodel} implies the validity of the KS relation and thus universality of the charge-relaxation resistance $R_{\rm q}$. The proof was originally devised for spin-fluctuations~\cite{garst_energy-resolved_2005} and we extend it here to the case of charge fluctuations.  For drives of the form~\eqref{eq:edt}, the power dissipated by the system is proportional to the imaginary part of the dynamic charge susceptibility, see Appendix~\ref{app:power},
\begin{equation}\label{eq:powerhigh}
\mathcal P=\frac12\varepsilon_\omega^2\omega\mbox{Im} \chi_{\rm c}(\omega)\,.
\end{equation} 
A direct calculation of $\mbox{Im}\chi_{\rm c}$ is a difficult task and this is where  the low-energy model~\eqref{eq:lowmodel} becomes useful. Similarly as for Eq.~\eqref{eq:powerhigh}, the LFL theory~\eqref{eq:lowmodel} predicts the dissipated power 
\begin{equation}\label{eq:bastapower}
\mathcal P=\frac12 \varepsilon_\omega^2\omega\mbox{Im} \chi_{ A}(\omega)\,,
\end{equation}
where the linear response function $\chi_{ A}(t-t')=\frac i\hbar \theta(t-t')\av{\comm{ A(t)}{ A(t')}}_0$ 
is a correlator at different times of the potential scattering operator 
\begin{equation}\label{eq:basta}
 A=\frac{\chi_{\rm c}}{\nu_0}\sum_{kk'}a^\dagger_k a_{k'}\,,
\end{equation}
responsible  for the creation of particle-hole pairs. The Fourier transform of the response function reads
\begin{equation}\label{eq:linlowint}
\chi_{ A}(\omega)=-\frac1\hbar\frac{\chi_{\rm c}^2}{\nu_0^2}\sum_{pp'}f(\varepsilon_p)[1-f(\varepsilon_{p'})]\left[\frac1{\omega+\frac{\varepsilon_p-\varepsilon_{p'}}{\hbar}+i0^+}-\frac1{\omega+\frac{\varepsilon_{p'}-\varepsilon_{p}}{\hbar}+i0^+}\right]\,,
\end{equation}
in which $f(\varepsilon_p)=1/(e^{\beta\varepsilon_p}+1)$ is the Fermi distribution. We consider the electron lifetime as infinite, i.e. much longer than the typical time scales $\tau_{\rm c}$ and $\tau_{\rm f}$. Taking the imaginary part and the continuum limit for the spectrum in the wide-band approximation, one finds, at zero temperature,
\begin{equation}\label{eq:powlow}
\mbox{Im}\chi_{ A}(\omega)=\pi\hbar \omega\chi_{\rm c}^2\,.
\end{equation}
The two dissipated powers~\eqref{eq:powerhigh} and~\eqref{eq:powlow} have to be identical, implying the Korringa-Shiba relation~\eqref{eq:KS}, enforcing then a universal value for the charge relaxation resistance $R_{\rm q}=h/2e^2$.

\subsection{The LFL theory of large quantum dots. The mesoscopic crossover $R_{\rm q}=h/2e^2\rightarrow h/e^2$.}\label{sec:continuum}

\begin{figure}[t!]
\begin{center}
\includegraphics[width=.7\textwidth]{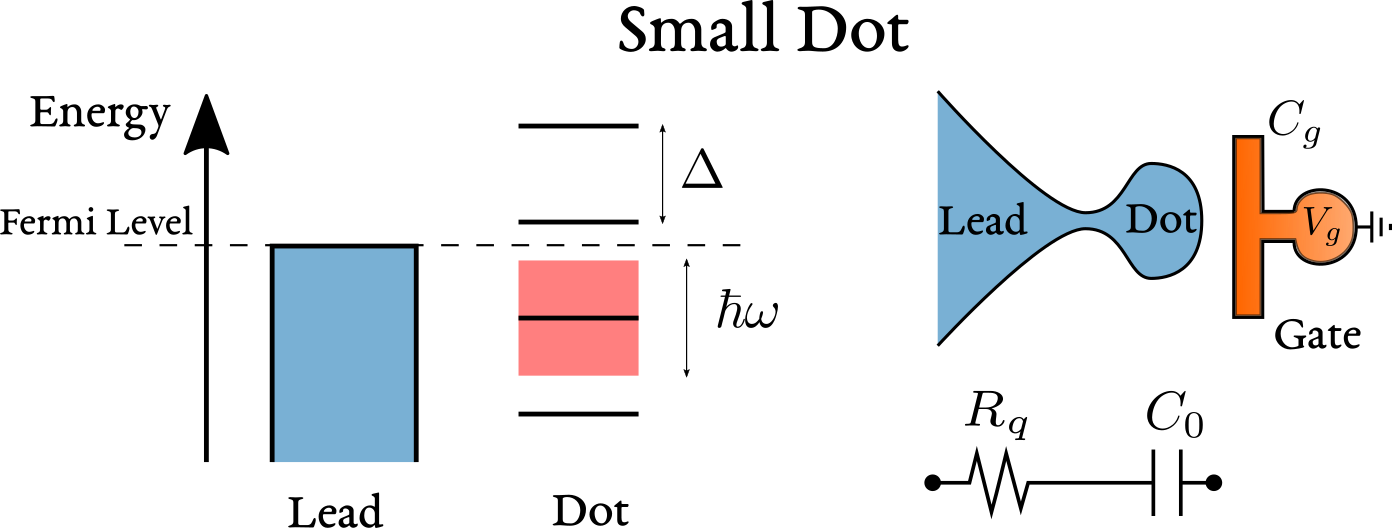}\\~\vspace{3mm}\\
\includegraphics[width=.7\textwidth]{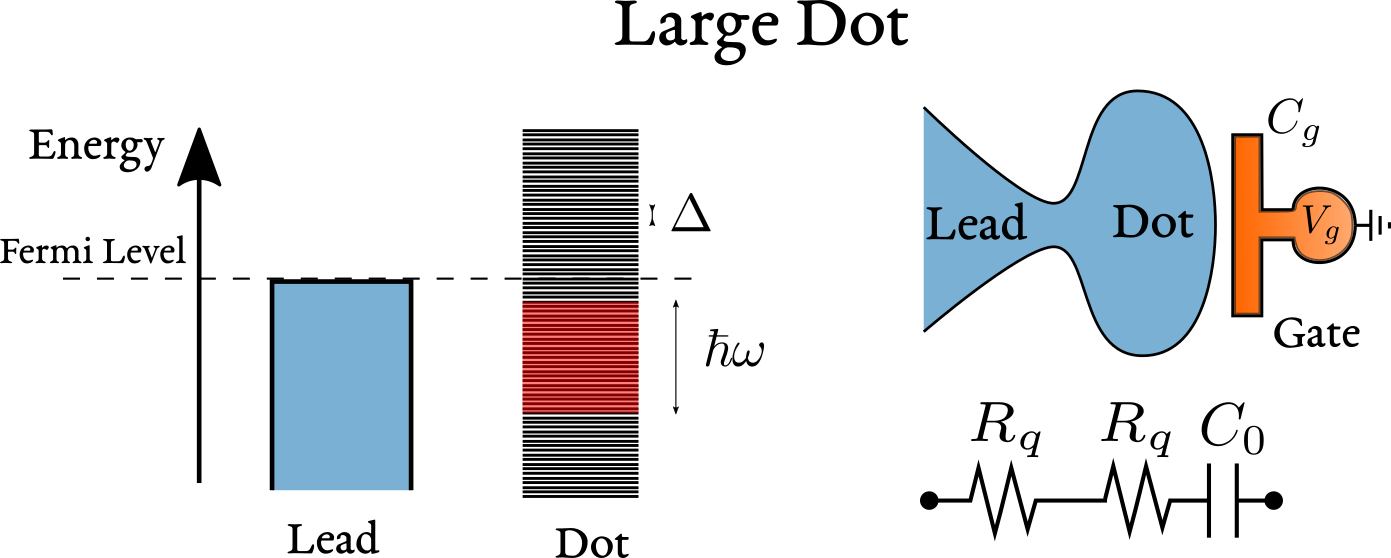}
\caption{Mesoscopic crossover in the charge relaxation resistance.  Top -- In a small dot, the level spacing $\Delta$ is larger than the driving energy $\hbar\omega$ and  energy levels in the dot are not excited. The universal resistance $R_{\rm q}=h/2e^2$ of the equivalent {\it RC} circuit is furnished exclusively by the lead electron reservoir. Bottom -- Excitation of energy levels inside the dot are permitted in the large dot limit, which acts as a further dissipative reservoir in series to the lead. }\label{fig:series}
\end{center}
\end{figure}

The above demonstration has to be slightly adapted to show the mesoscopic crossover $R_{\rm q}=h/2e^2\rightarrow h/e^2$. This crossover takes place for the CBM~\eqref{eq:cbm}, in the infinite-size limit of the dot. As implicit in the effective description~\eqref{eq:effintmat} of the CBM, the dot and the lead constitute two separate Fermi liquids.  Sections~\ref{sec:coherence} and~\ref{part:overview} illustrated how the energy cost $E_{\rm c}$ prevents the low-energy transfer  of electrons between the dot and the lead \cite{aleiner_mesoscopic_1998}. The electrons of both these gases are then only  backscattered at the lead/dot boundary with opposite amplitudes. In the quasi-static approximation, all the steps carried in the previous discussion apply for the Hamiltonian~\eqref{eq:effintmat}. In this case,  the time variation of the orbital energy $\varepsilon_d$ also drives particle-hole excitations  in the dot. The operator responsible for energy dissipation becomes 
\begin{equation}
A=\frac{\chi_{\rm c}}{\nu_0}\left(\sum_{kk'}c^\dagger_k c_{k'}-\sum_{ll'}d^\dagger_l d_{l'}\right),
\end{equation}
in which  the operators $c^\dagger_k$ and $d^\dagger_l$ create lead and dot electrons of energy $\varepsilon_{k,l}$ respectively. This formulation of  the operator $A$ adds a further contribution to Eq.~\eqref{eq:linlowint},  analogous to the contribution of particle-hole pairs excited in the lead, namely
\begin{equation}\label{eq:linlowdot}
-\frac1\hbar\frac{\chi_{\rm c}^2}{\nu_0^2}\sum_{ll'}f(\varepsilon_l)[1-f(\varepsilon_{l'})]\left[\frac1{\omega+\frac{\varepsilon_l-\varepsilon_{l'}}{\hbar}+i0^+}-\frac1{\omega+\frac{\varepsilon_{l'}-\varepsilon_{l}}{\hbar}+i0^+}\right]\,.
\end{equation}
The limits $\omega\rightarrow0$ and $\Delta\rightarrow0$ do not commute in the above expression. This fact has a clear physical interpretation: if the frequency is sent to zero before the level spacing, energy cannot be dissipated in the cavity  and no additional  contribution to $\mbox{Im}\chi_{\rm c}(\omega)$ is found. If the opposite limit is taken, the condition~\eqref{eq:condition} is met and the Korringa-Shiba relation is then modified by a factor two
\begin{equation}\label{eq:kscontinuum}
\mbox{Im}\chi_{c}(\omega)=2\pi\hbar \omega\chi_{\rm c}^2\,,
\end{equation}
which doubles the universal value of the single-channel charge relaxation resistance  $R_{\rm q}=h/e^2.$ The relation~\eqref{eq:kscontinuum} was originally shown by explicit perturbation theory in the tunneling amplitude, close and away from  charge degeneracy points~\cite{mora_universal_2010}.  As summarized in Fig.~\ref{fig:series}, driving at a frequency higher than the dot level spacing induces the creation of particle/hole pairs inside the dot as well, enhancing energy dissipation with respect to the small dot limit $\hbar \omega<\Delta$. As energy can be {\it coherently} dissipated in two fermionic baths (dot and lead),  the dot acts effectively as a further (Joule) resistor in series with the lead, leading to a doubled and still universal charge relaxation resistance. 

\subsection{The multi-channel case and universal effects triggered by Kondo correlations}\label{sec:rckondo}

The above discussion also extends to the $M$ channels case, leading to a generalized  KS relation
\begin{equation}\label{eq:ksgen}
\left.\mbox{Im}\chi_{\rm c}(\omega)\right|_{\omega\rightarrow0}=\hbar\pi \omega\sum_\sigma\chi_\sigma^2\,,
\end{equation}
which corresponds to a non-universal expression for the charge relaxation resistance~\cite{filippone_admittance_2013,dutt_strongly_2013}
\begin{equation}\label{eq:rqgen}
R_{\rm q}=\frac{h}{2e^2}\frac{\sum_\sigma\chi_\sigma^2}{\left(\sum_\sigma \chi_\sigma\right)^2}\,.
\end{equation}
This expression is analogous to  the one obtained by Nigg and B\"uttiker~\cite{nigg_mesoscopic_2006}. In their derivation leading to  Eq.~\eqref{eq:rqdwell}, the densities of states, or dwell-times $\tau_\sigma$, of the $\sigma$ channel in the dot, replace the susceptibilities $\chi_\sigma$. The single channel case is remarkable in that the numerator simplifies with the denominator in Eq.~\eqref{eq:rqgen}, leading to the universal value $h/(2e^2)$, which is thus physically robust. Otherwise, in the fine-tuned case that all the channel susceptibilities are equal, one finds $R_{\rm q}=h/2e^2M$.

Spinful systems in the presence of a magnetic field are the simplest ones to study how the charge-relaxation resistance is affected by breaking the symmetry between different conduction  channels. Indeed, lifting the orbital level degeneracy by a magnetic field breaks the channel symmetry and the charge relaxation resistance is no longer universal, as it was  originally realized in studies of the AIM~\eqref{eq:aim} relying on the Hartree-Fock approximation~\cite{nigg_mesoscopic_2006}. 

\begin{figure}
\begin{center}
\includegraphics[width=.49\textwidth]{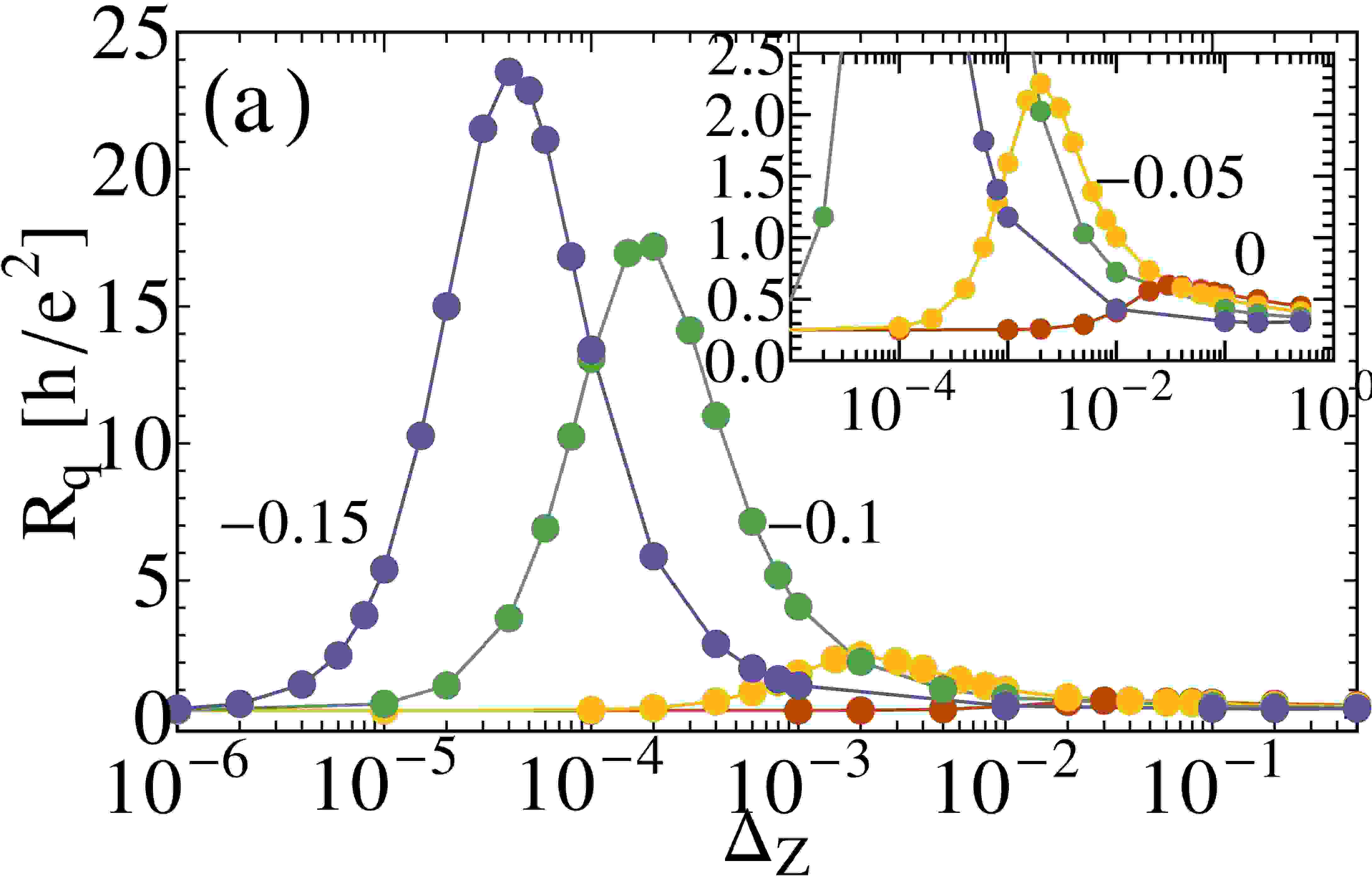}
\includegraphics[width = .49\textwidth]{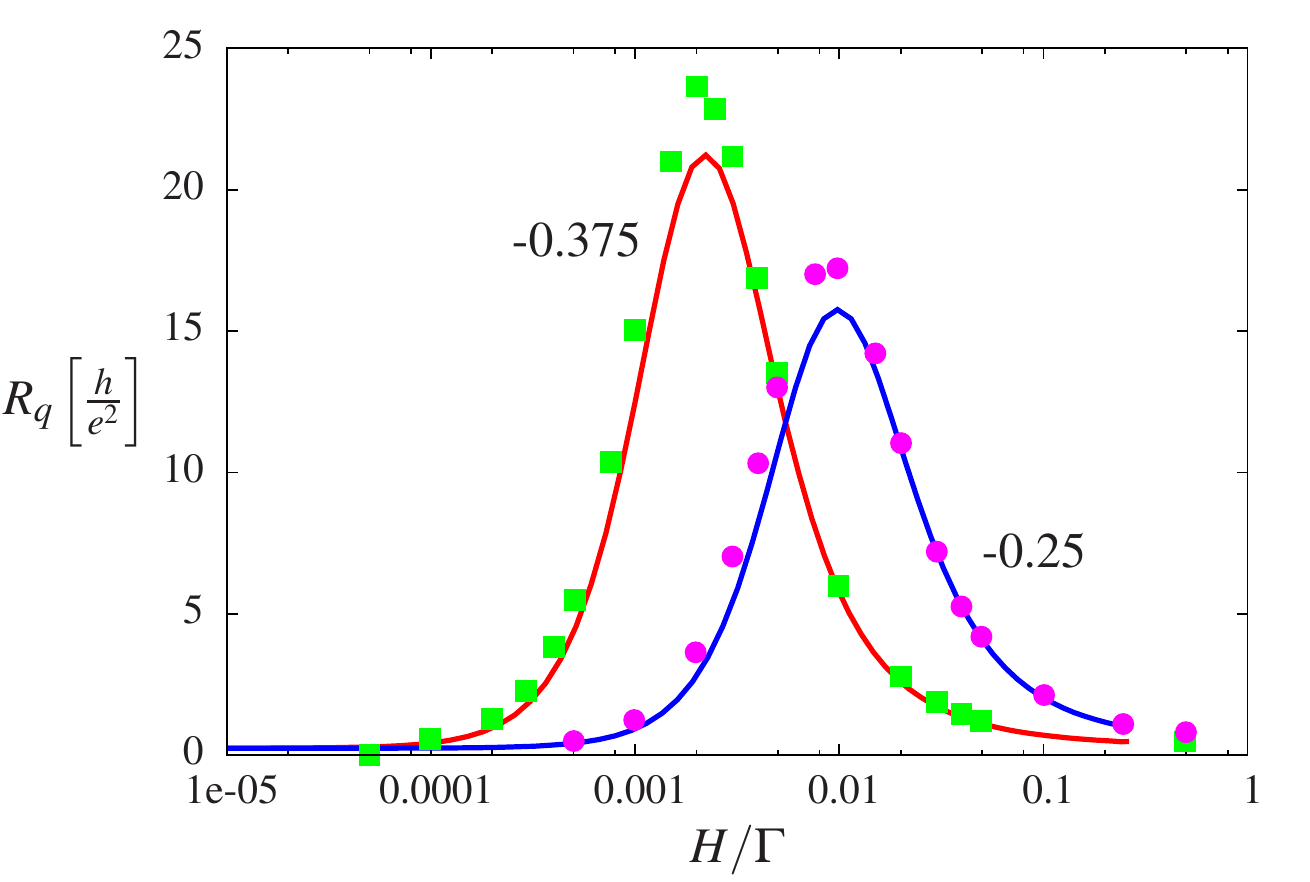}
\caption{{ Left --}  Dependence of $R_{\rm q}$ on the Zeeman splitting $\Delta_Z$ in the Kondo regime from Ref.~\cite{lee_effect_2011}. These results have been obtained by NRG calculations with $\Gamma=0.02$ and $E_{\rm c}=0.2$ [both quantities are measured in units of the contact bandwidth $D$ and the definition of the hybridization energy $\Gamma$ is provided in App.~\ref{sec:res_lev}]. They show that, for Zeeman energies of the order of the Kondo temperature, a giant non-universal peak appears in the charge relaxation resistance. {Right --} Comparison of $R_{\rm q}$ as a function of the magnetic field between NRG calculations (dots) (extracted from Ref.~\cite{lee_effect_2011}) and our Bethe ansatz results (solid lines) for different $\varepsilon_d/U$ and $U/\Gamma=20$, showing excellent agreement~\cite{filippone_giant_2011,filippone_admittance_2013}.}\label{fig:nigg}
\end{center}
\end{figure}

Nevertheless, the self-consistent approach misses important and sizable effects triggered by strong Kondo correlations. These were originally observed  relying on the numerical renormalization group (NRG)~\cite{lee_effect_2011}. The numerical results, reported in Fig.~\ref{fig:nigg}, showed that, for Zeeman splittings of the order of the Kondo temperature $T_{\rm K}$, the charge relaxation resistance can reach up to 100 times the universal value of $R_{\rm q}=h/(4e^2)$, which would be expected in the two-fold spin degenerate case. 
	
\begin{figure}[h!!]
\begin{center}
\includegraphics[width=.485\textwidth]{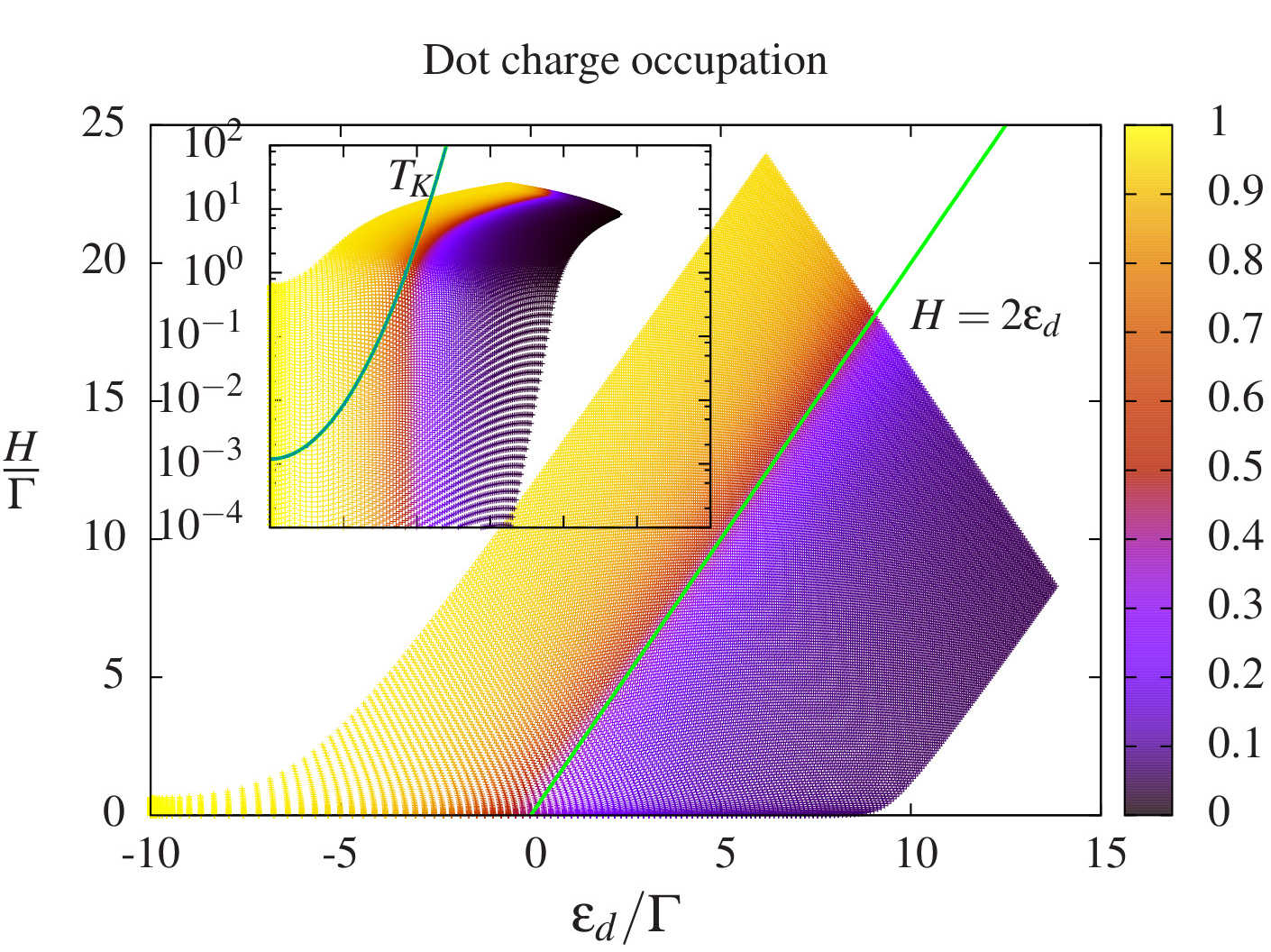}
\includegraphics[width=.485\textwidth]{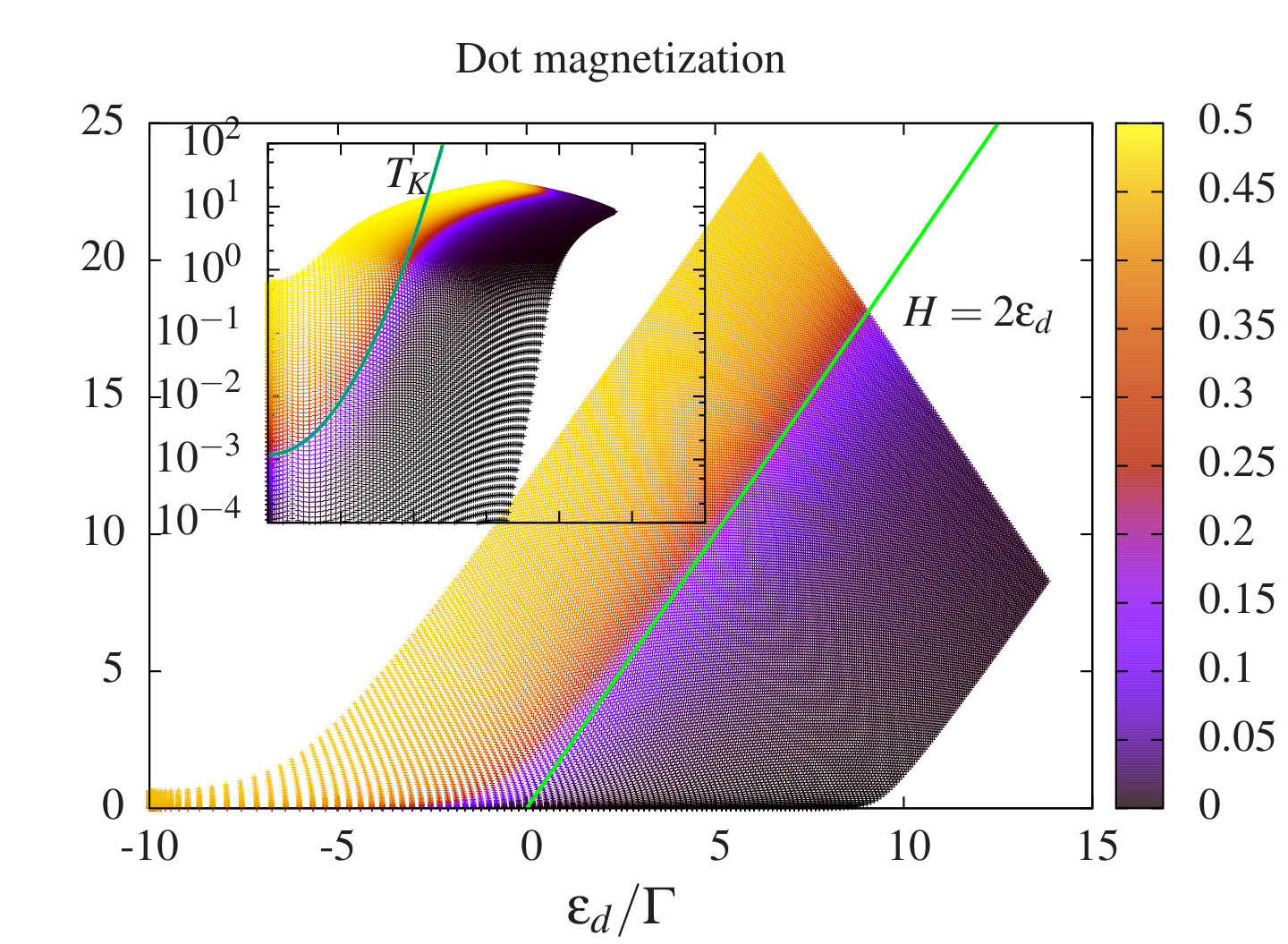}
\includegraphics[width=0.48\textwidth]{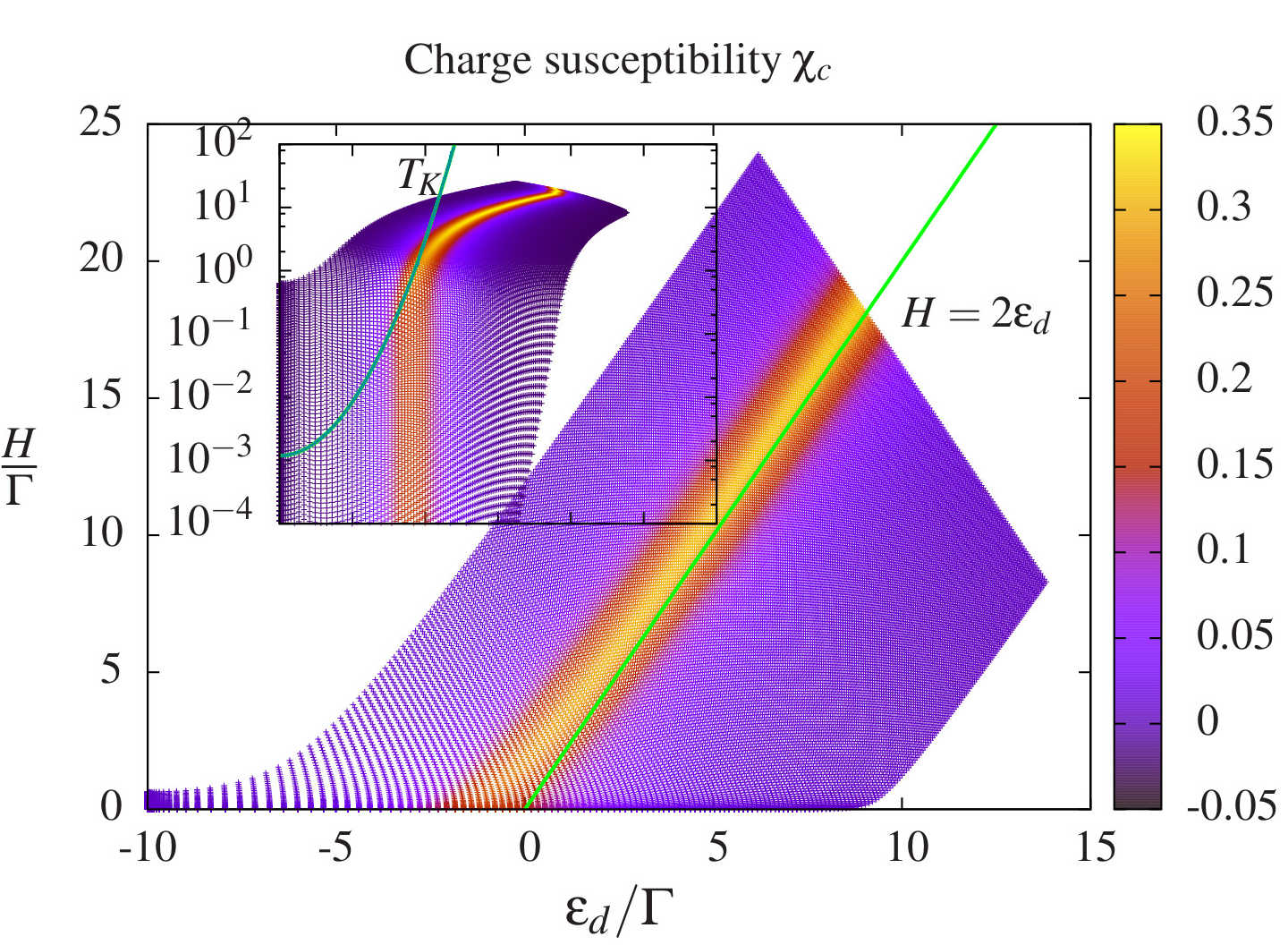}
\includegraphics[width=.48\textwidth]{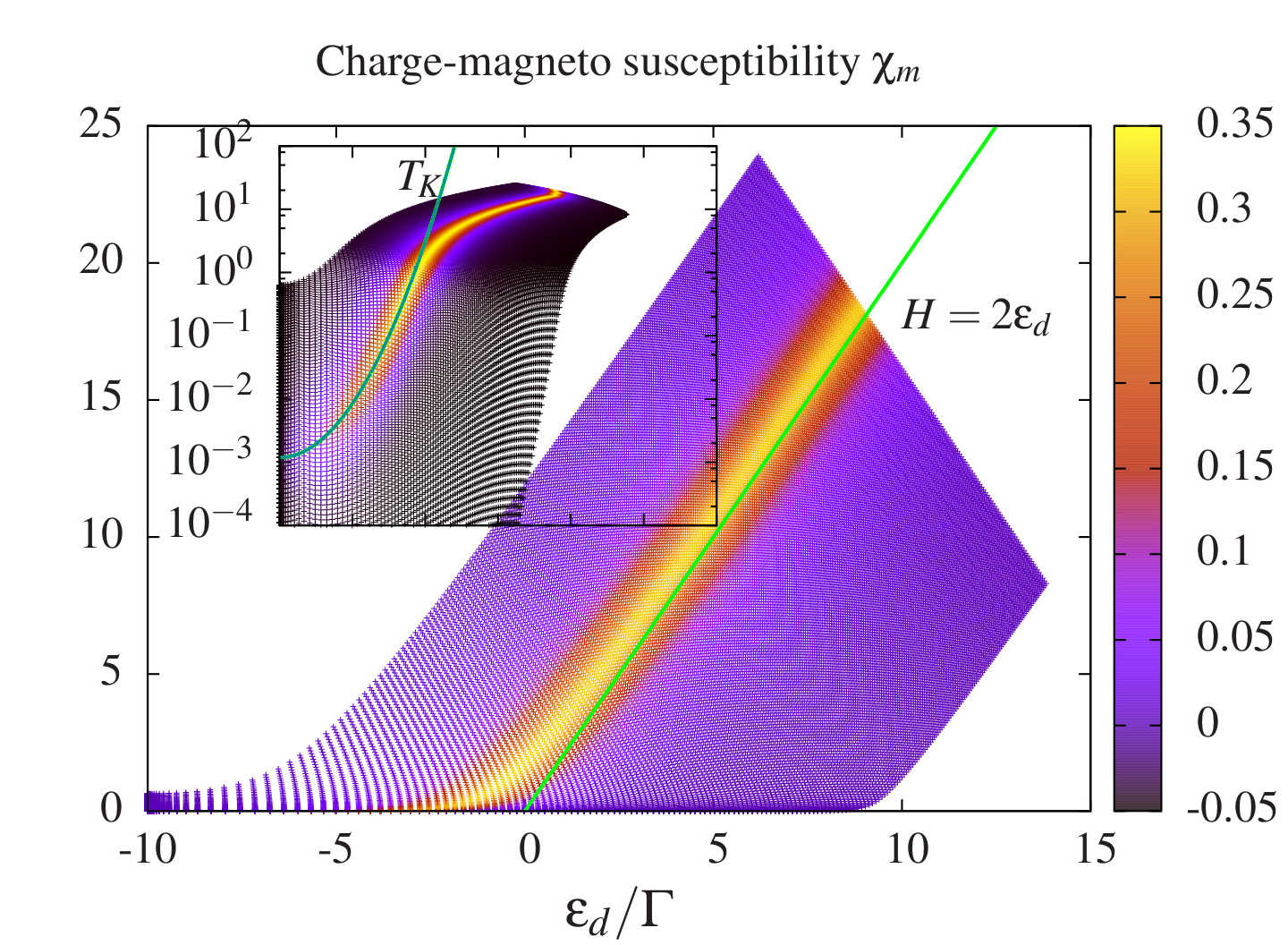}
\caption{ From Ref.~\cite{filippone_admittance_2013}. Top -- Charge occupation and magnetization of the dot for $U/\Gamma=20$ as function of the orbital energy $\varepsilon_d$ and magnetic field $H$. The insets show the same quantities on a logarithmic scale. The light green lines in the linear plots correspond to $H=2\varepsilon_d$, and separate regions with different charge occupations, while the dark green lines in the insets correspond to $T_{\rm K}$,  Eq.~\eqref{eq:tkoverview}, and separate regions with different magnetization.  The charge is not sensitive to the formation of the Kondo singlet for Zeeman energies below the Kondo temperature (green line), while the magnetization becomes zero.  Bottom -- Corresponding charge susceptibility and charge-magneto susceptibility.  The susceptibilities are in units of $1/\Gamma$. In the insets the same quantities are plotted on a logarithmic scale and the zone of appearance of the giant peak of the charge relaxation resistance can be appreciated. It is the region, following $T_{\rm K}$, in which $\chi_{\rm c}$ is close to zero, while $\chi_m$  acquires important values because of the formation of the Kondo singlet.}\label{fig:nm}
\end{center}
\end{figure}

The LFL approach allows the analytical quantification  and physical interpretation of such giant dissipative phenomenon~\cite{filippone_giant_2011,filippone_admittance_2013}. For two spin channels, the total charge on the dot is the sum of the two spin occupation $\langle N\rangle =\langle N_\uparrow\rangle+\langle N_\downarrow\rangle$.  Equation~\eqref{eq:rqgen} can then  be  recast in the useful form 
\begin{equation}\label{eq:rqspin}
R_{\rm q}=\frac h{4e^2}\left(1+\frac{\chi_m^2}{\chi_{\rm c}^2}\right)\,,
\end{equation}			
in which we  introduce the usual charge susceptibility $\chi_{\rm c}=-\partial \langle N\rangle/\partial\varepsilon_d$ and  the \textit{charge-magneto} susceptibility
\begin{equation}\label{eq:chim}
\chi_m=-2\frac{\partial \av {m}}{\partial \varepsilon_d}\,.
\end{equation}
This quantity is  twice the derivative of the dot magnetization $\av { m}=(\av { N_\uparrow}-\av{ N_\downarrow})/2$, with respect to the orbital energy $\varepsilon_d$. The charge-magneto susceptibility is an atypical object to study quantum dot systems, where the \emph{magnetic} susceptibility $\chi_H=-\partial \av{ m}/\partial H$ is rather considered to study the sensitivity of the local moment of the quantum dot to variations of the magnetic field $H$. Equation~\eqref{eq:rqspin} shows that the susceptibility of the \emph{magnetization} of the dot, and not its charge, is responsible for the departure from the universal quantization $h/(4e^2)$ of the charge relaxation resistance. Equation~\eqref{eq:rqspin} also separates explicitly charge and spin degrees of freedom of the electrons in the quantum dot. They can display very different behaviors in correlated  systems, as  illustrated in Fig.~\ref{fig:nm} in the Kondo regime, defined for one charge blocked on the dot and Zeeman energies below the Kondo temperature~\eqref{eq:tkoverview}.

In particular, Kondo correlations strongly affect the dot magnetization, but not its  occupation. The points where $\chi_m$ differs from $\chi_{\rm c}$ correspond to  non-universal charge relaxation resistances. In Fig.~\ref{fig:nigg}, the values derived with the LFL approach~\eqref{eq:rqspin} are compared to those obtained with NRG~\cite{lee_effect_2011}, showing excellent agreement. Additionally, the LFL approach also allows  to derive an exact analytical description of this peak, showing a genuinely  giant dissipation regime: simultaneous breaking of the SU(2) ($H\neq0$) and particle-hole symmetry ($\varepsilon_d\neq-U/2$) trigger a peak in $R_{\rm q}$ which  scales as the 4th(!) power of $U/\Gamma$ and has its maximum  for Zeeman splittings of the order of the Kondo temperature. This effect is caused by the fact that breaking the Kondo singlet by a magnetic field activates spin-flip processes  which dissipate energy through creation of particle-hole pairs~\cite{lee_effect_2011}.

\begin{figure}[t]
\begin{center}
\includegraphics[width=\textwidth]{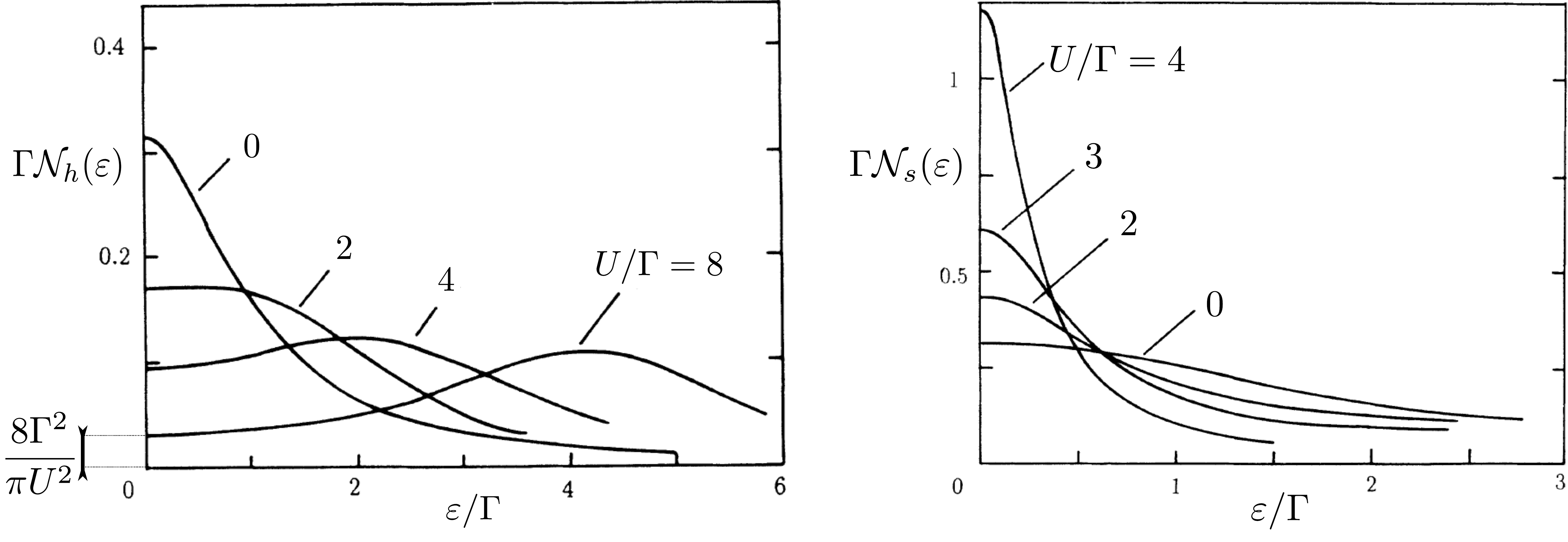}
\caption{From Ref.~ \cite{kawakami_density_1990}. Left -- Density of states of local holons on the dot $\mathcal N_h(\varepsilon)$. $\varepsilon$ is the excitation energy. A Coulomb peak emerges increasing the interaction parameter $U/\Gamma$ and vanishes at zero energy as $8\Gamma/\pi U^2$. This quantity coincides with  $\chi_c$,  plotted in Fig. \ref{fig:nm}. Right -- Density of states of local spinons $\mathcal N_s(\varepsilon)$. It behaves as the holonic one for $U/\Gamma=0$ and develops the Abrikosov-Suhl resonance at zero energy, the signature of the formation of the strongly correlated Kondo singlet state.}\label{fig:okiji}
\end{center}
\end{figure}

We conclude by discussing the deviations of the differential capacitance $C_0$ from the local density of states of the cavity, which is clearly apparent in Kondo regimes. The spin/charge separation arising in the AIM allows to observe important physical effects on the differential capacitance of strongly interacting systems. Charge and spin on the dot are carried by different excitations: holons and spinons. We report in Fig.~\ref{fig:okiji} the density of states of these excitations in the particle-hole symmetric case $\varepsilon_d=-U/2$. In the absence of interactions ($U/\Gamma=0$), they have the same shape, but they start to strongly  differ as   the interaction parameter $U/\Gamma$ is increased. They develop well pronounced peaks, but at different energies, signaling the appearance of separated charge and spin states.  In the case of holons, the excited charge state appears close to $\varepsilon=U/2$,  the energy required to change the dot occupation at particle-hole symmetry. In Ref.~\cite{kawakami_density_1990}, it is shown that the density of states of the holons  equals the static charge susceptibility $\chi_c$, coinciding  then with the differential capacitance $C_0$. At particle-hole symmetry, this quantity scales to zero as $8\Gamma/\pi U^2$, see Eq.~\eqref{eq:chicba}. Instead, the spinon density of states develops a sharp peak  at zero energy, known as the Abrikosov-Suhl resonance~\cite{suhl_dispersion_1965}, signaling the emergence of the strongly correlated Kondo singlet. . The differential capacitance $C_0$ is completely insensitive to this resonance, which dominates  the \textit{total} density of states on the dot.   Such effect was distinctly observed in carbon nanotube devices coupled to high-quality-factor microwave cavities~\cite{desjardins_observation_2017}. These systems efficiently probe the admittance~\eqref{eq:admittance} also in quantum dots with more than two internal degrees of freedom~\cite{schiro_tunable_2014,le_hur_driven_2018}, such as extensions of the AIM to SU(4) regimes, relevant for quantum dots realized with carbon nanotubes~\cite{lansbergen_tunable_2010,tettamanzi_magnetic-field_2012,borda_su4_2003,le_hur_smearing_2003,zarand_kondo_2003,lopez_probing_2005,filippone_admittance_2013,filippone_kondo_2014}.

\vspace{.2cm}

The above discussion completes the review of the application of the LFL theory to study the low-energy dynamics of quantum impurity driven systems. Further applications could be envisioned to  describe  various correlation effects on different aspects of weakly driven interacting quantum-dot systems, as long as they can be described by an effective theory of the form~\eqref{eq:potscatt}. An important case involves  the driving of the coupling term $\mathcal H_{ \rm res-dot}\rightarrow \mathcal H_{ \rm res-dot}(t)$ in the Hamiltonian~\eqref{eq:gen}, which has been implemented experimentally, with important metrologic applications~\cite{leicht_generation_2011,battista_spectral_2012,fletcher_clock-controlled_2013,waldie_measurement_2015,kataoka_time--flight_2016,johnson_ultrafast_2017}. An other interesting perspective concerns the application of the LFL theory to  energy transfer~\cite{ludovico_dynamical_2014,ludovico_adiabatic_2016,romero_nonlinear_2017},  or coupling quantum-dot systems to mechanical degrees of freedom~\cite{benyamini_real-space_2014,bode_scattering_2011,micchi_mechanical_2015,micchi_electromechanical_2016,avriller_andreev_2015,pistolesi_bistability_2018,schaeverbeke_single-photon_2019}, which are described by similar models as quantum-dot devices embedded in circuit-QED devices~\cite{delbecq_coupling_2011,delbecq_photon-mediated_2013,schiro_tunable_2014,liu_photon_2014,bruhat_cavity_2016,mi_circuit_2017,viennot_towards_2016}.

Deviations from universal and coherent behaviors are expected in non-LFL regimes, arising when the reservoirs are Luttinger Liquids~\cite{hamamoto_dynamic_2010,hamamoto_quantum_2011} or in over-screened Kondo impurities, in which the internal degrees of freedom of the bath surpass those of the impurity~\cite{mora_probing_2013,burmistrov_charge_2015}.

We have thus illustrated how a coherent and effectively non-interacting LFL theory  accounts for strong correlation effects in the dynamics of quantum dot devices. It has to be clarified how interaction are supposed to affect proper out-of-equilibrium regimes.  As a direct example, consider again the admittance~\eqref{eqn:rccircuit1}. Its expansion to low-frequencies completely reproduces the self-consistent predictions of Refs.~\cite{buttiker_dynamic_1993,buttiker_mesoscopic_1993,pretre_dynamic_1996}, but it `hides' the qualitative difference between the two time scales $\tau_{\rm c}$ and $\tau_{\rm f}$, associated to interactions and free-coherent propagation respectively. Higher-frequency driving will inevitably unveil this important difference, as we are going to demonstrate by giving a new twist to past experimental data in the next conclusive section.


\section{What about out-of-equilibrium regimes? A new twist on experiments. }\label{sec:ooe}

We conclude this review by showing how  interaction inevitably  dominate proper out-of-equilibrium or fastly driven regimes. We will focus, also in this case,  on the mesoscopic capacitor. In particular, we will show that past experimental measurements, showing fractionalization effects in out-of-equilibrium
charge emission from a driven mesoscopic capacitor~\cite{freulon_hong-ou-mandel_2015},  also manifest previously overlooked signatures of non-trivial many-body dynamics induced by interactions in the cavity.

As a preliminary remark,  notice that the circuit analogy~\eqref{eqn:rccircuit2} does not apply for a non-linear response to a gate voltage change or to fast (high-frequency) drives. An important example is a large step-like change in the gate voltage $V_{\rm g}(t)=V_{\rm g}\theta(t)$, $\theta(t)$ being the Heaviside step function, which is relevant to achieve triggered emission of quantized charge \cite{parmentier_current_2012}. Such a non-linear high-frequency response has been considered extensively for non-interacting cavities \cite{feve_-demand_2007,parmentier_current_2012,keeling_coherent_2008,olkhovskaya_shot_2008,moskalets_quantized_2008,sasaoka_single-electron_2010,moskalets_single-electron_2013}, where the current response to a gate voltage step at time $t=0$ was found to be of the form of simple exponential relaxation \cite{feve_-demand_2007,keeling_coherent_2008,moskalets_quantized_2008,moskalets_single-electron_2013}
\begin{equation}\label{eq:iscatt}
  I(t)\propto e^{-t /\tau_{\rm R}} \theta(t).
\end{equation}
For a cavity in the quantum Hall regime the relaxation time $\tau_{\rm R} = \tau_{\rm f}/(1-|r|^2)$, where $\tau_{\rm f}$ is the time of flight around the edge state of the cavity, see Figs.~\ref{fig:emission}-\ref{fig:c0}, and $r$ the reflection amplitude of the point contact.

There have been relatively few studies of the out-of-equilibrium behavior of the mesoscopic capacitor in the presence of interactions. The charging energy leads to an additional time scale $\tau_{\rm c} = 2 \pi \hbar C_{\rm g}/e^2$ for charge relaxation. The limit $1-|r|^2 \ll 1$ of a cavity weakly coupled to the lead, such that it can effectively be described by a single level, was addressed in Refs.\ \cite{splettstoesser_charge_2010,contreras-pulido_time_2012,kashuba_nonlinear_2012,alomar_time-dependent_2015,alomar_coulomb-blockade_2016,vanherck_relaxation_2017}. 

The full characterization of the out-of-equilibrium dynamics behavior of the mesoscopic capacitor, with a close-to-transparent point contact, was carried out in Ref.~\cite{litinski_interacting_2017}, extending the analysis of Ref.~\cite{mora_universal_2010} to non-linear response in the gate voltage $V_{\rm g}$.  A main result, spectacular in its simplicity, is that for a fully transparent contact ($r=0$) the linear-response admittance~\eqref{eqn:rccircuit1} also describes the non-linear response, {\it i.e.}, the correction terms in Eq.~\eqref{eq:admittance} vanish for an ideal point contact connecting cavity and lead \cite{safi_transport_1995,maslov_landauer_1995,ponomarenko_renormalization_1995,cuniberti_ac_1998}. The Fourier transform of the admittance~\eqref{eqn:rccircuit1} describes the real-time  evolution of the charge $Q(t)$ after a step change in the gate voltage.

\begin{figure}[ht!!]
\begin{center}
\includegraphics[width=1.\textwidth]{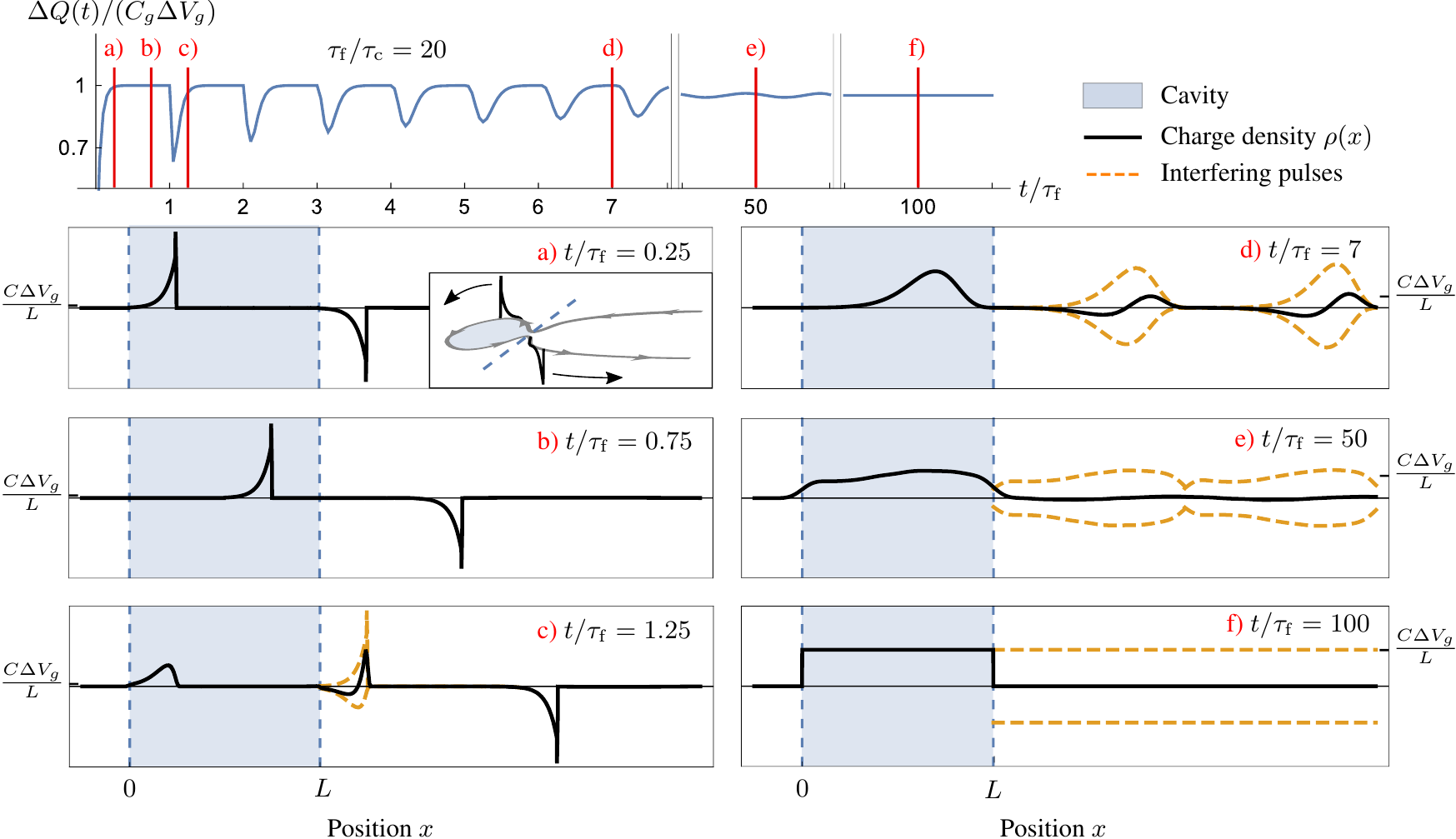}
\caption{From Ref.~\cite{litinski_interacting_2017}. Time-evolution of the current/charge density following a sudden gate voltage shift at time $t=0$ for large interaction strength, $\tau_{\rm f} = 20 \tau_{\rm c}$. Top -- Charge response $\Delta Q(t)$ as a function of time $t$. Bottom -- Series of snapshots of the current/charge density $j(x,t)$ at different times. In the inset of panel \textit{(a)}, the real-space representation (reproducing the one adopted in Fig.~\ref{fig:c0} with the dot site $\ell=L$) of the mesoscopic capacitor with the profiles of the emitted charge pulses is given. The times at which the snapshots are taken are indicated by vertical dashed lines in the top panel.  Notice that the scale changes along the vertical axis in the different panels.  At time $t=0$, two charge pulses of width $\sim v_F \tau_{\rm c}$ and opposite sign emerge from the point contact ($a$, $b$), one pulse entering the cavity and one pulse entering the chiral edge of the bulk two-dimensional electron gas. Both pulses have a net charge approaching $C_{\rm g} \Delta V_{\rm g}$. The pulse that is emitted into the cavity returns to the point contact at time $t=\tau_{\rm f}$. As that pulse leaves the cavity, a second pulse-antipulse pair is generated ($c$), partially canceling the original charge pulse that leaves the cavity at $t = \tau_{\rm f}$. The resulting pulse exiting the cavity is the sum of the dashed profiles. The repetition of this mechanism leads to the widening and lowering of successive pulses ($d$ and $e$) (notice the change of scale between snapshots). Finally, the asymptotic configuration is attained with a charge $C \Delta V_{\rm g}$ uniformly distributed along the cavity edge ($f$).}\label{fig:ooe}
\end{center}
\end{figure}

Figure~\ref{fig:ooe} illustrates  that initially, for times up to $\tau_{\rm f}$, $Q(t)$ relaxes exponentially with time $\tau_{\rm c}$, whereas at time $t = \tau_{\rm f}$ the capacitor abruptly enters a regime of exponentially damped oscillations, the period and the exponential decay of which are controlled by a complex function of $\tau_{\rm f}$ and $\tau_{\rm c}$, which does not correspond to any time scale extracted from low-frequency circuit analogies. This behavior is not captured by Eq.~\eqref{eq:iscatt}, derived in the non-interacting limit. These oscillations correspond to the emission of initially sharp charge density pulses, which are damped and become increasingly wider after every charge oscillation. Such complex dynamics is exquisitely coherent, but totally governed by interactions.

Additionally, it is also interesting to consider the effect of a small reflection amplitude $r$ in the point contact. In this case,  the charge $Q_r$ acquires nonlinear terms in the gate voltage $V_{\rm g}$,
\begin{equation}\label{eqn:backscattering5}
  Q_r(t)= Q(t) - \frac{e \tilde{r}}{\pi C} \int dt'\,  \mathcal A(t-t') \sin[2 \pi Q(t')/e],
\end{equation}
in which $\mathcal A(t)$ and $Q(t)$ are the Fourier transform of the admittance and charge for the case of a point contact with perfect transparency, $r=0$, see Eqs.\ \eqref{eq:admittance} and (\ref{eqn:rccircuit1}). The parameter $\tilde r$ involves both the (weak) backscattering amplitude $r$ and temperature $T$, details can be found in Ref.~\cite{litinski_interacting_2017}.

\subsection{Experimental signatures of the effects of interaction in quantum cavities driven out of equilibrium }

\begin{figure}[h!!]
\begin{center}
\includegraphics[width=.9\textwidth]{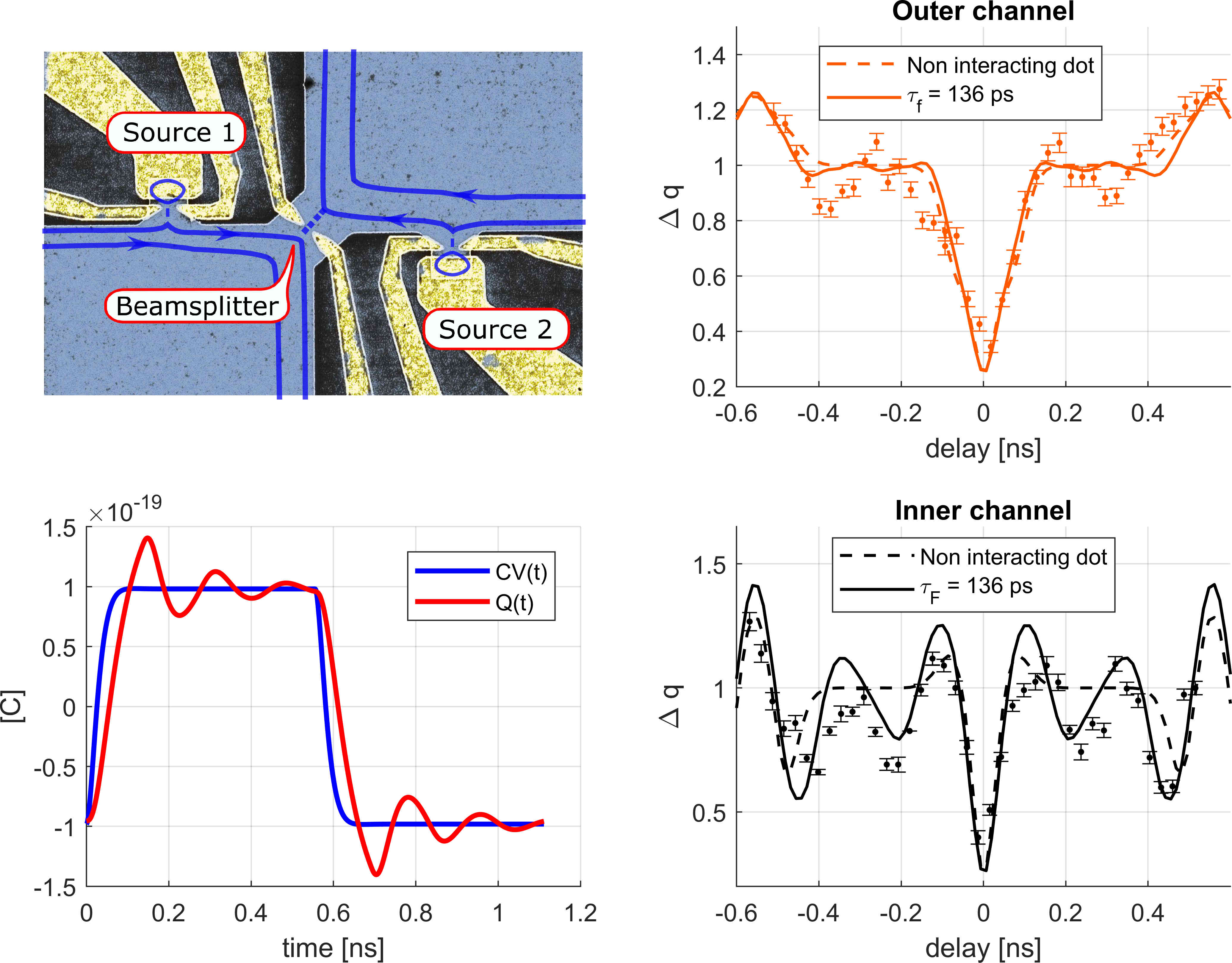}
\caption{ Top-Left -- Hong-Ou-Mandel experiment from Ref.~\cite{freulon_hong-ou-mandel_2015}: two single electron sources, as that shown in Fig.~\ref{fig:emission}, inject single electron towards the same QPC, which works as a beamsplitter. Bottom-left -- Simulation based on Ref.~\cite{litinski_interacting_2017} of the charge exiting the open dot when applying a square pulse sequence $V(t)$ with a rise time of 30 ps, one clearly sees the additional pulses coming from interactions.  Top-Right -- Normalized Hong-Ou-Mandel current noise $\Delta q$ of the outer edge as a function of the time delay $\tau$ with which charge arrives on the QPC from different sources. Noise is suppressed for $\tau=0$ because of anti-bunching effects, but additional oscillations were observed for $\tau\neq0$, which the theory in Ref.~\cite{litinski_interacting_2017} contributed to explain. The points are the experimental data while the solid and dashed lines are theoretical curves with different fittings for the time of flight $\tau_{ \rm f}$ for electrons in the cavity.  Bottom-Right --  Same as in the central panel but for the inner edge.}\label{fig:hom}
\end{center}
\end{figure}

The prediction that, in the open dot limit, interactions trigger the emission of a series of subsequent charge density pulses led to the possible explanation of additional effects that relate to a, so far not satisfactorily explained, part of the Hong-Ou-Mandel current noise measurements at the LPA~\cite{freulon_hong-ou-mandel_2015,marguerite_two-particle_2017}. The experimental setup is the solid state realization of the Hong-Ou-Mandel experiment, see left panel Fig.~\ref{fig:hom}: when two electrons collide at the same time on the QPC from different sources (states 1 and 2), they cannot occupy the same state because of Pauli's exclusion principle and their probability to end up in different leads (states 3 and 4) is increased. As a consequence, the current noise is suppressed~\cite{bocquillon_coherence_2013}, see the right panels in Fig.~\ref{fig:hom}. More generally, $\Delta q(\tau)$ measures the cross-correlation (or overlap) in time of the two incoming current at the level of the QPC. If the two incoming currents are identical in each input, one should get $\Delta q(\tau=0)=0$ and the rest of the curve will reflect on the time trace of the current. However, because of small asymmetries in the two electronic paths and the two electron sources, the noise suppression is not perfect~\cite{marguerite_two-particle_2017}. In Ref.~\cite{freulon_hong-ou-mandel_2015}, the current noise $\Delta q$ as a function of the time delay $\tau$ with which electrons arrive at the QPC from different sources was measured in more detail for the outer and inner edge of the filling factor $\nu=2$ (central and right panel in Fig.~\ref{fig:hom}). The current in the inner edge channel is induced by inter-edge Coulomb interactions and can be computed with a plasmon-scattering formalism~\cite{degiovanni_plasmon_2010,freulon_hong-ou-mandel_2015}. In addition to what this plasmon scattering model predicted, unexpected oscillations as a function of $\tau$ were observed. These could be satisfactorily explained by our prediction \cite{litinski_interacting_2017} of further charge emission triggered by interactions in the electron sources  \cite{marguerite_two-particle_2017}. 
From independent calibration measurements, the total RC time constant of the source could be measured to set the constrain $(\tau_{\rm f }^{-1}+\tau_c^{-1})^{-1}/2=\tau_{RC}= 21$ ps. Combining Eqs.\ \eqref{eq:admittance} and (\ref{eqn:rccircuit1}) with the plasmon scattering formalism one can compute the current noise in the ouput of the Hong-Ou-Mandel interferometer $\Delta q(\tau)$ with only one fitting parameter: the ratio $\tau_{\rm f }/\tau_{\rm c}$. The minimization procedure gave one most-likely result: $\tau_{\rm f } = 136$ ps. The comparison is shown on the right panels of Fig.~\ref{fig:hom}, where the model for $\tau_{\rm f } = 2\tau_{RC}$ [$\tau_{\rm c }\rightarrow\infty$, \textit{i.e.} no interaction within the dot] describes the fractionalization process due to interedge Coulomb interactions but not the interactions within the dot itself. This provides a reasonable qualitative and quantitative agreement with the experimental data reported in Ref.~\cite{freulon_hong-ou-mandel_2015}. On the bottom-left panel in Fig.~\ref{fig:hom} we compare, for $\tau_{\rm f} =136$ ps, the charge exiting the dot $Q(t)$ with the applied square pulse sequence on the top-gate $V(t)$ which has a finite rise time of 30 ps. In particular, it can explain the appearance of extra rebounds in $\Delta q$ for time delays $\tau$ between 70 and 450 ps which is not possible with a non-interacting dot ($\tau_{\rm f }=0$). This is directly due to the additional effects coming from the interactions within the dot itself and cannot be explained by the fractionalization mechanism. Indeed, relying exclusively on the model describing fractionalization, we could not reproduce the pronounced additional rebound for $|\tau|=200$ ps for $\tau_{\rm f }<100$ ps. This highlights the relevance of Coulomb interactions in the open dot dynamics.


\section{Conclusions}

This review addressed the importance of interactions for the investigation and control of dynamical quantum coherent phenomena in mesoscopic quantum-dot devices. 

In Section~\ref{sec:coherence}, we discussed how interactions are the essential ingredient allowing long-range quantum state transfer in mesoscopic devices. Notice that the same phenomenon has been suggested to enforce nonlocal phase-coherent electron transfer in wires supporting topologically protected Majorana modes at their edges~\cite{fu_electron_2010,shi_long_2020}. Such effect is currently being considered  in various Majorana  network  models  for  stabilizer  measurements in  corresponding  implementations  of  topological  quantum  error  correction  codes~\cite{vijay_majorana_2015,plugge_roadmap_2016,litinski_combining_2017}.

The local Fermi liquid approach, discussed in Section~\ref{part:overview}, provides the unifying theoretical framework to describe the low-energy dynamics of such various mesoscopic devices. Its application to the various experimental setups mentioned in the Introduction, will be definitively useful to bring further understanding in the complex and rich field of out-of-equilibrium many-body systems. The insight given on universal quantum dissipation phenomena, discussed for the mesoscopic capacitor in Section~\ref{sec:meso}, and, in particular, the novel interaction effects, unveiled in the experiment discussed in Section~\ref{sec:ooe}, give two clear examples of the utility of this approach. 

Beyond the already mentioned potential for quantum dot devices coupled to microwave cavities~\cite{schiro_tunable_2014,le_hur_many-body_2016,le_hur_driven_2018,delbecq_coupling_2011,delbecq_photon-mediated_2013,liu_photon_2014,bruhat_cavity_2016,mi_circuit_2017,viennot_towards_2016,deng_quantum_2015} and to  energy transfer~\cite{ludovico_dynamical_2014,ludovico_adiabatic_2016,romero_nonlinear_2017}, important extensions  of the LFL  approach should be envisioned for understanding the properties of  mesoscopic devices involving non-Fermi liquids at the place of normal metallic leads. The most important cases would  involve superconductors~\cite{van_zanten_single_2016,basko_landau-zener-stueckelberg_2017,moca_noise_2018,ferrier_universality_2016,delagrange_manipulating_2015,delagrange_0_pi_2016,delagrange_emission_2018,crepieux_emission_2018} or fractional Quantum Hall edges states, in which quantum noise measurements have been crucial to address and unveil the dynamics of fractionally charged excitations~\cite{saminadayar_observation_1997,reznikov_observation_1999,kapfer_josephson_2019,bisognin_microwave_2019,chamon_tunneling_1995,bena_emission_2007,safi_time-dependent_2011,safi_driven_2019,safi_partition_2001,guyon_klein_2002,kim_signatures_2005,bena_effects_2006,carrega_spectral_2012,ferraro_multiple_2014,ferraro_relevance_2008,crepieux_photoassisted_2004,roussel_perturbative_2016,bartolomei_fractional_2020}. Additionally, the recent realization of noiseless levitons ~\cite{levitov_electron_1996,ivanov_coherent_1997,keeling_minimal_2006,dubois_minimal-excitation_2013,jullien_quantum_2014} paves the way  to interesting perspectives to investigate flying anyons~\cite{rech_minimal_2017,kapfer_josephson_2019,c_design_2020} and novel interesting dynamical effects~\cite{hamamoto_dynamic_2010,hamamoto_quantum_2011,wagner_driven_2019}. 



\vspace{6pt} 



\authorcontributions{M.F. prepared and finalized the original draft of this review, C.M. gave substantial suggestion for its structure and all the authors equally contributed to revise the manuscript. A. M. and G. F. realized the original data analysis presented in Fig.~\ref{fig:hom}. }

\funding{M.F. acknowledges support from the FNS/SNF Ambizione Grant PZ00P2\_174038 and G.F. is funded by ERC consolidator grant ``EQuO'' (no. 648236).}

\acknowledgments{M.F. is indebted to Christopher B\"auerle, G\'eraldine Haack, Fr\'ed\'eric Pierre, In\`es Safi and Eugene Sukhorukov for important comments and suggestions. }

\conflictsofinterest{The authors declare no conflict of interest. The founders had no role in the design of the study; in the collection, analyses, or interpretation of data; in the writing of the manuscript, or in the decision to publish the results.} 

\abbreviations{The following abbreviations are used in this manuscript:\\

\noindent 
\begin{tabular}{@{}ll}
QPC & quantum point contact\\
LFL & local Fermi liquid\\
CBM & Coulomb blockade model\\
AIM & Anderson impurity model\\
SW & Schrieffer-Wolff\\
DC & direct current\\
2DEG & two-dimensional electron gas\\
KS & Korringa-Shiba\\
AC & alternate current\\
NRG & numerical renormalization group 
\end{tabular}}


\appendixtitles{no} 
\appendix


\section{Scattering theory, phase-shifts and the Friedel sum rule }\label{app:scattering}
In this Appendix we review some useful results from scattering theory. In Section~\ref{sec:general_scatt}, we provide the definition of the S- and T-matrix in scattering theory. In Section~\ref{sec:fsr}, we derive the Friedel sum rule for a non-interacting electron gas with an elastic impurity. In Sections~\ref{sec:res_lev}, we illustrate these concepts on the simple case of a chiral edge state tunnel coupled to a single resonant level. In Section~\ref{app:tmatrix}, we derive the phase-shift induced on lead electrons by the scattering potential of the LFL, Eq.~\eqref{eq:potscatt}.

\subsection{General definitions}\label{sec:general_scatt}

\begin{figure}[t]
\begin{center}
\includegraphics[width=0.7\linewidth]{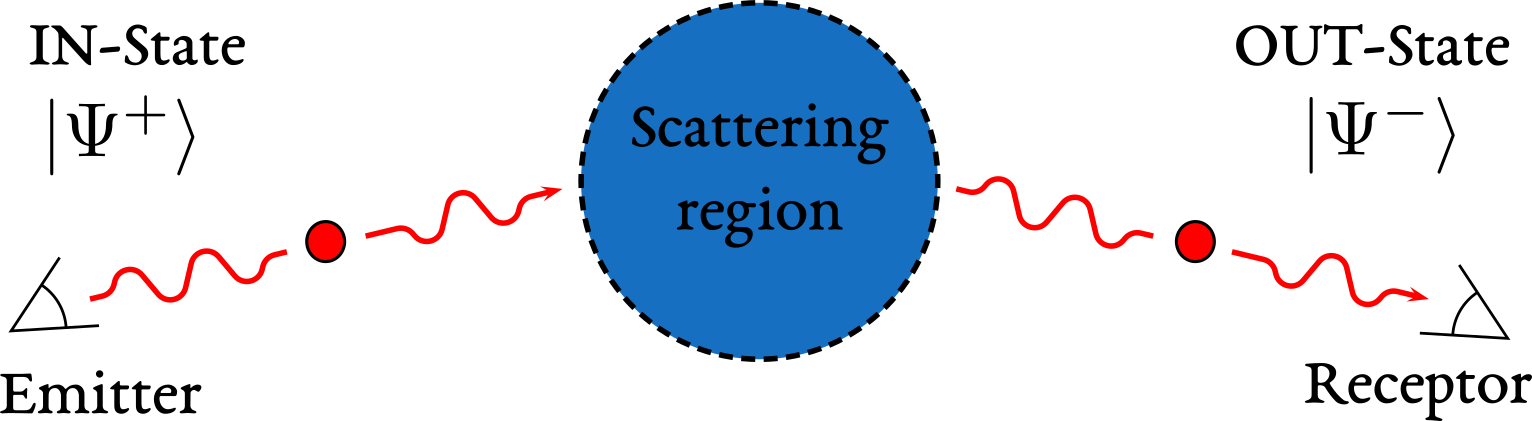}
\includegraphics[width=0.7\linewidth]{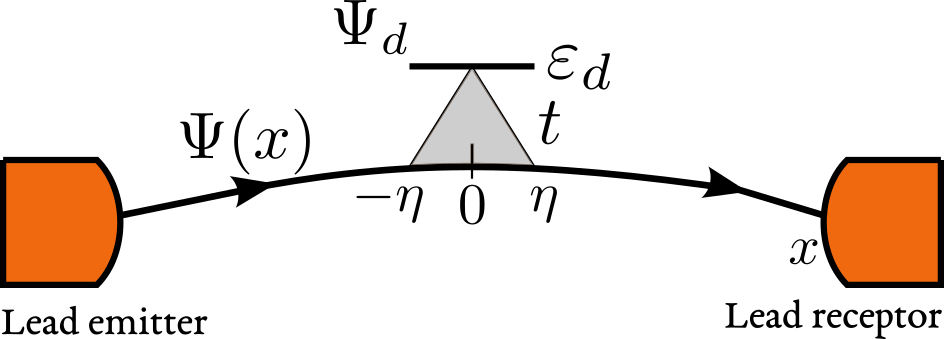}
\caption{Top -- Illustration of the physical situation described by the scattering formalism. Electron wave packets are emitted in the IN-State $\ket{\Psi^+}$ and then measured in the OUT-state $\ket{\Psi^-}$ once they have passed through the scattering region. Bottom -- Realization of the scattering setup with a quantum Hall chiral edge state tunnel coupled (in a region of size $2\eta$) with a resonant level of energy $\varepsilon_d$. }\label{fig:scattering}
\end{center}
\end{figure}

 We aim at describing the general situation  of Fig.~\ref{fig:scattering}, which is reproduced by mesoscopic settings, such as a single resonant level coupled to a chiral edge state, which also describes the mesoscopic capacitor in the non-interacting limit, see also Fig.~\ref{fig:c0} in the main text. Consider  a wave packet emitted and  detected in the distant past and future, namely $t=-\infty$ and $t=+\infty$, and which enters a scattering region at time $t=0$. Close to detection and emission, it is assumed that the wave packet does not feel the presence of the scatterer, whose interaction range  is delimited  inside the dashed line in Fig.~\ref{fig:scattering}. The system is described by a single-particle Hamiltonian of the form
\begin{equation}\label{eq:hamham}
\mathcal H=\mathcal H_0+\mathcal V\,,
\end{equation}
in which $\mathcal H_0$ describes the free propagation of a wave-packet and $\mathcal V$ the scatterer. The T-matrix describes the effects of a scatterer on the propagation of a free particle. It is an improper self-energy for the resolvent of the lead electrons, which  appears in a modified form of Dyson's equation
\begin{align}\label{eq:tdef}
G(z)&=G_0(z)+G_0(z)T(z)G_0(z)\,,&G(z)&=\frac1{z-\mathcal H}\,,
\end{align}
in which z is a complex number. $G_0$ is the free resolvent describing free electrons 
\begin{equation}
G^0_{kk'}(z)=\bra k G^0(z) \ket {k'}=\frac{\delta_{kk'}}{z-\varepsilon_k},
\end{equation}
the $\ket k$ states being the single particle eigenvectors of the unperturbed Hamiltonian $\mathcal H_0$. One can readily show that the T-matrix reads
\begin{equation}\label{eq:soltmat}
\mbox{T}(z)=V(\mathbb I-G_0(z)\mathcal V)^{-1}\,.
\end{equation}
The general definition of the phase-shift, a key quantity within scattering theory~\cite{weinberg_quantum_1996,newton_scattering_2002}, reads
\begin{equation}\label{eq:deltath}
\delta_\varepsilon=\arg\left[\mbox{T}(\varepsilon+i0^+)\right]\,.
\end{equation}

We define the IN and OUT states $\ket{\Psi^\pm}$ as the eigenvectors of energy $\varepsilon$ of the Hamiltonian of the whole system, including the scattering region, as the states coinciding asymptotically with free plane waves in the past and in the future respectively. The scattering matrix  S  gives the overlap between these two states
\begin{equation}
\mbox { S}_{kk'}(\varepsilon)=\langle\Psi^-_k|\Psi^+_{k'}\rangle\,,
\end{equation}
where $k/k'$ are the momenta of the OUT/IN states. The  T- and S-matrix are related by the relation~\cite{weinberg_quantum_1996,newton_scattering_2002}
\begin{equation}\label{eq:strel}
\mbox {S}_{kk'}=\delta_{kk'}-2\pi i \delta(\varepsilon_k-\varepsilon_{k'})\mbox{T}_{kk'}\,,
\end{equation}
or equivalently, in the energy representation,
\begin{equation}\label{eq:stmatrix}
\mbox{S}(\varepsilon)=\mathbb I-2\pi i\nu_0 \mbox{T}(\varepsilon)\,.
\end{equation}

The S-matrix is unitary and in the single channel case it is completely defined by a phase $\mbox{S}(\varepsilon)=e^{2i\delta_\varepsilon}$. The phase $\delta_\varepsilon$ is the phase-shift caused by  scattering and in general the condition $\mbox{S}(\varepsilon)=e^{2i\delta_\varepsilon}$ is always verified if we take the definition of the phase-shift directly from the T-matrix
\begin{equation}\label{eq:defphaset}
\mbox{T}(\varepsilon)=-\frac1{\pi\nu_0}\sin \delta_\varepsilon e^{i\delta_\varepsilon}\,.
\end{equation}
There is an interesting connection with the Kondo regime. In Section~\ref{sec:aim}, we illustrated how particle-hole symmetry enforces that the phase-shift is given by  $\delta=\delta_{\rm K}=\pi/2$. The scattering matrix thus equals the identity. This case is also known as the unitary limit of the Kondo model, in which the transmission probability through a Kondo correlated dot is unity.


\subsection{The Friedel sum rule}\label{sec:fsr}
We show here the Friedel sum rule for non-interacting electrons scattering on an elastic impurity.  We first consider the total electron occupation $\av{ \mathcal N }$ of an electron gas, which reads 
\begin{equation}\label{eq:totalnumber}
\av{\mathcal N}=\sum_\alpha\int_{-\infty}^\infty d\omega A_\alpha(\omega)f(\omega)\,,
\end{equation}
the sum on the label $\alpha$ running over all the eigenstates of the Hamiltonian~\eqref{eq:hamham}. $A_\alpha(\omega)$ is the spectral function of the state $\alpha$, defined as
\begin{equation}\label{eq:specral}
A_\alpha(\omega)=-\frac1\pi\mbox{Im}  G_{\alpha\alpha}(\omega+i0^+)=-\frac1\pi\mbox{Im} \langle\alpha| G(\omega+i0^+)|\alpha\rangle\,,
\end{equation}
$G_{\alpha\alpha}$ being the \emph{retarded} Green's function associated to the state $\alpha$ defined in Eq.~\eqref{eq:tdef}. In the absence of the scatterer, $G_{kk}(\omega+i0^+)=G_0(\omega+i0^+)=(\omega+i0^+-\varepsilon_k)^{-1}$ and Eq.~\eqref{eq:totalnumber} reduces to a sum over the Fermi function $\sum_k f(\varepsilon_k)$ giving the total number of electrons $\av{\mathcal N	}$ in the system. If the chemical potential is fixed, the introduction of the scatterer modifies the total average number of electrons. The difference with the initial  one gives the amount of electrons $\av N$ displaced by the scatterer. In the case of a single channel, one obtains 
\begin{equation}\label{eq:nres}
\av{\mathcal N}_{\rm with~scatterer}-\av{ \mathcal N}_{\rm without~scatterer }=\av{N}=-\frac1\pi\mbox{Im}\int_{-\infty}^\infty d\omega \mbox{Tr}[G(\omega)-G_0(\omega)]f(\omega)\,.
\end{equation}
Using Eq.~\eqref{eq:soltmat} and the fact that $\frac d{d\varepsilon}\mbox{Tr}\log[A(\varepsilon)]=\mbox{Tr}\left[A^{-1}(\varepsilon)\frac d{d\varepsilon}A(\varepsilon)\right]$, one finds
\begin{equation}
\begin{split}
\av N&=-\frac i {2\pi} \int_{-\infty}^\infty d\omega \,f(\omega)\frac d{d\omega}\log\det[\mathbb I-2\pi i\nu_0\,\mbox T(\omega+i0^+)]\\
&=-\frac i {2\pi} \int_{-\infty}^\infty d\omega\,f(\omega) \frac d{d\omega}\log\det\, \mbox S(\omega+i0^+)\,,
\end{split}
\end{equation}
in which we applied the definition~\eqref{eq:stmatrix} of the S-matrix. This is the general form of the  Friedel sum rule. For $\mbox{S}(\varepsilon)=e^{2i\delta_\varepsilon}$ and zero temperature it gives a direct relation between the charge displaced by the impurity and the phase-shift at the Fermi energy $E_F$: 
\begin{equation}\label{eq:friedel_app}
\av{N}=\frac {\delta_{E_F} }\pi\,,
\end{equation}
The extension to $M$ channels requires to add an overall sum over the channel label $\sigma$ and leads to  Eq.~\eqref{eq:friedel} in the main text.   

\subsection{Illustration on the resonant level model}\label{sec:res_lev}

We illustrate now the above concepts on a simple situation, sketched in Fig.~\ref{fig:scattering}, in which  the scatterer is a single  resonant level tunnel coupled to chiral electrons propagating on a an edge state. Such situation is an effective representation of the mesoscopic capacitor, see Figs.~\ref{fig:c0} and~\ref{fig:setup}, and it is described by the Coulomb Blockade Model (CBM)~\eqref{eq:cbm}, that we remind here to the reader
\begin{equation}\label{eq:cbm_app}
\mathcal H_{\rm CBM}=\sum_{k}\varepsilon_{k}c^\dagger_kc_k+t\sum_{k,l}\Big[c^\dagger_k d_l+d^\dagger_l c_k\Big]+\sum_l(\varepsilon_d+\varepsilon_l) d^\dagger_ld_l+E_{\rm c}\Big(N-\mathcal N_{\rm g}\Big)^2\,.
\end{equation}
In this section, we neglect the last term, corresponding to interactions, and, for simplicity, we retain only a single fermionic level (annihilated by the fermion operator $d$) for the cavity, with $\varepsilon_l=0$. One thus obtains the Hamiltonian of a resonant level 
\begin{equation}\label{eq:res}
\mathcal H_{\rm Res}=\sum_{k}\varepsilon_k c^\dagger_{k}c_{k}+t\sum_{k}\left(c^\dagger_{k}d+d^\dagger c_{k}\right)+\varepsilon_{d} d^\dagger d\,.
\end{equation}
We first calculate the occupation of the cavity by calculating the retarded Green's functions, defined as  $G_{dk}(t-t')=-i\theta(t-t')\av{\{d(t),c_k^\dagger(t')\}}$.  They are derived by solving the equations of motion in frequency space
\begin{align}
(\omega-\varepsilon_d)G_{dd}(\omega)&=1+t\sum_kG_{kd}(\omega)\,, & (\omega-\varepsilon_k)G_{kk'}(\omega)&=\delta_{kk'}+tG_{dk'}(\omega)\,,\\
(\omega-\varepsilon_k)G_{kd}(\omega)&=tG_{dd}(\omega)\,,  & (\omega-\varepsilon_d)G_{dk}
&=t\sum_{k'}G_{k'k}(\omega)\,.
\end{align}
Solving the system, the Green's function for the lead electrons reads 
\begin{equation}\label{eq:gkres}
G_{kk'}(\omega)=\frac{\delta_{kk'}}{\omega-\varepsilon_k}+\frac1{\omega-\varepsilon_k}t^2G_{dd}(\omega)\frac1{\omega-\varepsilon_{k'}}\,.
\end{equation}
Writing Eq.~\eqref{eq:gkres} in the form $G=G^0+G^0TG^0$,   the T-matrix is found to be
\begin{align}\label{eq:phaseres}
\mbox T(\omega+i0^+)&=t^2G_{dd}(\omega+i0^+)=\frac{t^2}{\sqrt{(\omega-\varepsilon_d)^2+\Gamma^2}}e^{i\delta_\omega}\,, & \delta_\omega&=\frac\pi2-\arctan\left(\frac{\varepsilon_d-\omega}{\Gamma}\right)\,.
\end{align}
in which we introduced the hybridization constant 
\begin{equation}\label{eq:gamma}
\Gamma=t^2\sum_k(\omega+i0^+-\varepsilon_k)^{-1}\sim\pi\nu_0t^2\,,
\end{equation}
which corresponds to the width acquired by the resonant level  by coupling to the lead and which depends on  the density of states of the lead electrons  at the Fermi energy  $\nu_0$. 

We can now determine the number of displaced charges $\langle N\rangle$ as given by Eq.~\eqref{eq:nres} and show the validity of the Friedel sum rule~\eqref{eq:friedel_app} in this example.   In the wide-band limit, the contribution from the second term in Eq.~\eqref{eq:gkres} can be neglected.
 The number of displaced electrons is given solely by the quantum dot Green's function $G_{dd}$:
\begin{equation}\label{eq:nres}
\av{ \mathcal N}_{\rm with~dot}-\av{ \mathcal N}_{\rm without ~dot }=\av{ N}=-\frac1\pi\mbox{Im}\int_{-\infty}^\infty d\omega G_{dd}(\omega)f(\omega)=\frac12-\frac1\pi\arctan\left(\frac{\varepsilon_d}\Gamma\right)\,,
\end{equation}
which is consistent with the Friedel sum rule~\eqref{eq:friedel_app} with the phase-shift~\eqref{eq:phaseres}. Equation~\eqref{eq:nres} is also meaningful because: {\it i)} it shows that, in the wide-band approximation, the number of displaced electrons $\av N$ is given by the local Green's function $G_{dd}$, which  can be interpreted as the charge occupation of the quantum dot; {\it ii)} the number of displaced electrons depends on the orbital energy $\varepsilon_d$. As a consequence, a time-dependent variation of $\varepsilon_d$    drives a current in the system by  displacing electrons in the leads.

As an additional illustration,  clarifying how the scattering phase-shift appears on single particle wave-functions, we also solve  explicitly the same problem in its real-space formulation. We consider  chiral fermions $\Psi(x)$ which  tunnel on the resonant level of energy $\varepsilon_d$ (and wave-function amplitude $\Psi_d$), from  a region of size $2\eta$ centered around $x=0$, useful to properly regularize the calculation and to  be sent to zero at the end~\cite{filippone_tunneling_2016}. For lead electrons at the Fermi energy, that we set to zero ($\omega=0$ in Eq.~\eqref{eq:phaseres}), the Schr\"odinger equation reads (for $x\in[-\eta,\eta]$)
\begin{align}\label{eq:scatt1}
0&=-i\hbar v_F\partial_x\Psi(x)+\frac{t}{2\eta}\Psi_d\,, & 0&=\varepsilon_d\Psi_d+\frac t{2\eta}\int_{-\eta}^\eta dx'\Psi(x')\,.
\end{align}
By integrating the first equation in the interval $[-\eta,x<\eta]$ and inserting the result for $\Psi(x)$ in the second one, one finds
\begin{align}\label{eq:scatt2}
\Psi(x)&=\Psi(-\eta)+\frac{t}{2\eta i\hbar v}(x+\eta)\Psi_d\,, & \Psi_d&=\frac{-t}{\varepsilon_d-i\Gamma}\Psi(-\eta)\,.
\end{align}
Notice that $\eta$ does not appear on the last equality and can be sent safely to zero. We consider, as boundary condition for $\Psi(x)$, incoming scattering states of the form $\Psi(x<0)=e^{ikx}/\sqrt{2\pi\hbar v_F}=\sqrt{\nu_0}e^{ikx}$. One thus finds 
\begin{align}
|\Psi_d|^2&=\frac{1}{\pi}\frac{\Gamma}{\varepsilon_d^2+\Gamma^2}\,, & \Psi(0^+)&=\sqrt{\nu_0}e^{2i\delta_0}\,,&\delta_0&=\frac\pi2-\arctan\left( \frac{\varepsilon_d}\Gamma\right)\,,
\end{align}
in agreement with Eq.~\eqref{eq:phaseres}. This short calculation illustrates how, just after scattering with the resonant level, the electron wave-packet at the Fermi energy acquires a phase $e^{2i\delta_0}$ which is fixed by the resonant level occupation via the Friedel sum rule.  Notice that, at the resonance condition for the orbital energy ($\varepsilon_d=0$), $\delta_0=\pi/2$ and $\Psi(0^+)+\Psi(0^-)=0$, as for the unitary limit in the Kondo model, in which $\delta_{\rm K}=\pi/2$, see also the discussion in Section~\ref{sec:aim}.


\subsection{$T$-matrix in the potential scattering Hamiltonian}\label{app:tmatrix}
We derive now  the phase-shift caused by the potential scattering term on lead electrons in a local Fermi liquid (LFL). It is useful to recall here the LFL Hamiltonian~\eqref{eq:potscatt} 
\begin{equation}\label{eq:potscatt_app}
\mathcal H_{\rm LFL}=\sum_{k\sigma}\varepsilon_k c^\dagger_{k\sigma}c_{k\sigma}+W(\varepsilon_d,E_{\rm c},\ldots)\sum_{k\neq k'\sigma}c^\dagger_{k\sigma}c_{k'\sigma}\,.
\end{equation}
We focus on the the single-channel case for simplicity ($M=1$). The generalization to $\sigma=1,\ldots M$ channels is straightforward. The Hamiltonian~\eqref{eq:potscatt_app} is quadratic and the Green's function of the  lead electrons can be readily obtained relying on the path integral formalism~\cite{altland_condensed_2006}. The partition function corresponding  to Eq.~\eqref{eq:potscatt_app} reads
\begin{equation}\label{eq:zlfl}
\mathcal Z=\int \mathcal D\left[c,c^\dagger\right]e^{-\mathcal S_{\rm LFL}\left[c,c^\dagger\right]}\,,
\end{equation}
where $\mathcal S_{\rm LFL}\left[c,c^\dagger\right]$ is the action of the system, which reads
\begin{equation}\label{eq:actionres_potscatt}
\mathcal S_{\rm LFL}=\int_0^\beta d\tau\left\{ -\sum_{k}c^\dagger_{k}(\tau)G_k^{-1}(\tau)c_{k}(\tau)+W\sum_{k\neq k'}c^\dagger_k(\tau)c_{k'}(\tau)\right\}\,,
\end{equation}
where we introduced the free propagator $G_k^{-1}(\tau)=-\partial_\tau-\varepsilon_k$ and in which  $c_{k}$ is a Grassmann variable. It is practical to switch to the frequency representation $c_{k\sigma}(\tau)=\frac1\beta\sum_{i\omega_n}e^{-i\omega_n\tau}c_{k\sigma}(i\omega_n)$, where we defined the fermionic Matsubara frequencies $i\omega_n=(2n+1)\pi/\beta$, $n\in \mathbb Z$. They satisfy the anti-periodicity property $c(\beta)=-c(0)$ and lead to
\begin{equation}\label{eq:actionresfreq_potscatt}
\mathcal S_{\rm LFL}=\sum_{i\omega_n}\left\{ -\sum_{k}c^\dagger_{k}(i\omega_n)G_k^{-1}(i\omega_n)c_{k}(i\omega_n)+W\sum_{k\neq k'}c^\dagger_k(i\omega_n)c_{k'}(i\omega_n)\right\}\,,
\end{equation}
with  $G_k^{-1}(i\omega_n)=i\omega_n-\varepsilon_k$, which recovers the usual retarded/advanced Green's functions by performing  the analytical continuation $i\omega_n\rightarrow\omega\pm i0^+$. The full Green's function $G_{kk'}(i\omega_n)=-\av{c_k(i\omega_n)c^\dagger_k(i\omega_n)}$ is derived by expanding the partition function~\eqref{eq:zlfl}  in the coupling $W$ and by applying Wick's theorem~\cite{wick_evaluation_1950}. The pertubation expansion of $G_{kk'}(i\omega_n)$ has the simple form  
\begin{align}
G_{kk'}(i\omega_n)&=\frac{\delta_{kk'}}{i\omega_n-\epsilon_k}+\frac1{i\omega_n-\epsilon_k}\frac1{i\omega_n-\epsilon_{k'}}W\Big[1+\Sigma(i\omega_n)+\Sigma^2(i\omega_n)+\Sigma^3(i\omega_n)+\ldots\Big]\,,
\end{align}
in which we introduced the self-energy
\begin{align}
\Sigma(i\omega_n)&=\sum_p\frac {W(\varepsilon_d)}{i\omega_n-\epsilon_p}.
\end{align}
Using the definition~\eqref{eq:tdef}, the T-matrix  thus reads 
\begin{equation}
\mbox T(z)=\frac{W}{1-\Sigma(z)}.
\end{equation}
Making the  analytical continuation $i\omega_n\rightarrow \omega+i0^+$ and considering a constant density of states $\nu_0$ for the lead electrons, we obtain
\begin{align}\label{eq:tapp}
\mbox T(\omega+i0^+)&=\frac {W(\varepsilon_d)}{1+i\pi\nu_0W}=\frac{W(\varepsilon_d)}{\sqrt{1+[\pi\nu_0W(\varepsilon_d)]^2}}e^{i\delta_W}\,,
\end{align}
in which we introduced the phase-shift
\begin{equation}\label{eq:deltapotscatt_app}
\delta_W=-\arctan\left(\pi\nu_0W\right)\,.
\end{equation}
Applying the definition of the phase-shift given in Eq.~\eqref{eq:deltath}, one finds Eq.~\eqref{eq:deltapotscatt} in the main text. As a consistency check, substituting Eq. \eqref{eq:tapp} in Eq.~\eqref{eq:stmatrix}, we find  that the scattering matrix reads $\mbox{S}=e^{2i\delta}$.

\setcounter{equation}{0}    
\renewcommand{\theequation}{\thesection\arabic{equation}}
\renewcommand{\thesection}{\Alph{section}}

\setcounter{figure}{0}    
\renewcommand{\thefigure}{\thesection\arabic{figure}}

\section{Self-consistent description- of a 2DEG quantum {\it RC} circuit}\label{app:nonint}

In Section~\ref{sec:self}, we shortly review the self-consistent scattering theory of the mesoscopic capacitor and, in Section~\ref{sec:resonant}, we show how equivalent results can be derived with a Hamiltonian formulation. 


\begin{figure}[ht!!]
\begin{center}
\includegraphics[width=.84\textwidth]{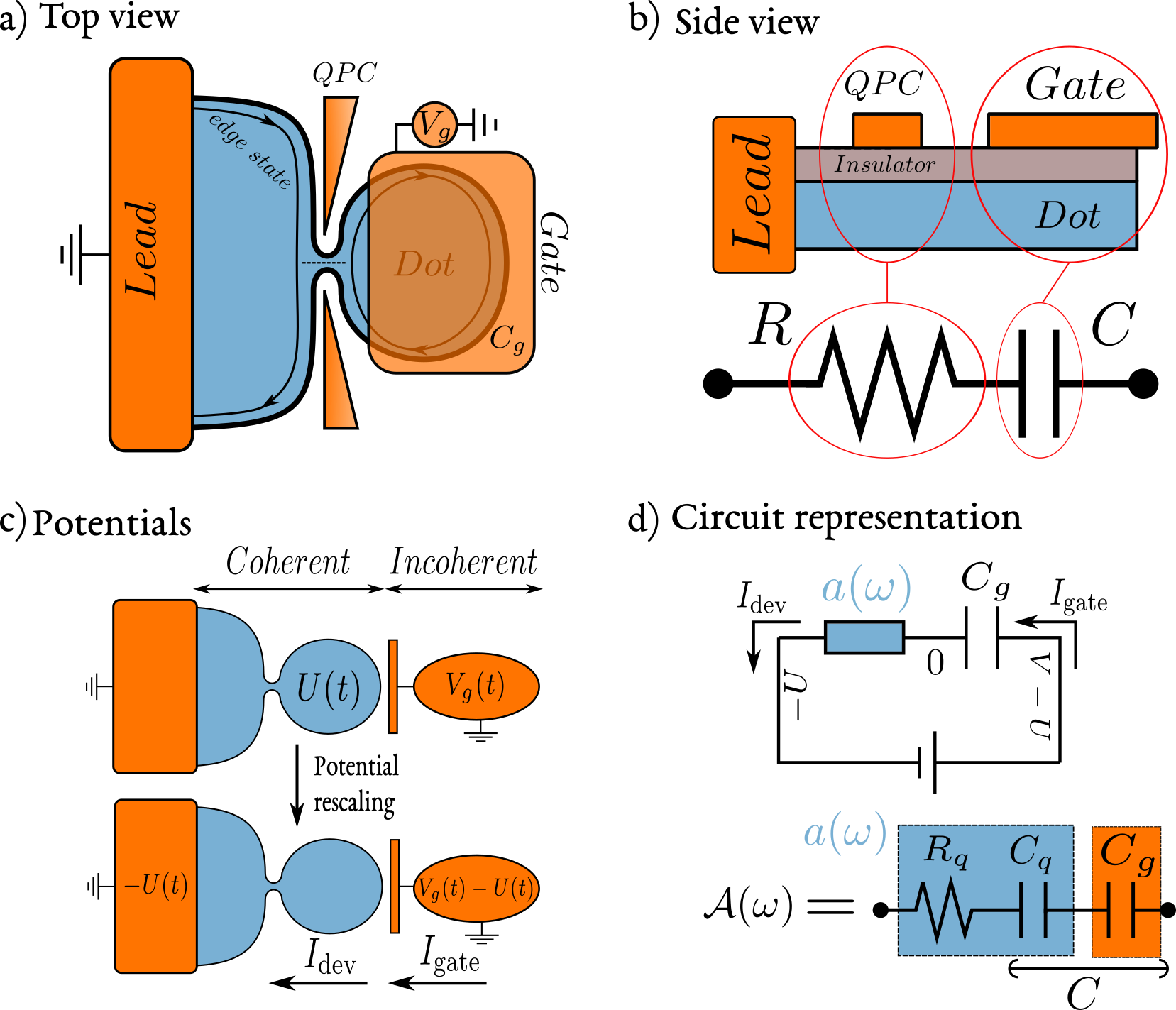}\vspace{5mm}
\caption{{ a-b)} Schematic representation of the quantum {\it RC} circuit.  Electrons in the edge states of a 2DEG in the integer quantum Hall regime tunnel inside a quantum dot through a QPC. The dot is driven by a top metallic gate. The dot and the gate are separated by an insulator and cannot exchange electrons, thus forming the two plates of a capacitor $C$. The QPC is a resistive element of resistance $R$. These two circuit elements are in series and define a quantum coherent {\it RC} circuit. { c-d)} The blue region of the two-dimensional gas 
is phase coherent . The top metallic gate is an incoherent metal, driven by a time-dependent gate potential $V_g(t)$, which induces an unknown uniform potential $U(t)$ on the dot.  The classical circuit analogy in {(d)} is made possible by shifting all energies by $-U(t)$. The whole device behaves as a charge relaxation resistance $R_{\rm q}$ in series to a total capacitance $C$: series of a quantum and geometrical capacitance $C_{\rm q}$ and $C_{\rm g}$.}\label{fig:setup}\label{fig:gauge}
\end{center}
\end{figure}

\subsection{Self-consistent theory}\label{sec:self}
In this section, we shortly review the description of the mesoscopic capacitor in the seminal works by B\"uttiker, Thomas and Pr\^etre~\cite{buttiker_dynamic_1993,buttiker_mesoscopic_1993,pretre_dynamic_1996,gabelli_coherentrccircuit_2012}, based on a self-consistent extension of the Landauer-B\"uttiker scattering formalism~\cite{landauer_electrical_1970,landauer_spatial_1988,buttiker_generalized_1985}. The following discussion is also inspired from Refs. \cite{gabelli_violation_2006,gabelli_mise_2006,feve_quantification_2006}.

Figure~\ref{fig:setup} illustrates the intuition behind the interpretation of the mesoscopic capacitor as a quantum analog of a classical {\it RC} circuit.  The top metallic gate (a classical metal) and the quantum dot cannot exchange electrons. These two components make up the two plates of a capacitor on which electrons accumulate according to variations of the gate potential $V_{\rm g}$. The value of the capacitance depends on the geometry of the contact and, in  the experiment of Ref. \cite{gabelli_violation_2006}, the \textit{geometrical capacitance} $C_{\rm g}$ was estimated $\sim 10-100~ \mbox{fF}$. This capacitance is in series with a quantum point contact. As mentioned in the main text, direct transport measurements \cite{van_wees_quantized_1988,wharam_addition_1988} consider QPCs as  resistive elements of resistance  $R_{\rm DC}= h/e^2D$, where $D$ is the QPC transparency.  The above considerations suggest the interpretation of the device in Fig.~\ref{fig:setup} as an {\it RC} circuit. The admittance of a classical {\it RC} circuit reads 
\begin{equation}\label{eq:GRC}
\mathcal A(\omega)=\frac{-i\omega C}{1-i\omega RC}=-i\omega C\big(1+i\omega CR\big)+O\big(\omega^3\big)\,.
\end{equation}
The admittance~\eqref{eq:GRC}  can be calculated with the scattering formalism~\cite{landauer_electrical_1970,landauer_spatial_1988,buttiker_generalized_1985}, which requires to be  adapted to describe the mesoscopic capacitor in Fig.~\ref{fig:setup}. The main problem is the ``mixed'' nature of the quantum {\it RC} device: The mesoscopic capacitor is composed of a phase-coherent part (two-dimensional electron gas +  quantum dot) in  contact to an incoherent  top metallic gate. As  these constituents do not exchange electrons, preventing a direct current, electron transport is only possible by driving the system. We focus on the case of a gate potential oscillating periodically, as Eq.~\eqref{eq:vgt} in the main text,
\begin{equation}\label{eq:vgt_app}
V_{\rm g}(t)=V_{\rm g}+\varepsilon_\omega\cos(\omega\,t)\,. 
\end{equation}
For small oscillation amplitudes $\varepsilon_\omega$, the Landauer-B\"uttiker formalism allows to derive the circuit admittance within linear response theory. In the case of a single conduction mode, the admittance reads \cite{buttiker_dynamic_1993}
\begin{equation}\label{eq:grc}
a(\omega)=\frac{e^2}h\int d\varepsilon \mbox{Tr}\left[1-S^\dagger(\varepsilon)S(\varepsilon+\hbar\omega)\right]\cdot\frac{f(\varepsilon)-f(\varepsilon+\hbar\omega)}{\hbar\omega}\,,
\end{equation} 
in which $f(\varepsilon)$ is the Fermi distribution function, in which we fix to zero the value of the Fermi energy. For one channel, the elastic scattering assumption implies that the matrix $S(\varepsilon)$ reduces to a pure phase $S(\varepsilon)=e^{2i\delta_\varepsilon}$, as electrons entering the dot return to the lead with unit probability. The phase-shift $\delta_\varepsilon$ is related to the dot electron occupation, via the Friedel sum rule~\eqref{eq:friedel_app}. Additionally, this phase is also related to the \textit{dwell-time} that electrons spend in the quantum dot, or Wigner-Smith delay time \cite{wigner_lower_1955,ringel_delayed_2008} 
\begin{equation}\label{eq:wigner}
\frac{\tau(\varepsilon)} h=\frac1{2\pi i}S^\dagger(\varepsilon)\frac{dS(\varepsilon)}{d\varepsilon}=\frac1{\pi}\frac{d\delta_\varepsilon}{d\varepsilon}\,.
\end{equation}
The interpretation of $\tau$ as a dwell-time is illustrated in Section~\ref{sec:cq}. In the limits $T\rightarrow0$ and $\hbar\omega\rightarrow0$, Eq.~\eqref{eq:grc} becomes
\begin{equation}\label{eq:gldwell}
a(\omega)=-i\omega\frac{e^2}h\left[\tau+\frac12i\omega\tau^2+O\big(\omega^2\big)\right]\,.
\end{equation}
The dwell-time $\tau=\tau(0)$ is considered at the Fermi energy. Notice that this expression has the same frequency expansion as Eq.~\eqref{eq:GRC}. Matching term by term, one finds
\begin{align}\label{eq:cr}
C_{\rm q}&=\frac{e^2}{h}\tau\,,&R_{\rm q}&=\frac h{2e^2}\,.
\end{align} 
Such relation between time-delays and circuit elements comes from the fact that electrons arriving  on the dot   at different times are differently phase-shifted, because of the variations in time of the gate potential $V_{\rm g}$. This effect causes a local accumulation of charges, which is responsible for the emergence of quantum capacitive effects, corresponding to $C_{\rm q}$. The time delay of the electron phase $\delta$ with respect to the driving potential $U(t)$ is responsible for energy dissipation, controlled by $R_{\rm q}$.  The characteristic time that an electron spends in the quantum dot is given by $\tau=2R_{\rm q}C_{\rm q}$, twice the {\it RC} time because it includes the charging and relaxation time of the {\it RC} circuit. Notice the emergence of a universally quantized relaxation resistance, regardless of any microscopic detail of the quantum {\it RC} circuit, in contrast with the resistance $R_{\rm DC}=h/e^2D$, sensitive to the transparency $D$ of the QPC and which would be measured in a DC experiment, see also Eq.~\eqref{eq:dcr} in the main text.

Notice that the geometrical capacitance $C_{\rm g}$ does not appear in the previous discussion. The admittance Eq.~\eqref{eq:grc} has been derived by applying linear response theory for the driving potential $U(t)$ in the quantum dot, see Fig. \ref{fig:gauge}.
The potential $U(t)$ does not coincide with the actual driving gate-potential $V_g(t)$. The situation is pictured in Fig.~\ref{fig:gauge}: the geometric capacitance $C_{\rm g}$ leads to a potential drop between gate and dot. In a mean-field/Hartree-Fock treatment, the potential $U(t)$ on the dot  is assumed to be uniform for each electron. This assumption is equivalent to a random phase approximation (RPA), valid for weak interactions or to leading order in a $1/M$ expansion, with $M$ the number of channels connected to the dot~\cite{brouwer_nonequilibrium_2005}. The potential $U$ can then be determined self-consistently from the constraint of charge/current conservation in the whole device. The current $I_{\rm{dev}}$ flowing  in the coherent part of the device has to equal the current $I_{\rm{gate}}$ flowing in the incoherent metallic gate
\begin{equation}\label{eq:current}
I=I_{\rm{dev}}=I_{\rm{gate}}\,.
\end{equation}
As all potentials are defined with respect to an arbitrary energy, they can be shifted by $-eU(t)$, setting the potential to zero in the quantum dot. Thus, the currents in the device and in the metallic gate read
\begin{align}
I_{\rm{dev}}&=-U(\omega)g(\omega)\,,&I_{\rm{gate}}&=-iC_{\rm g}\omega\big[U(\omega)-V_g(\omega)\big]\,.
\end{align}
Applying the current conservation condition \eqref{eq:current}, the potential $U$ can be eliminated, and the admittance of the total device is derived
\begin{equation}\label{eq:totalseries}
\mathcal A(\omega)=-\frac{I}{V_g}=\frac1{\frac1{a(\omega)}+\frac1{-i\omega C_{\rm g}}}\,.
\end{equation} 
Recalling the low frequency behavior of $a(\omega)$ in Eq. \eqref{eq:gldwell}, the above expression shows that the whole device behaves as an RC circuit. Albeit with two capacitances in series, Eq. \eqref{eq:totalseries} still gives a universally quantized $R_{\rm q}=h/2e^2$. The series of $C_{\rm q}$ and $C_{\rm g}$ gives the total capacitance $C$, originally denoted as  \textit{electro-chemical capacitance}~\cite{buttiker_dynamic_1993,buttiker_mesoscopic_1993,pretre_dynamic_1996,gabelli_coherentrccircuit_2012}.

\begin{figure}[ht!!]
\begin{center}
\includegraphics[width=0.8\textwidth]{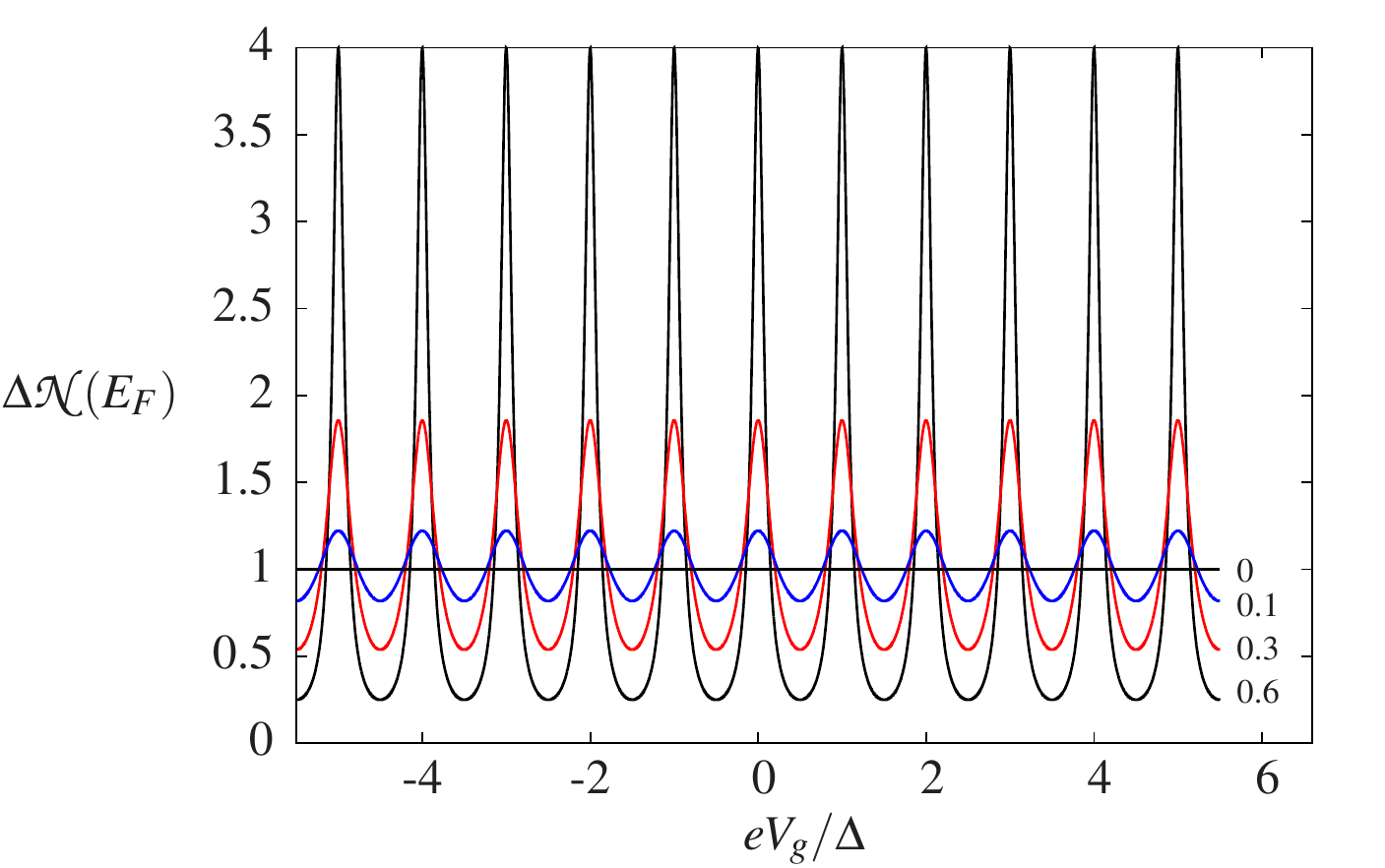}
\caption{Peaked structure of the local density of states $\mathcal N(\varepsilon)$  on the dot as a function of the orbital energy shift  controlled by the gate potential $eV_g$ from Eq. \eqref{eq:scattdos}. $\mathcal N(E_F)$  is plotted for different values of the backscattering amplitude $r$. The progressive opening of the dot drives a transition from a Lorentzian to an oscillatory behavior of $C_{\rm q}$, coherent with the experimental measurements illustrated in Fig. \ref{fig:gabellic}. For a completely transparent dot ($r=0$) the density of states is uniform, which implies $C_0=e^2/\Delta$.}\label{fig:oscillations}
\end{center}
\end{figure}

We can consider a simple model for the mesoscopic capacitor to estimate the behavior of the dwell-times $\tau$ setting the quantum capacitance~\eqref{eq:cr}. We consider the case of Fig.~\ref{fig:setup}, in which electrons propagate in the integer quantum Hall edges inside the quantum dot, see also Fig.~\ref{fig:c0} in the main text.  We label  $\ell$ the length  of the edge state in the quantum dot and $v_F$ the Fermi velocity of the electron.
The dwell-time for electrons of velocity $v_F$ inside the dot is $\tau_{\rm f}=\ell/v_F$. An electric wave of energy $\varepsilon$ acquires  a phase $\phi(\varepsilon)=(\varepsilon-eU)\tau_{\rm f}/\hbar$, when making a tour of the dot. Notice that we had to shift the energy $\varepsilon$ of the electron by $-eU$ because of the potential shift schematized in Fig. \ref{fig:gauge}. The chiral nature of the edge states allows for a one-dimensional representation of the problem, pictured in the right-top of Fig. \ref{fig:c0}. For a quantum well of size $\ell/2$, close to the Fermi energy, the  level spacing $\Delta=h v_F/\ell$ is constant. Thus $\tau_{\rm f}=h/\Delta$ and, substituting in Eq.~\eqref{eq:cr}, leads to the uniform quantum capacitance $C_{\rm q}=e^2\tau_{\rm f}/h$, derived heuristically in the main text, see Eq.~\eqref{eq:c0pauli}. If the reflection amplitude at the entrance  of the dot is $r$ and  $D=1-|r|^2$ the transmission probability, the dot can be viewed as a Fabry-Perot cavity and the phase of the out-coming electron is 
\begin{equation}
\mbox S(\varepsilon)=r-De^{i\phi(\varepsilon)}\sum_{q=0}^\infty r^qe^{iq\phi(\varepsilon)}=\frac{r-e^{i\phi(\varepsilon)}}{1-re^{i\phi(\varepsilon)}}=e^{i2\delta_\varepsilon}\,.
\end{equation}
Applying Eq. \eqref{eq:wigner} we obtain the local density of states 
\begin{equation}\label{eq:scattdos}
\mathcal N(\varepsilon)=\frac{\tau_{\rm f}}h \frac{1-r^2}{1-2r\cos\left[\frac{2\pi}h\big(\varepsilon-eU\big)\tau_{\rm f}\right]+r^2}\,.
\end{equation}
This quantity is plotted in Fig. \ref{fig:oscillations} and reproduces the oscillatory behavior of the capacitance in Fig. \ref{fig:gabellic}. In the limit of small transmission ($D\ll1$ and $r\approx1$), Eq. \eqref{eq:scattdos} reduces to a sum of Lorentzian peaks of width $\hbar\gamma$, $\gamma=D/\tau_{\rm f}$:
\begin{equation}\label{eq:lollo}
\mathcal N(\varepsilon)=\frac2{\pi\hbar\gamma}\sum_n\frac{1}{1+\left(\frac{\varepsilon-eU-n\Delta}{\hbar\gamma/2}\right)^2}\,.
\end{equation}
These peaks are the discrete spectrum of the dot energy levels.  

The above arguments can be readily generalized to the case with  $M$ channels in the lead and in the cavity. In this case, every single channel $\sigma$ can be considered independently.  The admittance~\eqref{eq:gldwell} can be then cast in the form
\begin{equation}\label{eq:gldwellM}
a(\omega)=-i\omega\frac{e^2}h\sum_{\sigma=1}^M\left[\tau_\sigma+\frac12i\omega\tau_\sigma^2+O\big(\omega^2\big)\right]\,.
\end{equation}
The low-frequency expansion of the RC circuit admittance~\eqref{eq:GRC} is then recovered by defining
\begin{align}\label{eq:rqdwell}
C_{\rm q}&=\frac{e^2}h\sum_{\sigma=1}^M\tau_\sigma\,,&R_{\rm q}&=\frac{h}{2e^2}\frac{\sum_\sigma\tau_\sigma^2}{\left(\sum_\sigma\tau_\sigma\right)^2}\,,
\end{align}
in which $\tau_\sigma$ are the dwell-times in the quantum dot of electrons in the $\sigma$-th mode. 


\subsection{Hamiltonian description of the quantum {\it RC} circuit with a resonant level model }\label{sec:resonant}
In this appendix, we study the resonant level model~\eqref{eq:res} as an RC circuit. The generalization of the following calculations to the many-channel case is straightforward and, in particular, we extend to the multi-level case. We thus recover the self-consistent scattering theory  analysis discussed in App.~\ref{sec:self}, corresponding to the limit $C_{\rm g}\rightarrow\infty$ (non-interacting limit).  Our aim is the calculation of the dynamical charge susceptibility
\begin{equation}\label{eq:dyncharge_app}
\chi_{\rm c}(t-t')=\frac i\hbar \theta(t-t')\av{\comm{ N(t)}{ N(t')}}_0\,,
\end{equation}
which leads to the admittance of the circuit $\mathcal A(\omega)=-i\omega e^2\chi_{\rm c}(\omega)$, see also Eq.~\eqref{eq:gcdyn} in the main text. We make use of the path integral formalism as in App.~\ref{app:tmatrix}. The partition function corresponding  to Eq.~\eqref{eq:res} reads
\begin{equation}
\mathcal Z=\int \mathcal D\left[c,c^\dagger, d, d^\dagger\right]e^{-\mathcal S_{\rm Res}\left[c,c^\dagger,d,d^\dagger\right]}\,,
\end{equation}
where $\mathcal S\left[c,c^\dagger,d,d^\dagger\right]$ is the action of the system, which reads
\begin{equation}\label{eq:actionres}
\mathcal S_{\rm Res}=\int_0^\beta d\tau\left\{ -\sum_{k}c^\dagger_{k}(\tau)G_k^{-1}(\tau)c_{k}(\tau)-d^\dagger (\tau)D^{-1}(\tau)d(\tau)+t\sum_{k}\left[c^\dagger_k(\tau)d(\tau)+\mbox{c.c.}\right]\right\}\,,
\end{equation}
with the free propagators 
\begin{align}
G_k^{-1}(\tau)&=-\partial_\tau-\varepsilon_k\,,&
D^{-1}(\tau)&=-\partial_\tau-\varepsilon_d\,,
\end{align}
in which  $c_{k}$ and $d_{l}$ are Grassmann variables. It is practical to switch to the Matsubara frequency representation, which leads to 
\begin{multline}\label{eq:actionresfreq}
\mathcal S_{\rm Res}=\sum_{i\omega_n}\left\{ -\sum_{k}c^\dagger_{k}(i\omega_n)G_k^{-1}(i\omega_n)c_{k}(i\omega_n)-d^\dagger (i\omega_n)D^{-1}(i\omega_n)d(i\omega_n)\right.\\
\left.+t\sum_{k}\left[c^\dagger_{k}(i\omega_n)d(i\omega_n)+d^\dagger(i\omega_n) c_{k}(i\omega_n)\right]\right\}\,,
\end{multline}
with  $G_k^{-1}(i\omega_n)=i\omega_n-\varepsilon_k$ and $D^{-1}(i\omega_n)=i\omega_n-\varepsilon_d$. Reestablishing dimensions, the Fourier transform of Eq.~\eqref{eq:dyncharge_app} reads  
\begin{equation}\label{eq:chicnu}
\chi_{\rm c}(i\nu_n)=\frac1\hbar\int_0^{\hbar\beta}d(\tau-\tau')e^{i\nu_n(\tau-\tau')}\av{ N(\tau) N(\tau')}_0\,,
\end{equation}
 and $\chi_{\rm c}(\omega)$ is recovered  by performing the analytical continuation $i\nu\rightarrow\omega+i0^+$. This function is periodic in imaginary time and its Fourier transform is a function of the bosonic Matsubara frequencies $i\nu_n=2n\pi/\beta$. $N(\tau)=d^\dagger(\tau)d(\tau)$ counts the number of charges on the dot. The cyclic invariance property of the trace implies that $\av{ N(\tau)N(\tau')}_0=f(\tau-\tau')$, allowing us to recast~\eqref{eq:chicnu} in the form
\begin{equation}\label{eq:chiccalc}
\chi_{\rm c}(i\nu_n)=\frac 1\beta\sum_{i\omega_{1,2}}\av{d^\dagger(i\omega_1)d(i\omega_1+i\nu_n)d^\dagger(i\omega_2)d(i\omega_2-i\nu_n)}\,.
\end{equation}
To calculate this expression, we first perform the Gaussian integral of the lead modes in Eq.~\eqref{eq:actionresfreq}, leading to the effective action $\mathcal S'_{\rm Res}$ of the resonant level model 
\begin{align}\label{eq:acres}
\mathcal S_{\rm Res}'&=-\sum_{i\omega_n}d^\dagger(i\omega_n)\mathcal D(i\omega_n)d(i\omega_n)\,,
&
\mathcal D^{-1}(i\omega_n)&=i\omega_n-\varepsilon_d-t^2\sum_{k}G_k(i\omega_n)\,,
\end{align}
In the wide-band approximation the propagator can be written as $
\mathcal D^{-1}(i\omega_n)=i\omega_n-\varepsilon_d+i\Gamma\mbox{sgn}(\omega_n)$,
where we introduced the hybridization constant $\Gamma=\pi\nu_0t^2$. The action~\eqref{eq:acres} is quadratic and the application of Wick's theorem~\cite{wick_evaluation_1950} in Eq.~\eqref{eq:chiccalc} leads to
\begin{equation}\label{eq:kfree}
\begin{split}
\chi_{\rm c}(i\nu_n)&=-\frac1\beta\sum_{i\omega_n}\mathcal D(i\omega_n)\mathcal D(i\omega_n+i\nu_n)\rightarrow-\frac 1{\pi \Gamma}\int_{-\infty}^\infty dxf(\Gamma x+\varepsilon_{d})\frac{2x}{(x^2+1)[x^2-(\omega/\Gamma+i)^2]}\,,
\end{split}
\end{equation}
where the analytical continuation $i\nu\rightarrow\omega+i0^+$ has been performed. At zero temperature, the integral can be calculated analytically, leading to
\begin{equation}\label{eq:chifinal}
\chi_{\rm c}(\omega)=\frac 1{\pi \Gamma}\frac1{\frac\omega{\Gamma}\left(\frac \omega{\Gamma}+2i\right)}\ln\frac{\varepsilon_{d}^2+\Gamma^2}{\varepsilon_{d}^2-(\omega+i\Gamma)^2}\,.
\end{equation}
The low frequency expansion  of this expression matches the one of a classical RC circuit~\eqref{eq:GRC}. Reestablishing correct dimensions $\omega\rightarrow\hbar\omega$, the result recovers Eq. \eqref{eq:cr} obtained within scattering theory
\begin{align}\label{eq:c0rq}
C_0&= \frac{e^2}h\nu(\varepsilon_d)\,,&R_{\rm q}&=\frac h{2e^2}\,,
\end{align}
where $\nu(\varepsilon_d)$ is the density of states associated to the single orbital $\varepsilon_d$ 
\begin{equation}\label{eq:lorenzo}
\nu(\varepsilon_d)=\frac{ 1}{\pi}\frac{\Gamma }{\varepsilon_{d}^2+\Gamma^2}\,.
\end{equation}
The extension  to $M$ channels is straightforward. One can consider an Hamiltonian of the form
\begin{equation}
H_{\rm Res-M ch}=\sum_{k\sigma}\varepsilon_k c^\dagger_{k\sigma}c_{k\sigma}+t\sum_{k\sigma}\left(c^\dagger_{k\sigma}d_\sigma+d^\dagger_\sigma c_{k\sigma}\right)+\sum_\sigma\varepsilon_{\sigma} d^\dagger_\sigma d_\sigma\,,
\end{equation}
with $\sigma=1,\ldots,M$ the number of channels. In this model, each channel can be treated independently and one finds a a generalization of Eq.~\eqref{eq:chifinal}
\begin{equation}\label{eq:chifinalM}
\chi_{\rm c}(\omega)=\sum_{\sigma=1}^M\frac 1{\pi \Gamma}\frac1{\frac\omega{\Gamma}\left(\frac \omega{\Gamma}+2i\right)}\ln\frac{\varepsilon_{\sigma}^2+\Gamma^2}{\varepsilon_{\sigma}^2-(\omega+i\Gamma)^2}\,.
\end{equation}
This expression, when expanded to low frequency, has also the form~\eqref{eq:gldwellM}, where the dwell-times are substituted by the density of states of the channels~\eqref{eq:lorenzo}: $\tau_\sigma\rightarrow\nu_\sigma=\nu(\varepsilon_\sigma)$. One thus finds  expressions for the differential capacitance and the charge relaxation resistance analog to Eq. \eqref{eq:rqdwell}
\begin{align}
C_0&=\frac{e^2}h\sum_{\sigma=1}^M\nu_\sigma\,,&R_{\rm q}&=\frac{h}{2e^2}\frac{\sum_\sigma\nu_\sigma^2}{\left(\sum_\sigma\nu_\sigma\right)^2}\,.
\end{align}

\subsubsection{Multi-level case}\label{app:multilevel}
In this section we carry out the calculation of the quantum dot density of states in the case of a single channel and an infinite number of equally spaced levels in the quantum dot. The  action  reads
\begin{equation}
\mathcal S=\sum_{i\omega_n}\left\{ -\sum_{k}c^\dagger_{k}G_k^{-1}(i\omega_n)c_{k}-\sum_l d_l^\dagger D_l^{-1}(i\omega_n)d_l+t\sum_{kl}\left[c^\dagger_{k}d_l+d^\dagger_l c_{k}\right]\right\}\,,
\end{equation}
with  $G_k^{-1}(i\omega_n)=i\omega_n-\varepsilon_k$ and $D_l^{-1}(i\omega_n)=i\omega_n-\varepsilon_l$. The Gaussian integration of the lead electron modes leads to the effective action
\begin{equation}
\mathcal S'=\sum_{i\omega_n}\left\{ -\sum_l d_l^\dagger(i\omega_n) D_l^{-1}(i\omega_n)d_l(i\omega_n)+t^2\sum_k G_k(i\omega_n)\sum_{ll'}d^\dagger_l (i\omega_n)d_{l'}(i\omega_n)\right\}\,.
\end{equation}
Applying Wick's theorem~\cite{wick_evaluation_1950} the full propagator of the dot electrons is readily obtained 
\begin{equation}\label{eq:fullapp}
\mathcal D_{ll'}(i\omega_n)=\delta_{ll'}D_l(i\omega_n)+D_l(i\omega_n)D_{l'}(i\omega_n)\frac{\gamma(i\omega_n)}{1-\gamma(i\omega_n)\Theta(i\omega_n)}\,,
\end{equation}
where we defined
\begin{align}
\gamma(i\omega_n)&=t^2\sum_k G_k(i\omega_n)\,,&\Theta(i\omega_n)&=\sum_l D_l(i\omega_n)\,.
\end{align}
In the wide band limit $\gamma(i\omega_n)=-i\Gamma\mbox{sgn}(i\omega_n)$. The charge $\av {Q}=e\sum_l\av{d^\dagger_l d_l}$ on the dot is given by
\begin{equation}
\begin{split}
\av {Q}=\frac e\beta\sum_{l,i\omega_n}e^{i\omega_n0^+}\mathcal D_{ll}(i\omega_n)=\frac e{2\pi i}\sum_l\int_{-\infty}^\infty d\varepsilon f(\varepsilon)\left[\mathcal D_{ll}(\varepsilon+i0^+)-\mathcal D_{ll}(\varepsilon-i0^+)\right]\,.
\end{split}
\end{equation}
We write the energy spectrum on the dot as $\varepsilon_l=-eV_g+l\Delta$, with $l\in\mathbb Z$ and $\Delta$ the level spacing. Equation~\eqref{eq:fullapp} is then a function of $\varepsilon+eV_g$. Shifting all energies by  $eV_g$, the differential capacitance $C_0=-\partial \av Q/\partial V_g$ is readily obtained at zero temperature 
\begin{equation}\label{eq:minusan}
C_0=\frac{e^2}{2\pi i}\sum_l\Big[\mathcal D_{ll}(eV_g+i0^+)-\mathcal D_{ll}(eV_g-i0^+)\Big]\,,
\end{equation}
with
\begin{equation}
\begin{split}
\mathcal D_{ll}(eV_g\pm i0^+)&=\frac1{eV_g-l\Delta}\mp\frac1{(eV_g-l\Delta)^2}\frac{i\Gamma}{1\pm i\Gamma\left[\sum_p\frac1{eV_g-p\Delta}\right]}\\
&=\frac1\Delta\left\{\frac1{x+l}\mp\frac{i\Gamma/\Delta}{(x+l)^2}\frac1{1\pm i\pi\frac\Gamma\Delta\coth(\pi x)}\right\}\,,
\end{split}
\end{equation}
where $x=eV_g/\Delta$ and we exploited the fact that $\sum_l\frac1{x+l}=\Psi_0(1-x)-\Psi_0(x)=\pi\coth(\pi x)$, in which $\Psi_0(x)$ is the digamma function. Substituting this expression in Eq. \eqref{eq:minusan}, the sum over levels can be also carried out, leading to
\begin{equation}
C_0=e^2\frac{\pi\Gamma}{2\Delta^2}\frac1{\sin^2\left(\pi \frac{eV_g}\Delta\right)}\left[\frac1{1+i\pi\frac \Gamma\Delta\coth\left(\pi \frac{eV_g}\Delta\right)}+\frac1{1-i\pi \frac\Gamma\Delta\coth\left(\pi \frac{eV_g}\Delta\right)}\right]\,,
\end{equation}
where we relied on the identity: $\sum_l\frac1{(l+x)^2}=\Psi_1(1-x)-\Psi_1(x)=\frac{\pi^2}{\sin^2(\pi x)}$, in which $\Psi_n(x)$ is the polygamma function. Some algebra leads to
\begin{equation}\label{eq:poly}
C_0=\frac{e^2}{\Delta}\frac{2}{\frac\Delta{\pi\Gamma}+\frac{\pi\Gamma}\Delta-(\frac\Delta{\pi\Gamma}-\frac{\pi\Gamma}\Delta)\cos\left(\frac{2\pi e V_g}{\Delta}\right)}\,.
\end{equation}
This quantity is plotted in Fig.~\ref{fig:oscillations} as a function of the gate potential $V_g$ and reproduces the oscillations of the capacitance observed in Fig.~\ref{fig:gabellic}, in the main text.  As we did not consider any many-body interaction to derive $C_0$, this quantity corresponds to the quantum capacitance $C_{\rm q}=e^2\mathcal N(E_F)$ corresponding to the density of states at the Fermi level, see also discussion in Sec.~\ref{sec:cq}. Such density of states was also  derived within scattering theory in App.~\ref{sec:self}. Indeed, Eqs.~\eqref{eq:scattdos} and~\eqref{eq:poly} coincide if one makes  the identification~\footnote{It is useful to recall here that $\tau_{\rm f}=h/\Delta.$}
\begin{align}\label{eq:mapping}
\frac{\pi\Gamma}\Delta&=\frac{(1-r)^2}{1-r^2}=\frac{1-r}{1+r}& \Leftrightarrow&~& r&=\frac{1-\frac{\pi\Gamma}\Delta}{1+\frac{\pi\Gamma}\Delta}\,.
\end{align}
Notice that the fully transparent limit coincides with $\pi\Gamma/\Delta=1$, corresponding to a change of sign of the reflection amplitude  $r$ (remind that we assumed $r$ to be a real number). Additionally, if we consider the tunneling limit $\pi\Gamma/\Delta\ll1$, we can write $r=\sqrt{1-D}$ and, in the low-transparency limit $D\ll1$ one recovers $\pi\Gamma/\Delta=D/4$, which is consistent with the expectation $D\propto t^2$ in the tunneling limit of the Hamiltonian~\eqref{eq:res}. Notice also that the relation~\eqref{eq:mapping} implies $r=-1$ in the $\Gamma\rightarrow \infty$ limit, which can be explained by the formation of bonding and anti-bonding states at the junction between electrons in the lead and in the dot, suppressing tunneling in the dot~\cite{filippone_tunneling_2016}. For a single level and one channel, we recover the universal charge relaxation resistance $R_{\rm q}=h/2e^2$.


\setcounter{equation}{0}    

\section{Useful results of linear response theory}\label{app:power}
In this appendix we remind some useful  properties of linear response theory following Ref.~\cite{cohen-tannoudji_photons_1989}. In Section~\ref{secapp:realodd}, we show that the real/imaginary parts of the dynamical charge susceptibility~\eqref{eq:dyncharge_app} are respectively even/odd functions of the frequency, leading to
\begin{equation}\label{eq:lowfrecgdyn_app}
\mathcal A(\omega)=-i\omega e^2\left\{\chi_{\rm c}+i\mbox{Im}\left[\chi_{\rm c}(\omega)\right]\right\}+\mathcal O(\omega^2)\,,
\end{equation}
that is Eq.~\eqref{eq:lowfrecgdyn} in the main text. In Section~\ref{secapp:power}, we demonstrate that the power dissipated by the quantum {\it RC} circuit in the linear response regime is given  by 
\begin{equation}\label{eq:powerhigh_app}
\mathcal P=\frac12\varepsilon_\omega^2\omega\mbox{Im} \chi_{\rm c}(\omega)\,,
\end{equation} 
that is Eq.~\eqref{eq:powerhigh} in the main text.

\subsection{Parity of the dynamical charge susceptibility}\label{secapp:realodd}
The Lehman representation \cite{bruus_many-body_2004} of the dynamical charge susceptibility $\chi_{\rm c}(\omega)$~\eqref{eq:dyncharge_app} makes explicit its real and imaginary parts. This is obtained from the Fourier transform  of Eq.~\eqref{eq:dyncharge_app}
\begin{equation}\label{eq:fourierchic}
\chi_{\rm c}(\omega)=\frac i\hbar \int_{-\infty}^\infty d(t-t')e^{i(\omega+i0^+)(t-t')}\theta(t-t')\av{\comm{N(t)}{N(t')}}_0\,,
\end{equation}
where the factor $i0^+$ is inserted to regularize retarded functions. Inserting the closure relation with the eigenstates $\ket n$ of energy $E_n$ of the time independent Hamiltonian $\mathcal H_0$, the average can be written as
\begin{equation}
\av{ N(t) N(t')}_0=\sum_{n,m}p_n e^{i\omega_{nm}(t-t')}N_{nm}N_{mn}\,,
\end{equation} 
where $p_n=e^{-\beta E_n}/Z$ is the Boltzmann weight, $\hbar \omega_{nm}=E_n-E_m$ and $N_{nm}=\bra n N\ket m$ the matrix elements of the dot occupation. In this representation,  the Fourier transform~\eqref{eq:fourierchic} reads
\begin{equation}
\chi_{\rm c}(\omega)=-\frac 1\hbar\sum_{nm}p_n N_{nm}N_{mn}\left(\frac1{\omega+i0^++\omega_{nm}}-\frac1{\omega+i0^+-\omega_{nm}}\right)\,.
\end{equation}
Applying the relation  $\frac1{x\pm i0^+}=\mbox P\left[\frac1x\right]\mp i\pi\delta(x),$ with $\mbox P[f(x)]$ the principal value of the function $f(x)$, the real and imaginary part of $\chi_{c}(\omega)$ are readily obtained
\begin{align}
\mbox{Re}\left[\chi_{\rm c}(\omega)\right]&=-\frac 1\hbar\sum_{nm}p_n N_{nm}N_{mn}\left\{\mbox P\left[\frac1{\omega+\omega_{nm}}\right]-\mbox P\left[\frac1{\omega-\omega_{nm}}\right]\right\}\,,\\
\mbox{Im}\left[\chi_{\rm c}(\omega)\right]&=\frac {i\pi}\hbar\sum_{nm}p_n N_{nm}N_{mn}\Big\{\delta\big(\omega+\omega_{nm}\big)-\delta\big(\omega-\omega_{nm}\big)\Big\}\,,
\end{align}
which are respectively an even and odd function of $\omega$. As a consequence, in the low frequency expansion of the dynamical charge susceptibility $\chi_{\rm c}(\omega)=\chi_{\rm c}(0)+\omega\partial_\omega\chi_{\rm c}(\omega)|_{\omega=0}+\mathcal O(\omega^2)$, the linear term in $\omega$ has to coincide with the imaginary part of $\mbox{Im}[\chi_{\rm c}(\omega)]$, leading to Eq.~\eqref{eq:lowfrecgdyn_app}.

\subsection{Energy dissipation in the linear response regime}\label{secapp:power}
In the situation addressed in Section~\ref{sec:meso}, the time dependence of orbital energies in the dot is given by $\varepsilon_d(t)=\varepsilon_d^0+\varepsilon_{\omega}~\cos(\omega t)$. In the time unit,  the systems dissipates the energy
\begin{equation}\label{eq:deltawp}
\delta W=\delta\av{N}\varepsilon_\omega~\cos\big(\omega t\big)\,.
\end{equation}
In the stationary regime, the average power $\mathcal P$ dissipated by the system during the time period $T$ reads
\begin{equation}\label{eq:disp}
\mathcal P=\frac {\varepsilon_\omega}T\int_0^T dt~\frac{d\av{N(t)}}{dt}~\cos(\omega t)\,.
\end{equation}
Neglecting constant contributions,  $\av{N(t)}$ is given by the dynamical charge susceptibility~\eqref{eq:dyncharge_app}
\begin{equation}
\av{N(t)}=\varepsilon_\omega\int_{-\infty}^\infty dt'\chi_{\rm c}(t-t')\cos(\omega t)\,.
\end{equation} 
Substituting this expression in Eq. \eqref{eq:disp}, we obtain
\begin{equation}\label{eq:calcp}
\mathcal P=-i\omega\frac{\varepsilon_\omega^2}4\left[\chi_{\rm c}(\omega)-\chi_{\rm c}(-\omega)\right]+i\omega\frac{\varepsilon_\omega^2}T\int_0^T dt\frac{\chi_{\rm c}(-\omega)e^{2i\omega t}-\chi_{\rm c}(\omega)e^{-2i\omega t}}4\,.
\end{equation}
Expressing $\chi_{\rm c}(\omega)=\mbox{Re}\left[\chi_{\rm c}(\omega)\right]+i\mbox{Im}\left[\chi_{\rm c}(\omega)\right]$ as the sum of its real and imaginary part and applying the parity properties demonstrated in \ref{secapp:realodd}, the first term recovers Eq.~\eqref{eq:powerhigh_app} for the dissipated power
\begin{equation}
\mathcal P=\frac12\varepsilon_\omega^2\omega\mbox{Im}\left[\chi_{\rm c}(\omega)\right]\,,
\end{equation}
while the second term in Eq. \eqref{eq:calcp} reduces to vanishing integrals of $\sin (2\omega t)$ and $\cos(2\omega t)$ over their period. In the case of Section~\ref{sec:quasistatic}, describing the energy dissipated by the LFL effective low-energy theory, we can apply the same considerations by replacing  $\delta\av {  N}$ in Eq. \eqref{eq:deltawp} with the average of the operator $A$, defined in Eq.~\eqref{eq:basta}. One thus derives Eq.~\eqref{eq:bastapower} in the main text.


\reftitle{References}


\externalbibliography{yes}
\bibliography{biblio,new_biblio}



\end{document}